\documentclass[a4paper,11pt]{article}
\pdfoutput=1

\usepackage{jcappub} 
\usepackage[T1]{fontenc} 
\usepackage{multirow}
\usepackage{array, makecell}

\newcommand{\be}{\begin{eqnarray}}
\newcommand{\ee}{\end{eqnarray}}

\title{\boldmath Quantifying baryon effects on the matter power spectrum and the weak lensing shear correlation}

\author[a]{Aurel Schneider,}
\author[b]{Romain Teyssier,}
\author[b]{Joachim Stadel,}
\author[c]{Nora Elisa Chisari,}
\author[d]{Amandine M. C. Le Brun,}
\author[a]{Adam Amara,}
\author[a]{and Alexandre Refregier}
\affiliation[a]{Institute for Particle Physics and Astrophysics, ETH Zurich, Wolfgang Pauli Strasse 27, 8093 Zurich, Switzerland}\affiliation[b]{Institute for Computational Science, University of Zurich, Winterthurerstrasse 190, 8057 Zurich, Switzerland}\affiliation[c]{Department of Physics, University of Oxford, Keble Road, Oxford, OX1 3RH, United Kingdom}\affiliation[d]{AIM, CEA, CNRS, Universit\'e Paris-Saclay, Universit\'e Paris Diderot, Sorbonne Paris Cit\'e, F-91191 Gif-sur-Yvette, France.}

\emailAdd{aurel.schneider@phys.ethz.ch}

\abstract{Feedback processes from baryons are expected to strongly affect weak-lensing observables of current and future cosmological surveys. In this paper we present a new parametrisation of halo profiles based on gas, stellar, and dark matter density components. This parametrisation is used to modify outputs of gravity-only $N$-body simulations (following the prescription of \citet{Schneider:2015wta}) in order to mimic baryonic effects on the matter density field. The resulting \emph{baryonic correction model} relies on a few well motivated physical parameters and is able to reproduce the redshift zero clustering signal of hydrodynamical simulations at two percent accuracy below $k\sim10$ h/Mpc. A detailed study of the baryon suppression effects on the matter power spectrum and the weak lensing shear correlation reveals that the signal is dominated by two parameters describing the slope of the gas profile in haloes and the maximum radius of gas ejection. We show that these parameters can be constrained with the observed gas fraction of galaxy groups and clusters from X-ray data. Based on these observations we predict a beyond percent effect on the power spectrum above $k=0.2-1.0$ h/Mpc with a maximum suppression of 15-25 percent around $k\sim 10$ h/Mpc. As a result, the weak lensing angular shear power spectrum is suppressed by 15-25 percent at scales beyond $\ell\sim 100-600$ and the shear correlations $\xi_{+}$ and $\xi_{-}$ are affected at the 10-25 percent level below 5 and 50 arc-minutes, respectively. The relatively large uncertainties of these predictions are a result of the poorly known hydrostatic mass bias of current X-ray observations as well as the generic difficulty to observe the low density gas outside of haloes.}

\begin{document}
\maketitle

\section{Introduction}
Upcoming galaxy and weak lensing surveys such as {\tt DES}\footnote{\tt www.darkenergysurvey.org}, {\tt LSST}\footnote{\tt www.lsst.org/lsst}, and {\tt Euclid}\footnote{\tt sci.esa.int/euclid} will observe billions of galaxies over a significant portion of the sky, allowing for unprecedented investigations of the standard cosmological model. In order to take full advantage of these exquisite data-sets for constraining fundamental physics, it is crucial to track down and quantify any systematic errors appearing in the theoretical predictions.

Gravity-only numerical simulations provide accurate predictions for collisionless structure formation. In terms of the matter power spectrum, they achieve a precision of better than five percent over the full relevant range in physical scales and redshift \citep{Heitmann:2007hr,Schneider:2015yka,Knabenhans:2018cng,Smith:2018zcj}. However, these simulations implicitly assume that baryons do not affect the large-scale structure formation process, an assumption that has been shown to be incorrect \citep{vanDaalen:2011xb}.

In the past decade, it has become more and more evident that the energy injection of active galactic nuclei (AGN) has the potential to either heat up or push out large amounts of gas, altering the matter clustering signal at cosmological scales. These baryonic effects pose an important systematic for current and future weak-lensing measurements \citep{Semboloni:2011aaa,Huang:2018wpy,Parimbelli:2018yzv} seriously threatening the anticipated progress in terms of cosmology and fundamental physics.

Full hydrodynamical simulations of cosmological volumes are in principle the ideal tool to predict the weak-lensing signal. However, such simulations are unable to resolve and self-consistently calculate the black-hole energy injection governing the interplay between AGN and the galactic gas. AGN feedback energy is therefore dumped into the surrounding gas particles or cells following simple semi-analytical recipes. As a result, there is a variety of different results from hydrodynamical simulations that are inconsistent with each other. In terms of the matter power spectrum, some hydrodynamical simulations predict very strong effects starting at $k\sim0.1$ h/Mpc, right at the scale where modes become nonlinear \citep{vanDaalen:2011xb,Mummery:2017lcn}, while others show no effect until $k\gtrsim 1$ h/Mpc at more than ten times smaller scales \citep[][]{Hellwing:2016ucy,Springel:2017tpz,Chisari:2018prw}.

Next to hydrodynamical simulations, several analytical and semi-analytical approaches have been proposed to quantify the effects of baryons on the weak-lensing signal \citep[e.g.][]{Mohammed:2014lja,Dai:2018vvv}. For example, Refs.~\citep{Semboloni:2011aaa,Fedeli:2014gja,Mohammed:2014mba} show that the qualitative trend of the signal can be obtained with simple modifications of the halo model. More accurate results are obtained by the approach of Ref.~\citep{Mead:2015yca} who achieved good agreement with the OWLS hydrodynamical runs \citep{Schaye:2010aaa} by adding two parameters to their power spectrum estimator. The latter is based on a halo model approach subsequently fitted to $N$-body simulations.

In a previous paper \citep[henceforth ST15]{Schneider:2015wta}, we applied a novel method to displace particles in outputs of $N$-body simulations in order to mimic the effects of baryons on the total matter distribution. The method is based on the idea that the original NFW profiles of galaxy groups and clusters are altered by their gas and stellar components. Strong AGN activity pushes gas out of the halo centres, thereby flattening the total density profiles well beyond the virial radius. As a result, ST15 showed that the baryon-induced modification of the power spectrum is tightly coupled to the amount of observable gas within galaxy groups and clusters. The downside of the ST15 model is that it relies on an analytically motivated gas profile that shows some discrepancies with direct X-ray observations. Furthermore, it is unable to match the measured power spectra of hydrodynamical simulations at a quantitative level \citep{Chisari:2018prw}.

In the present paper, we build upon ST15 and present an updated version of the \emph{baryonic correction} (BC) model. While relying on the original algorithm to displace particles in outputs of $N$-body simulations, we adopt a completely new parametrisation for the model components. The gas profile is now described by a simple power-law with additional central core and with a steep truncation at the outskirts. We show that this parametrisation leads to good agreement with X-ray observations and is able to reproduce the results of hydrodynamical simulations.

The main advantage of the BC model is that it provides a realisation of the total matter density field based on a simple and physically motivated parametrisation for the baryonic effects. In the future, this will allow to perform full cosmological parameter estimates including baryon nuisance parameters that have priors from direct gas observations. Furthermore, the method operates on outputs of $N$-body simulations and is therefore not limited to two-point statistics as it is the case for approaches using the halo model.

The paper is organised as follows: In Sec.~\ref{sec:BCM} we describe the basics of the model, specifying the parametrisation of its different components. In Sec.~\ref{sec:PS} we quantify the effects of individual parameters on the matter power spectrum and we provide a comparison with the OWLS hydrodynamical runs. In Sec.~\ref{sec:constraints}, the model parameters are constrained with X-ray observations assuming three different values for the hydrostatic mass bias of X-ray observations. Based on the best fitting parameters, we then provide predictions for the matter power spectrum as well as the weak-lensing angular shear power spectrum and real-space correlation in Sec.~\ref{sec:predictions}. We conclude our work in Sec.~\ref{sec:conclusions}. The Appendices \ref{hydrosims} and \ref{systematics} are dedicated to a comparison with other hydrodynamical simulations and to a discussion of potential systematical uncertainties of the model.

\section{Baryonic correction model}\label{sec:BCM}
In this section we start by introducing the basic principles of the baryonic correction (BC) model. The parametrisation of each matter component (gas, stars, and dark matter) are discussed in separate subsections. Before starting, let us highlight that throughout this paper we define the virial radius of a halo ($r_{200}$) with respect to the over-density criterion of 200 times the \emph{critical density} of the universe ($\rho_{\rm crit}$). This means that $r_{200}$ is obtained by equating $\rho(<r_{200})=200\rho_{\rm crit}$ and the halo mass is therefore given by $M_{200}=4\pi 200\rho_{\rm crit}r_{200}^3$/3.

\subsection{Basic principle}\label{sec:basicprinciple}
In gravity-only $N$-body simulations both dark matter and baryons are assumed to only interact gravitationally. With this simplifying assumption, the profiles of haloes are well described by combining a truncated NFW profile ($\rho_{\rm NFW}$) \citep{Navarro:1995iw,Baltz:2007vq} with a 2-halo density component ($\rho_{\rm 2h}$) \citep{Diemer:2014gba}, i.e.,
\be\label{rhodmo}
\rho_{\rm dmo}(r)=\rho_{\rm nfw}(r)+\rho_{\rm 2h}(r).
\ee
The two terms of this total dark-matter-only (dmo) halo density profile are described in Sec.~\ref{sec:dmo}.

In a more realistic scenario including baryons, gas is allowed to cool and to form stars at the centres of haloes. At the same time, feedback effects from active galactic nuclei may heat up the gas and push it towards the outskirts of haloes. It is therefore important to separately model the dark matter, the gas, and the stellar halo components. We define a more realistic dark-matter-baryon (dmb) profile of the form
\be\label{rhodmb}
\rho_{\rm dmb}(r)= \rho_{\rm gas}(r) + \rho_{\rm cga}(r) + \rho_{\rm clm}(r) + \rho_{\rm 2h}(r),
\ee
where the subscripts stand for the gas (gas), the central galaxy (cga), and the collisionless matter components (clm), respectively. The latter is dominated by the dark matter component but also contains the stellar halo and the satellite galaxies, which are assumed to act as collisionless components following the same profile as the dark matter. The profiles of the different model components of Eq.~(\ref{rhodmb}) are described in Secs.~\ref{sec:stars} - \ref{sec:clm}.

With the density profiles at hand, we can readily obtain the mass profiles by integrating over the volume, i.e.,
\be\label{massprofile}
M_{\chi}(r)=4\pi\int_0^r ds s^2\rho_{\chi}(s),\hspace{1cm}\chi=\{\rm nfw,\,gas,\,cga,\,clm,\,2h,\,dmo,\,dmb\}.
\ee
Furthermore, it is possible to define the total halo mass as follows
\be\label{Mtot}
M_{\rm tot}=M_{\rm gas}(\infty) + M_{\rm cga}(\infty)+M_{\rm clm}(\infty)=M_{\rm nfw}(\infty),
\ee
implying that the mass profiles of all individual matter components converge towards large radii. This is of course not the case for the total dark-matter-only and dark-matter-baryon mass profiles which also contain the diverging 2-halo mass component ($M_{\rm 2h}$). However, both $M_{\rm dmo}(r)$ and $M_{\rm dmb}(r)$ are bijective functions that can be inverted to obtain $r_{\rm dmo}(M)$ and $r_{\rm dmb}(M)$. With this at hand, we can define a displacement function
\be\label{displ}
d(r_{\rm dmo}|M,c) \equiv r_{\rm dmb}(M)-r_{\rm dmo}(M)
\ee
for each halo of mass $M$ and concentration $c$. The displacement function corresponds to the distance that mass shells have to be displaced radially in order to recover the profile $\rho_{\rm dmb}(r)$ starting from an original profile $\rho_{\rm dmo}(r)$. Hence, applying Eq.~(\ref{displ}) to all simulation particles around haloes allows us to correct the outputs of $N$-body simulations, mimicking the effects of baryons on the total density field. More details about the method can be found in ST15.

\subsection{Dark-matter-only profile}\label{sec:dmo}
We start by defining the two components of the total dark-matter-only profile ($\rho_{\rm dmo}$) given by Eq.~(\ref{rhodmo}). The first component consists of a truncated NFW-profile
\be\label{rhoNFW}
\rho_{\rm nfw}(x)=\frac{\rho_{\rm nfw,0}}{x(1+x)^2}\frac{1}{(1+y^2)^2}\,,
\ee
where $x\equiv r/r_s$ and $y\equiv r/r_{t}$ \citep{Baltz:2007vq}. The scale radius $r_s$ is connected to $r_{200}$ via the halo concentration $c\equiv r_{200}/r_s$. The truncation radius $r_t$ denotes the edge of the halo and is well approximated by $r_t\equiv \varepsilon\times r_{200}$ with $\varepsilon=4$ \citep{Oguri:2011aaa,Schneider:2015wta}. We investigate different values of $\varepsilon$ in Appendix~\ref{systematics}.

The second component of the dark-matter-only profile (Eq.~\ref{rhodmo}) is the 2-halo term, which accounts for the fact that haloes are predominately located in high-density environments. The 2-halo term can be written as
\be
\rho_{\rm 2h}(r)=\left[b(\nu)\xi_{\rm lin}(r)+1\right]\Omega_{m}\rho_{\rm crit} ,
\ee
where $\xi_{\rm lin}(r)$ is the linear matter correlation function and $b(\nu)$ is the halo bias. The former is obtained via a Fourier transformation of the linear power spectrum ($P_{\rm lin}$)
\be
\xi_{\rm lin}(r)=\int d^3k P_{\rm lin}(k)\frac{\sin(kr)}{kr},
\ee
while the latter can be derived via the excursion-set formalism with the peak-background split approach \citep{Sheth:1999mn}
\be\label{halobias}
b(\nu) = 1+\frac{q\nu^2-1}{\delta_c}+\frac{2p}{\delta_c\left[1+(q\nu^2)^p\right]},\hspace{1cm}\nu=\delta_c\left[\int d^3k P_{\rm lin}(k)W^2(kR)\right]^{-0.5},
\ee
where $p=0.3$, $q=0.707$, $\delta_c=1.686/D(z)$, and $D(z)$ being the linear growth rate. See e.g. Ref. \citep[][]{Hayashi:2007uk} for more information about the 2-halo term.

\subsection{Stellar profile}\label{sec:stars}
We now consider the components of the total dark-matter-baryon profile ($\rho_{\rm dmb}$) defined by Eq.~(\ref{rhodmb}). The density profile of the bright galaxy in the halo centre can be described by a power-law profile with exponential cutoff, i.e.
\begin{equation}\label{rhocga}
\rho_{\rm cga}(r)=\frac{f_{\rm cga}M_{\rm tot}}{4\pi^{3/2}R_h}\frac{1}{r^2}\exp\left[-\left(\frac{r}{2R_h}\right)^2\right],\hspace{1cm}R_h=0.015\,r_{\rm 200},
\end{equation}
where $R_h$ is the stellar half-light radius \citep[see e.g.][]{Schneider:2015wta,Kravtsov:2014sra,Mohammed:2014mba}. Next to the central galaxy, satellite galaxies and the stellar halo emitting the intra-cluster light also contribute to the total stellar budget of a given halo. Since satellite galaxies and the stellar halo are collisionless components, they are expected to behave in the same way as the dark matter component, forming a NFW profile (which is, however, allowed to contract and expand under the influence of the central galaxy and the gas profile, see Sec.~\ref{sec:clm}).

The total abundance of stars within a halo is given by $f_{\rm star}\equiv f_{\rm cga}+f_{\rm sga}$, where $f_{\rm cga}$ refers to the stars of the central galaxy and $f_{\rm sga}$ to the satellite population including the stellar halo. The stellar fractions can be parametrised as follows:
\be\label{stellarfraction}
f_{\rm star}(M_{200})= A\left(\frac{M_1}{M_{200}}\right)^{\eta_{\rm star}},\hspace{0.9cm}f_{\rm cga}(M_{200})= A\left(\frac{M_1}{M_{200}}\right)^{\eta_{\rm cga}},
\ee
with $A=0.09$, $M_1=2.5\times10^{11}$ $M_{\odot}$/h, and with the consistency relation $\eta_{\rm star}<\eta_{\rm cga}$, that guarantees $f_{\rm star}$ to be larger than $f_{\rm cga}$ for all relevant scales. The functional form of Eq.~(\ref{stellarfraction}) corresponds to a simplified version of the fit provided by \citet{Moster:2012fv}. In Sec.~\ref{sec:constraints} we will show that these parametric functions provide a good match to results from the literature based on abundance matching.

\subsection{Gas profile}\label{sec:gas}
The gas profile is parametrised by the following function
\be\label{rhogas}
\rho_{\rm gas}(r)=\frac{\rho_{\rm gas,0}}{(1+u)^{\beta}(1+v^2)^{(7-\beta)/2}},
\ee
where $u\equiv r/r_{\rm co}$ and $v\equiv r/r_{\rm ej}$. The profile is characterised by a central core (with core radius $r_{\rm co}$) followed by a power-law decrease (of slope $\beta$) and a truncation at the maximum gas ejection radius ($r_{\rm ej}$). Beyond the gas ejection radius, the gas density is forced to decrease at the same rate as the truncated NFW profile (see Eq.~\ref{rhoNFW}). The normalisation parameter ($\rho_{\rm gas, 0}$) is given by
\be
\rho_{\rm gas, 0}=f_{\rm gas} M_{\rm tot}\left[4\pi\int_0^{\infty} dr \frac{r^2}{(1+u)^{\beta}(1+v^2)^{(7-\beta)/2}}\right]^{-1}
\ee
where $f_{\rm gas}\equiv\Omega_{\rm b}/\Omega_{\rm m} - f_{\rm star}$ is the universal gas fraction and $M_{\rm tot}$ is the total halo mass (see Eq.~\ref{Mtot}). The two characteristic radii of Eq.~(\ref{rhogas}) are defined as follows:
\be\label{characteristicradius}
r_{\rm co} \equiv \theta_{\rm co}r_{\rm 200},\hspace{1cm}r_{\rm ej} \equiv \theta_{\rm ej}r_{\rm 200},
\ee
where $\theta_{\rm co}$ and $\theta_{\rm ej}$ are free model parameters that are constrained to be within the bounds $\theta_{\rm co}<1$ and $\theta_{\rm ej}>1$ for consistency reasons. To further simplify the analysis of the present study, we fix the core parameter to
\be\label{thco}
\theta_{\rm co}=0.1\,.
\ee
This value is in agreement with observations (see e.g. the characteristic break at $\sim0.1\times r_{\rm 200}$ visible in the observed X-ray profiles from {\tt XMM-Newton} and {\tt Chandra} \citep{Croston:2008yr,Sanders:2017lce}). In Sec.~\ref{Xrayprofiles}, we furthermore show that Eq.~(\ref{thco}) leads to profiles in good agreement with stacked galaxy group and cluster data from Ref.~\citep{Eckert:2015rlr}.  The effects of other values for $\theta_{\rm co}$ on the cosmological density field are discussed in Appendix~\ref{systematics}. Here we focus on the ejection radius instead (parametrised by $\theta_{\rm ej}$), which affects the density profiles beyond the virial radius and is therefore highly relevant for large-scale statistics of the universe.

Finally, the slope of the gas profile ($\beta$) consists of another free model parameter. The slope is allowed to have both positive and negative values but is bound from above, i.e., $\beta \leq 3$. This means that the gas profile can be shallower than the NFW profile but never steeper. From observations it is well known that $\beta$ depends on halo mass, i.e., it is shallower for galaxy groups compared to clusters \citep{Eckert:2015rlr}. We therefore assume an explicit halo mass dependence of the form
\be\label{beta}
\beta(M_{200})=3-\left(\frac{M_{\rm c}}{M_{\rm 200}}\right)^{\mu}
\ee
with two free parameters $M_{\rm c}$ and $\mu$. The function $\beta(M_{\rm 200})$ approaches 3 at scales above $M_c$ and decreases towards smaller halo masses. In Sec.~\ref{sec:constraints} we show that this functional form provides a good match to data from X-ray observations.

\subsection{Collisionless matter profile}\label{sec:clm}
The collisionless matter component ($\rho_{\rm clm}$) is dominated by dark matter but it also contains all satellite galaxies and unbound stars within the halo. Based on results from gravity-only simulations, we expect the collisionless matter component to assemble building a NFW profile (as modelled in Sec.~\ref{sec:dmo}). However, the presence of a central galaxy and a gas component has a gravitational effect on the collisionless matter which is commonly referred to as adiabatic relaxation (i.e. adiabatic contraction or expansion). 

Early work on adiabatic relaxation assumed shells of collisionless matter to either contract or expand following angular momentum conservation, i.e., $r_i M_i=r_fM_f$, where $M_i$ and $M_f$ are the initial (dark-matter-only) and final (dark-matter-baryon) enclosed mass \citep{Barnes:1984aaa,Blumenthal:1985qy}. More recently, it has been shown that the effect of relaxation is more accurately captured by the relation
\be\label{ACmodel}
\frac{r_f}{r_i}-1=a\left[\left(\frac{M_i}{M_f}\right)^n-1\right],
\ee
where $a$ and $n$ are free model parameters. Ref.~\citep{Abadi:2009ve} found best agreement with simulations for the values of $a=0.3$ and $n=2$. Ref.~\citep{Teyssier:2011aaa}, on the other hand, found best results for $a=0.68$ and $n=1$. In the present work, we use the former as a default implementation for collisionless contraction and expansion, but we have checked that both models give nearly identical results in terms of clustering statistics (see also discussion in Appendix~\ref{systematics}). For the mass terms $M_i$ and $M_f$, we furthermore use
\be
\begin{split}
M_i&\equiv M_{\rm nfw}(r_i),\\
M_f&\equiv f_{\rm clm}M_{\rm nfw}(r_i)+ M_{\rm cga}(r_f)+M_{\rm gas}(r_f),
\end{split}
\ee
where $f_{\rm clm}=\Omega_{\rm dm}/\Omega_{\rm m}+f_{\rm sga}$ (with $f_{\rm sga}=f_{\rm star}-f_{\rm cga}$). It is possible to iteratively solve Eq.~(\ref{ACmodel}) for $\zeta\equiv r_f/r_i$ thereby obtaining the updated mass and density profiles of the collisionless component, i.e.,
\be\label{rhoclm}
M_{\rm clm}(r) = f_{\rm clm}M_{\rm nfw}(r/\zeta),\hspace{1cm}\rho_{\rm clm} (r)=\frac{f_{\rm clm}}{4\pi r^2}\frac{d}{dr}M_{\rm nfw}(r/\zeta).
\ee
The final collisionless matter profile ($\rho_{\rm clm}$) is steeper at small radii and somewhat shallower at large radii compared to the truncated NFW profile. This is due to both the presence of a exponential stellar component in the centre and an extended gas component in the outskirts of the halo.

\begin{table}
\small
\caption{Parameters of the \emph{baryonic correction model} including a short description and a reference to the corresponding matter component and to the equation in the text. The status specifies whether the parameter is kept free or is fixed to a given value in the model. \label{tab:paramvalues}}
    \begin{center}
    \setcellgapes{3pt}\makegapedcells
 \begin{tabular}{p{0.95cm}p{1.0cm}p{9.0cm}p{1.4cm}p{0.95cm}}
\hline
 Name & Comp. & Description & Equation & Status  \\
 \hline
 \hline
  $\theta_{\rm ej}$ & Gas & Parameter specifying the maximum radius of gas ejection relative to the virial radius. & (\ref{rhogas}) & free\\
 \hline
  $\theta_{\rm co}$ & Gas & Parameter specifying the core radius of the gas profile relative to the virial radius. & (\ref{rhogas}) & fixed\\
 \hline
 $M_c$ & Gas & Parameter related to the slope of the gas profile: defines the characteristic mass scale where the slope becomes shallower than minus three. &  (\ref{beta}) & free \\
 \hline
 $\mu$ & Gas & Parameter related to the slope of the gas profile: defines how fast the slope becomes shallower towards small halo masses. &  (\ref{beta})& free \\
\hline
 $A$, $M_1$ & Star & Parameters related to the stellar fractions: normalisation and slope of the power-law describing the halo mass dependence. & (\ref{stellarfraction}) & fixed \\
 \hline
 $\eta_{\rm star}$ & Star & Parameter specifying the total stellar fraction within a halo (including central galaxy, satellites, and halo stars). & (\ref{stellarfraction})& free \\
 \hline
 $\eta_{\rm cga}$ & Star & Parameter specifying the stellar fraction of the central galaxy. & (\ref{stellarfraction}) & free \\
 \hline
   $R_h$ & Star & Parameter specifying the truncation radius of the central galaxy. & (\ref{rhocga}) & fixed \\
 \hline
  $\varepsilon$ & DM & Parameter specifying the truncation radius of the NFW profile. & (\ref{rhoNFW}) & fixed \\
 \hline
 $a$, $n$ & DM & Parameters related to adiabatic relaxation of the dark matter (including galaxy satellites and halo stars). & (\ref{ACmodel}) & fixed \\
 \hline
 $q$, $p$ & 2-halo & Standard parameters specifying the 2-halo term (excursion-set modelling). & (\ref{halobias}) & fixed \\
 \hline\end{tabular}
\end{center}
\end{table}

\subsection{Summary and example case}
So far, we have defined a dark-matter-only ($\rho_{\rm dmo}$) and a dark-matter-baryon profile ($\rho_{\rm dmb}$) which fully specify the baryonic correction model. Each of these two profiles contains a number of parameters which are summarised in Table \ref{tab:paramvalues}. The table provides a short description of each parameter together with a link to the relevant equation in the text. Furthermore, it is explicitly specified whether a given parameter is allowed to vary freely or whether it is kept at a fixed value.

Before investigating the effects of the BC parameters on the large-scale structure, we will now illustrate the profile shapes using the example of a small cluster halo of $M_{\rm 200}=10^{14}$ M$_{\odot}$/h. The focus will be on the slope and the maximum ejection radius of the gas profile, which are expected have the strongest effect on cosmological observables. Regarding the slope, we directly relate to $\beta$ instead of the subsequent parameters $M_c$ and $\mu$ in this section. Note, however, that there is a set of values for $M_c$ and $\mu$ for each value of $\beta$ (see Eq.~\ref{beta}). The other free model parameters related to the stellar fractions are kept at default values of $\eta_{\rm star}=0.3$ and $\eta_{\rm cga}=0.6$.

\begin{figure}[tbp]
\center{
\includegraphics[width=.97\textwidth,trim={0.7cm 1.0cm 1.5cm 1.0cm}]{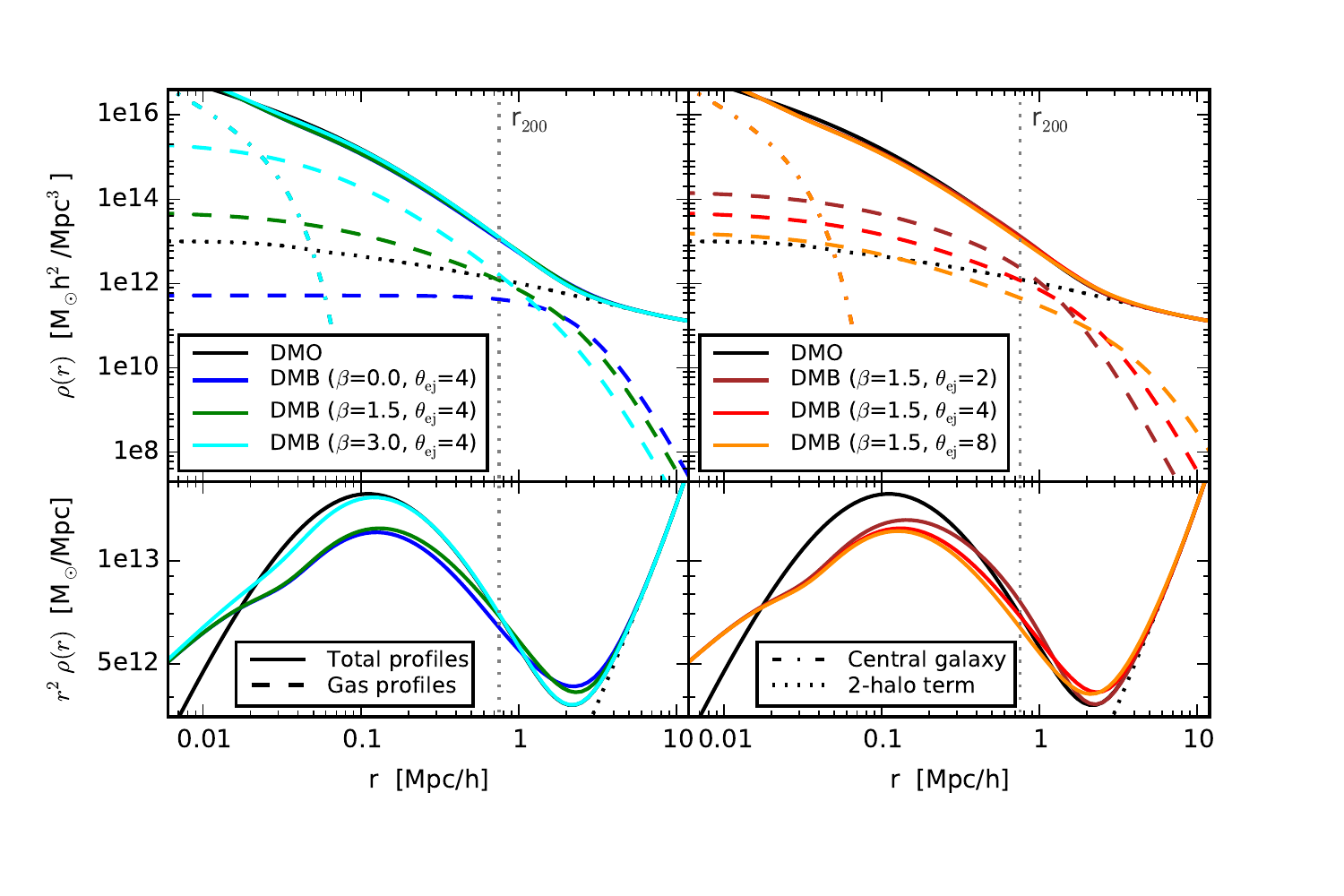}
\caption{\label{fig:profiles}Example profiles of a small cluster with $M_{200}=10^{14}$ M$_{\odot}$/h assuming $\eta_{\rm star}=0.3$, $\eta_{\rm cga}=0.6$, and the gas parameters listed in the caption. The top-panels show the total dark-matter-only (DMO) and dark-matter-baryon (DMB) profiles (solid lines), as well as the individual stellar and gas components (dash-dotted and dashed lines). The 2-halo term is illustrated as dotted black line. The bottom-panels focus on the total density profile multiplied by $r^2$ to improve visibility. The parameter $\beta$ can be related to the parameters $M_c$ and $\mu$ via Eq.~(\ref{beta}).}}
\end{figure}

The top panels of Fig.~\ref{fig:profiles} show the stellar, gas, and total profiles for different choices of $\beta$ (left) and $\theta_{\rm ej}$ (right). As expected, the value of $\beta$ sets the slope of the profile outside of the core radius (i.e. beyond 0.07 Mpc/h), while $\theta_{\rm ej}$ determines how far out the gas profile extends. The bottom panels of Fig.~\ref{fig:profiles} focus on the total profiles multiplied by radius squared to reduce the dynamic range and improve the visibility of the differences between the dark-matter-only and the dark matter-baryon cases.

At very small radii below $r\sim0.02$ Mpc/h, the total dark-matter-baryon density is strongly enhanced compared to the dark-matter-only case. This is due to the very steep profile of the stars in the central galaxy. Additionally, the central galaxy leads to adiabatic contraction of the dark matter component further steepening the inner profile. In the medium regime above $r\sim0.02$ Mpc/h but below the virial radius ($r_{\rm 200}=0.75$ Mpc/h), the dark-matter-baryon profile is suppressed compared to the dark-matter-only one. This suppression becomes stronger for lower values of $\beta$ and/or larger values of $\theta_{\rm ej}$. Beyond the virial radius, the dark-matter-baryon density lies above the dark-matter-only density again. This is a consequence of gas being ejected out of the halo. Although this effect looks rather small in Fig.~\ref{fig:profiles}, it is the main responsible for baryon effects on the large-scale structure of the universe. Depending on the parameters, there are relevant differences between the dark-matter-baryon and the dark-matter-only profiles at scales up to ten times the virial radius or more.


\section{Effects on the matter power spectrum}\label{sec:PS}
In this section we study the redshift dependence of the baryonic correction (BC) model and we investigate how the free model parameters affect the matter power spectrum. Furthermore, we show that the BC model is able to reproduce results from hydrodynamical simulations.

In order to determine the power spectrum with the BC model, we first displace the particles of the $N$-body output according to the method described in Sec.~\ref{sec:basicprinciple}, using a halo finder to determine the halo positions as well as their masses and concentrations. We then measure the power spectra of both the original and the modified $N$-body output and investigate their ratios in order to focus on the baryon effects.

The $N$-body simulation is run with {\tt Pkdgrav3} \citep{Stadel:2001aaa,Potter:2016ttn} based on the cosmological parameters $\Omega_m=0.32$, $\Omega_{\Lambda}=0.68$, $\Omega_b=0.048$, $n_s=0.96$, $\sigma_8=0.83$, $h_0=0.67$ from {\tt Planck} \citep{Planck:2015xua}. Regarding the box length and resolution, we use $L=256$ Mpc/h, and $N=512^3$ particles. This setup has been shown in ST15 to be fully converged at the relevant scales regarding the ratio of the dark-matter-baryon to dark-matter-only matter power spectrum. This means that with the present setup, the BC model provides indistinguishable results compared to simulations with larger volumes or higher resolutions. Finally, we use {\tt AHF} \citep{Knollmann:2009aaa} as our halo finder, assuming a minimum of 100 particles per halo which results in a minimum halo mass of $\sim10^{12}$ M$_{\odot}$/h.

\subsection{Varying model parameters}
In its simplest form presented in Sec.~\ref{sec:BCM}, the BC model contains five free parameters, three related to the gas profile ($M_{\rm c}$, $\mu$, $\theta_{\rm ej}$) and two related to the stellar profile ($\eta_{\rm star}$, $\eta_{\rm cga}$). Before constraining these parameters with observations, we will now investigate how each of them affects the matter power spectrum.

\begin{figure}[tbp]
\center{
\includegraphics[width=.32\textwidth,trim=0.4cm 0.1cm 1.5cm 0.8cm,clip]{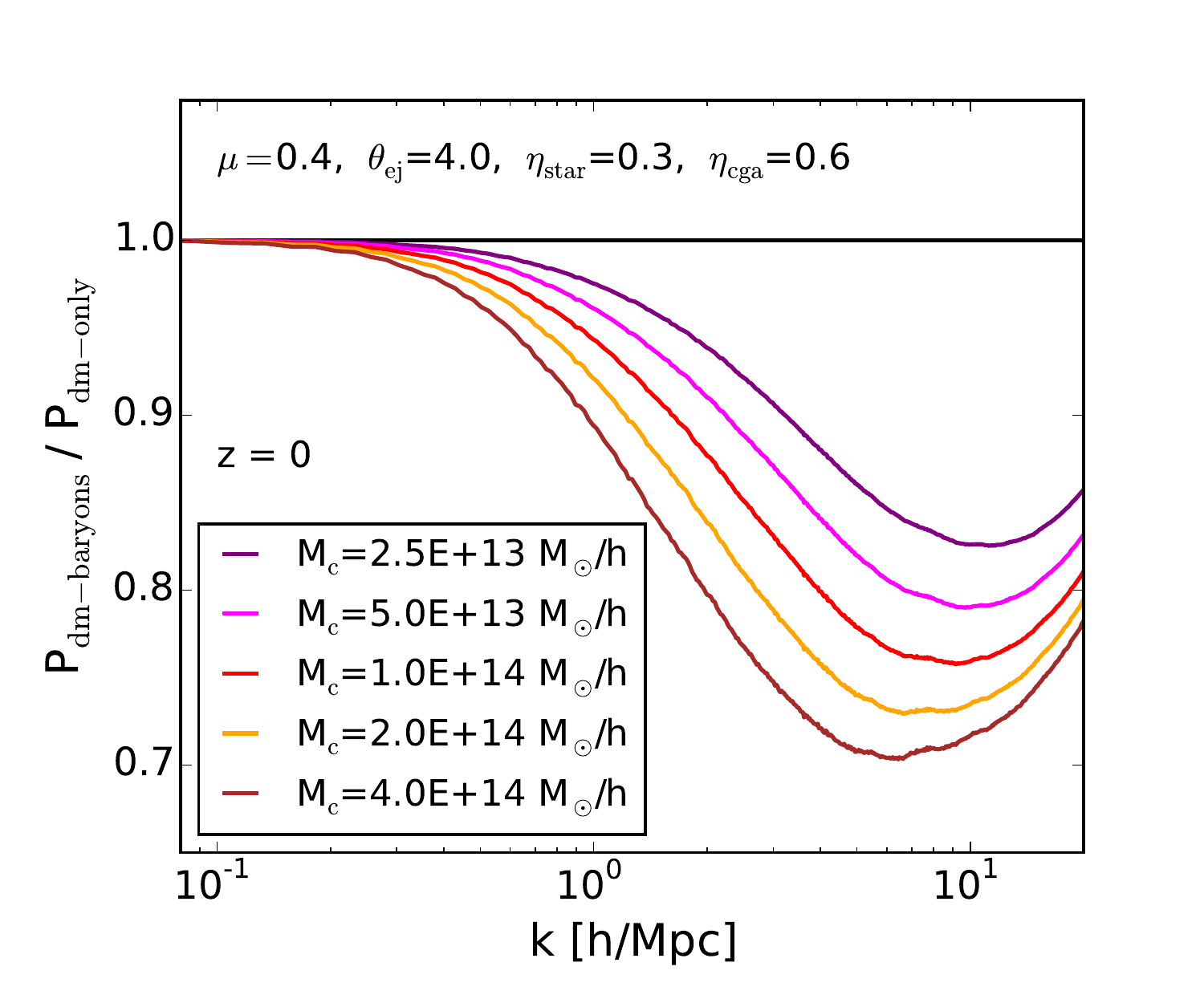}
\includegraphics[width=.32\textwidth,trim=0.4cm 0.1cm 1.5cm 0.8cm,clip]{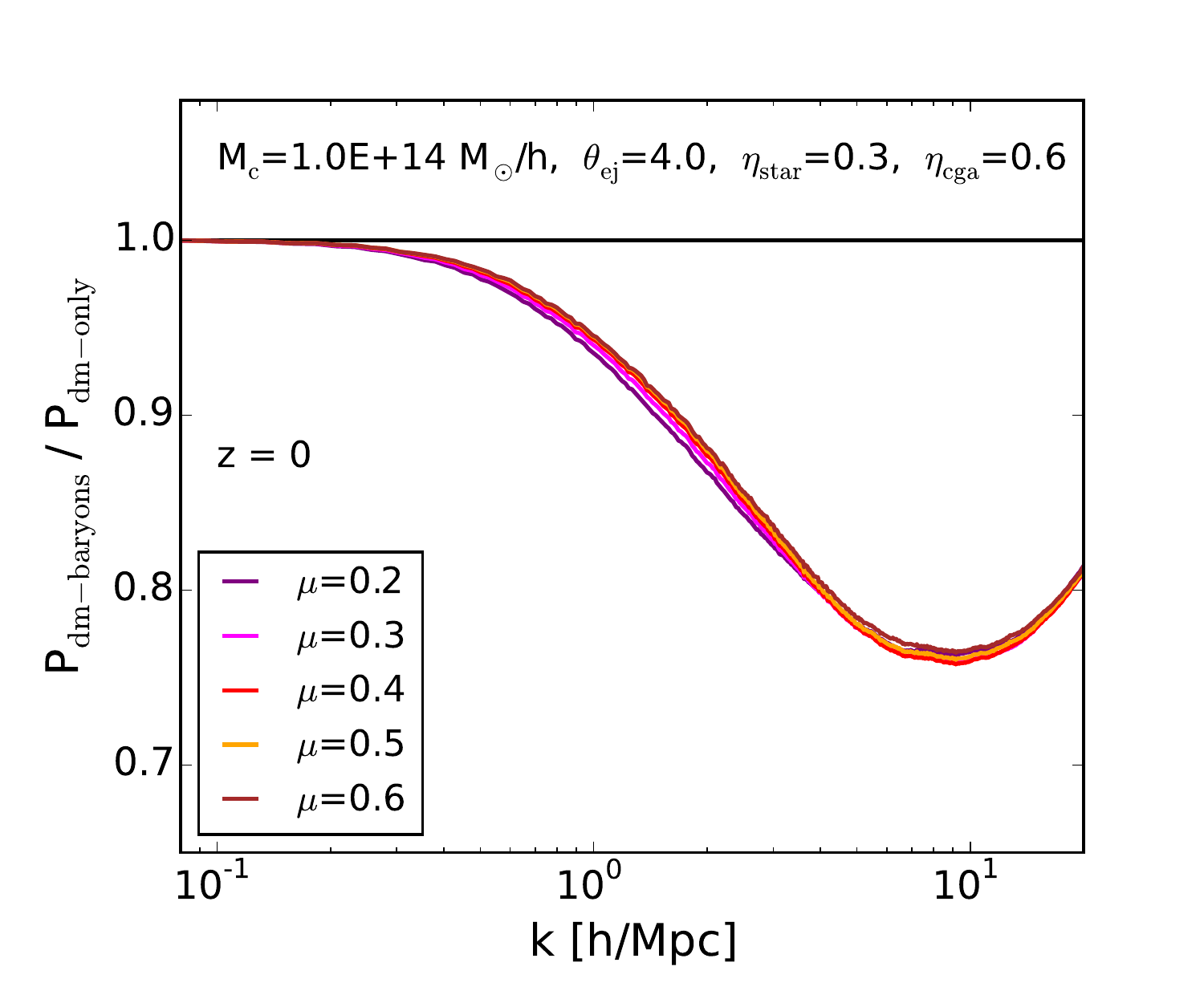}
\includegraphics[width=.32\textwidth,trim=0.4cm 0.1cm 1.5cm 0.8cm,clip]{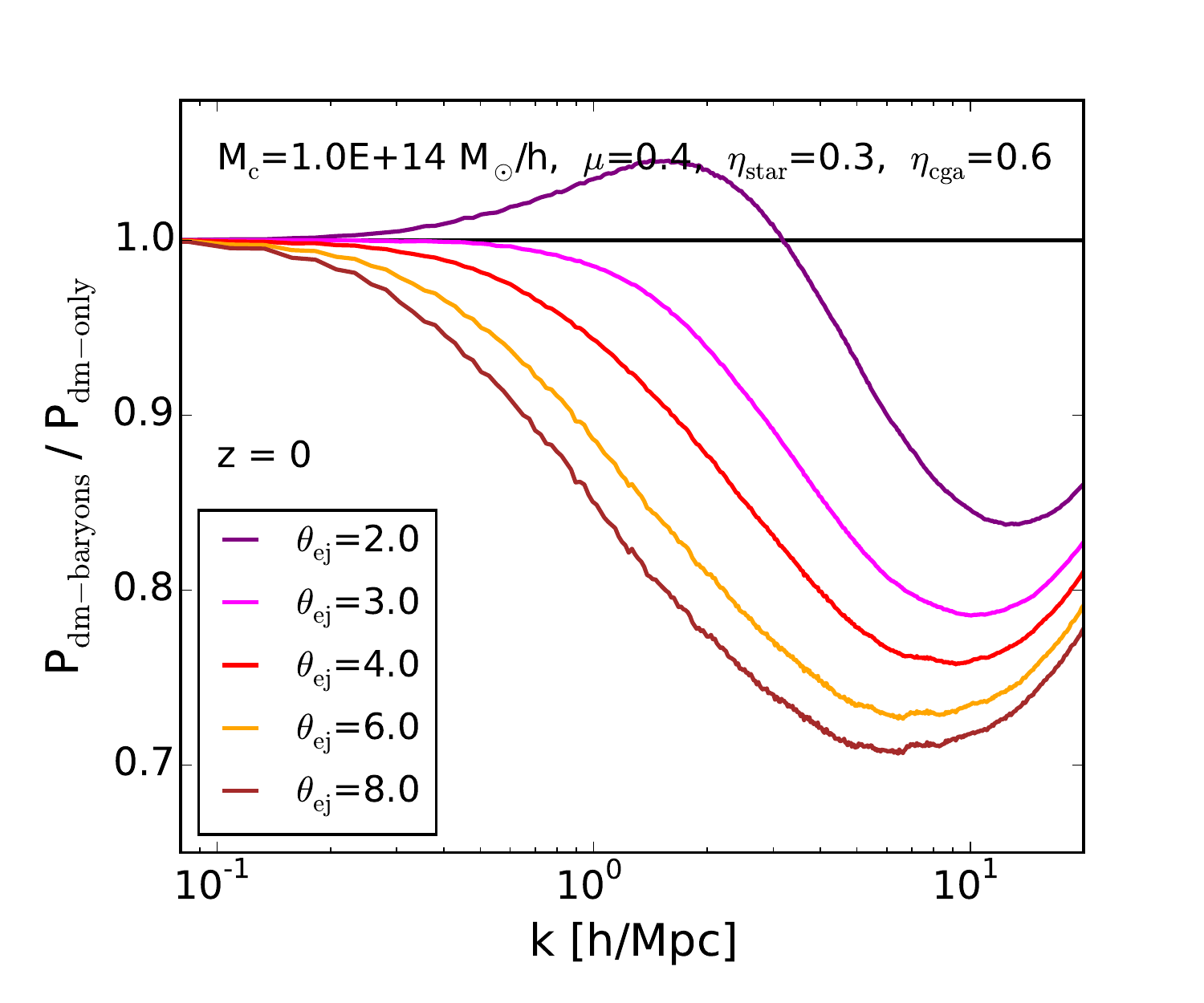}\\
\includegraphics[width=.32\textwidth,trim=0.4cm 0.1cm 1.5cm 0.8cm,clip]{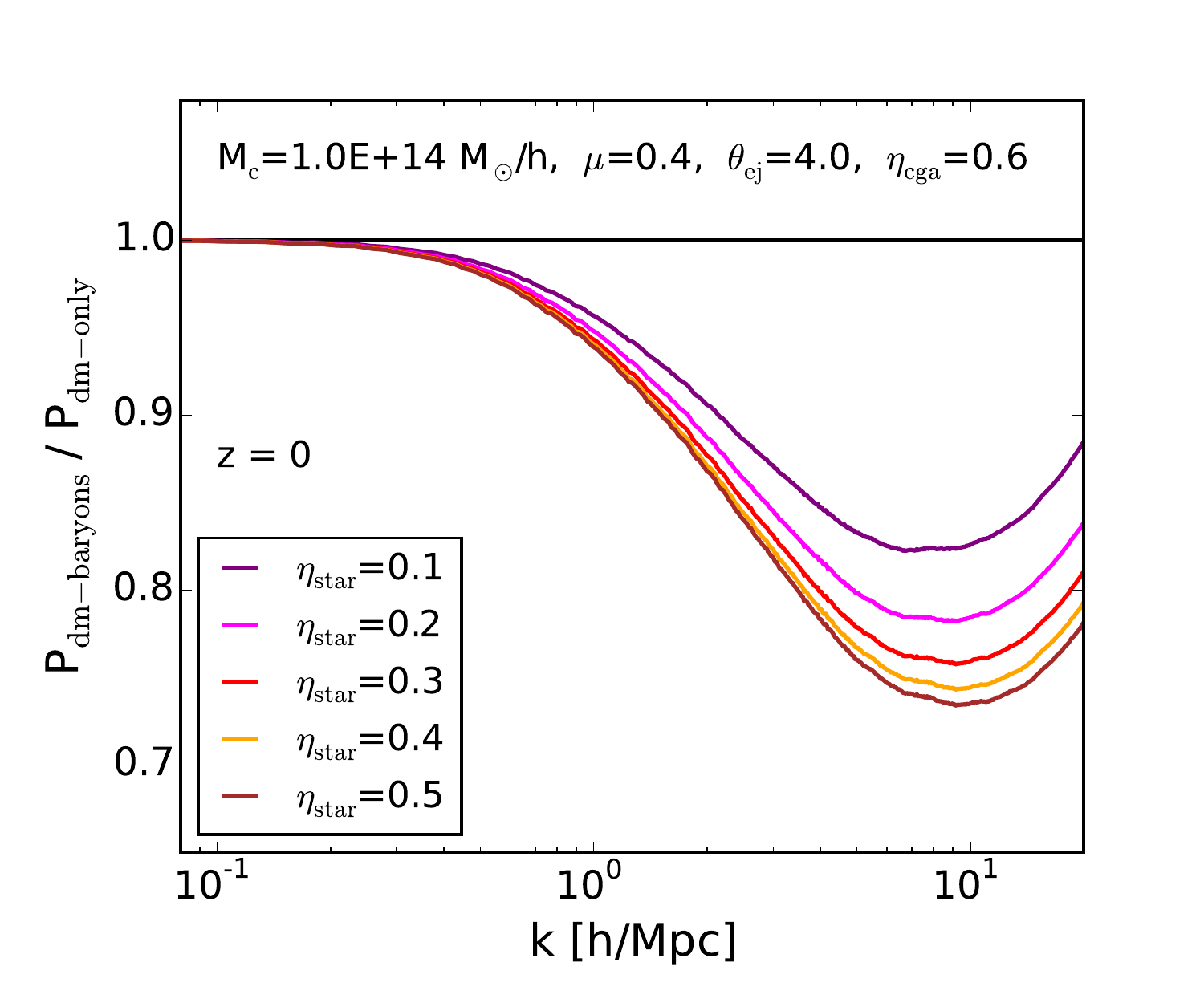}
\includegraphics[width=.32\textwidth,trim=0.4cm 0.1cm 1.5cm 0.8cm,clip]{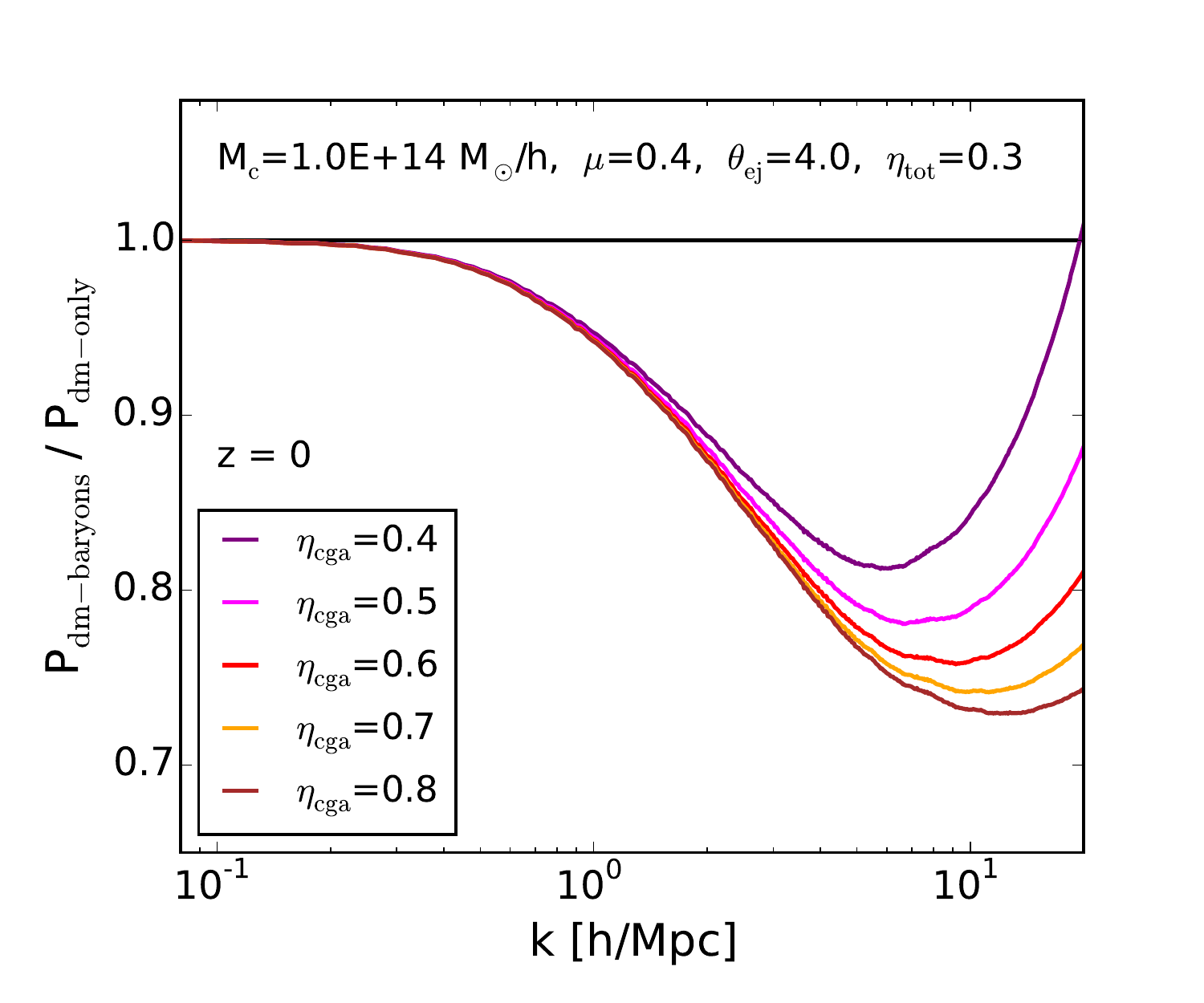}
\includegraphics[width=.32\textwidth,trim=0.4cm 0.1cm 1.5cm 0.8cm,clip]{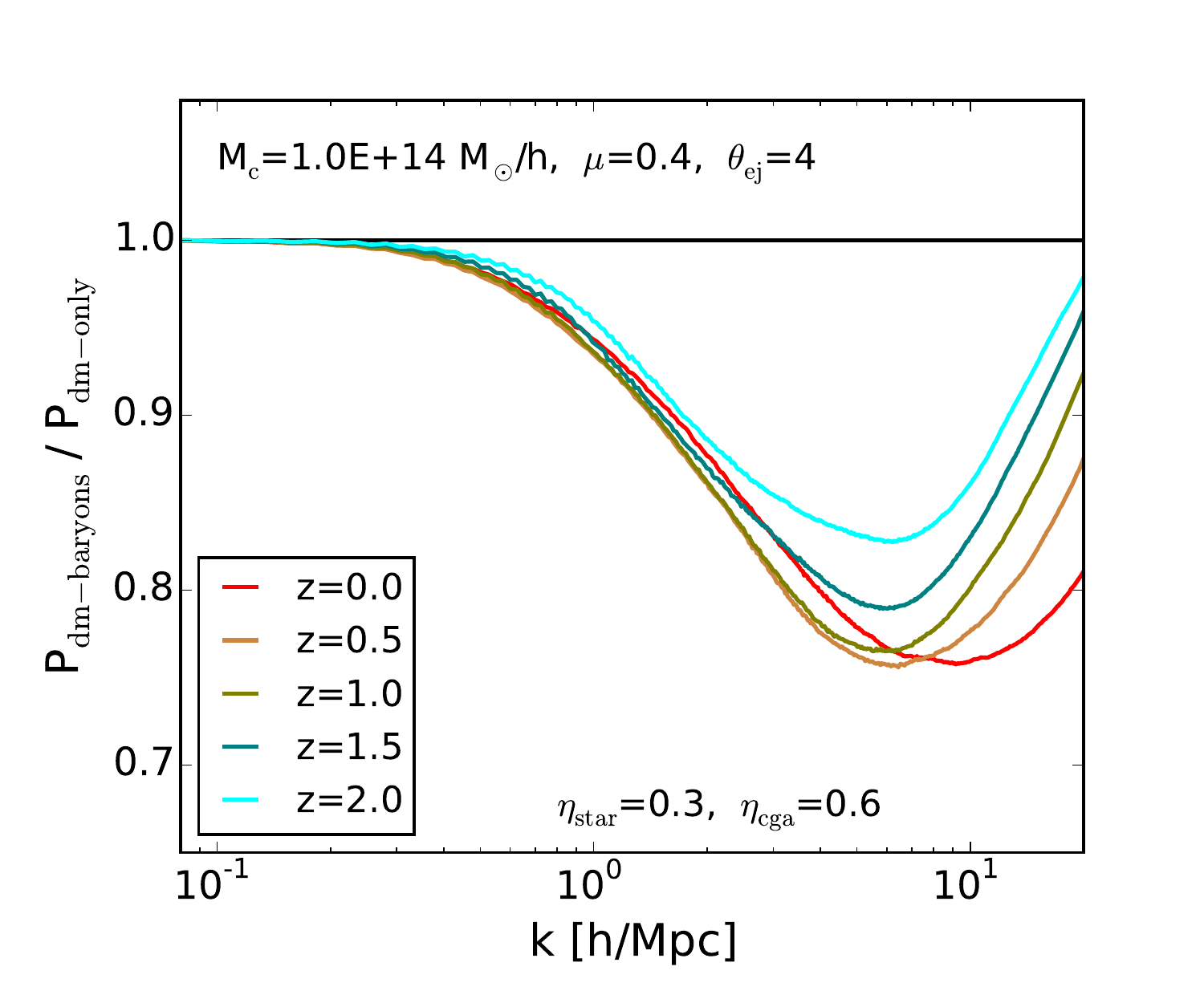}
\caption{\label{fig:PSvarparams}Baryon suppression of the matter power spectrum for different parameter choices of the BC model. In each panel one of five parameters is varied while the others are kept at a default value. The power spectrum is particularly sensitive to the gas parameters $M_c$ and $\theta_{\rm ej}$ (but not $\mu$). The stellar parameters  $\eta_{\rm star}$ and $\eta_{\rm cga}$ have a comparably smaller effect. \emph{Bottom-right:} Redshift dependence of the model.}}
\end{figure}

Fig.~\ref{fig:PSvarparams} shows the dark-matter-baryon power spectrum obtained with the BC model divided by the original dark-matter-only power spectrum. In each panel one of the five model parameters is varied while the others are kept constant. The first two panels (top-left and top-centre) show the effect of varying $M_c$ and $\mu$, the characteristic mass scale and rate at which the slope ($\beta$) of the gas density profile falls below $\beta=3$ towards small halo masses (see Eq.~\ref{rhogas}). The slope of the gas density profile is particularly important as it defines how much of the gas has been pushed out beyond the virial radius. Not surprisingly, this has a strong effect on the overall amplitude of the baryon power suppression. The suppression of the power spectrum becomes larger for increasing values of $M_c$, while it is not very sensitive to the parameter $\mu$.

The third panel of Fig.~\ref{fig:PSvarparams} (top-right) shows how different ejection radii (parametrised by $\theta_{\rm ej}$) affect the power spectrum. Larger values of $\theta_{\rm ej}$ both increase the overall amplitude of the suppression and push the signal towards smaller wave modes. This is not surprising as a larger ejection radius means that the gas is pushed further out in the intergalactic space, therefore affecting larger physical scales. Note that very small values of $\theta_{\rm ej}$ can lead to an increase of power around $k\sim 1$ h/Mpc. This is due to the fact that gas is pushed out of the centre of haloes forming an overdense ring around the virial radius.

The fourth and fifth panel of  Fig.~\ref{fig:PSvarparams} (bottom-left and bottom-centre) show the impact of the stellar parameters on the matter power spectrum. The slope of the total stellar fraction within the virial radius ($\eta_{\rm star}$) has a moderate but non-negligible impact on the power spectrum. A larger $\eta_{\rm star}$ (smaller stellar fraction $f_{\rm star}$, see Eq.~\ref{stellarfraction}) leads to a more pronounced power suppression. The reason for this behaviour is that a smaller stellar fraction automatically means a larger gas fraction which leads to an increase of the suppression signal. The slope of the central galactic stellar fraction $\eta_{\rm cga}$ only affects large wave numbers beyond $k\sim5$ h/Mpc, yielding a more prominent upturn of the small-scale power for lower values of $\eta_{\rm cga}$ (i.e. a larger fraction of stars in the central galaxy). The reason for this behaviour is a further steepening of the inner density profiles due to the presence of a more pronounced central galaxy reinforced by the subsequent adiabatic contraction effect.

In summary, Fig.~\ref{fig:PSvarparams} shows that the baryon suppression of the power spectrum is mainly driven by the gas parameters $M_c$ and $\theta_{\rm ej}$. This is not surprising as the baryon suppression is dominated by the AGN feedback mechanisms which controls how much of the gas is pushed into the intergalactic space. The stellar fraction also has a visible effect on the power spectrum. However, the fraction of stars in haloes is comparably well constrained by observations as we will see in Sec.~\ref{sec:constraints}.

Note that other choices related to the model parametrisation not discussed above could in principle also affect the power spectrum, adding potential systematics to the model. In Appendix~\ref{systematics} we discuss the main sources of systematics (such as for example the fixed core and truncation radii, the adiabatic contraction algorithm, or a potential introduction of scatter for certain model parameters) and we argue why we believe them to be subdominant.

\subsection{Varying redshift}\label{BCMredshift}
The parametrisation of the BC model does not feature any explicit redshift dependence, i.e. none of the parameters introduced in Sec.~\ref{sec:BCM} evolve with time. While this is an assumption to keep the model as simple as possible, there is observational evidence that the gas profiles do not significantly evolve with redshift, at least not below $z\sim1.5$ \citep{LaRoque:2006te,Morandi:2015pra,Sanders:2017lce,Chiu:2017nwm}. Regarding the stellar components, while the observed luminosity function shows some redshift evolution \citep[][]{Moster:2012fv,Leauthaud:2012aaa,Behroozi:2012iw,Chiu:2017nwm}, this does not seem to be the case for the total stellar fraction within $r_{500}$ \citep{Giodini:2009qf,Chiu:2017nwm}. However, at redshifts below $z\sim1.5$ the stellar redshift evolution is mild enough to justify our simplified model assumptions.

It is important to notice that the lack of explicit redshift evolution of the BC model parameters does not imply that the baryonic power suppression is constant with redshift. On the contrary, a redshift dependence is expected due to the fact that at higher redshifts there are fewer large haloes, which means that the power spectrum signal is influenced by smaller haloes with different stellar and gas profiles.

The lower-right panel of Fig.~\ref{fig:PSvarparams} shows the implicit redshift dependence of the power spectrum obtained with the BC model. Between $z=0-1$ the baryon-induced power suppression does not evolve much for $k\lesssim10$ h/Mpc. Above $z\sim1$, on the other hand, the overall amplitude of the power suppression starts to decrease substantially. The redshift evolution beyond $z=2$ is not shown, as the model assumptions are likely to degrade beyond this point due to expected redshift dependence of the gas and stellar fractions at high redshifts. Furthermore, present and future weak-lensing surveys are mostly limited to redshifts below 2.

\subsection{Comparing with hydrodynamical simulations}\label{comparingtohydro}
Before moving on and constraining the parameters of the baryonic correction model with observations, it is important to check if the model is in agreement with results from hydrodynamical simulations. In this section, we compare the BC model with results from OWLS (OverWhelmingly Large Simulation \citep[][]{Schaye:2010aaa,McCarthy:2010aaa} run with the {\tt Gadget3} code \citep{Springel:2005mi}) and based on the {\tt WMAP3} cosmology \citep{Spergel:2006hy}. Comparisons to other hydrodynamical simulations are shown in Appendix \ref{hydrosims}.

\begin{figure}[tbp]
\center{
\includegraphics[width=.49\textwidth,trim=0.9cm 0.8cm 2.0cm 0.9cm,clip]{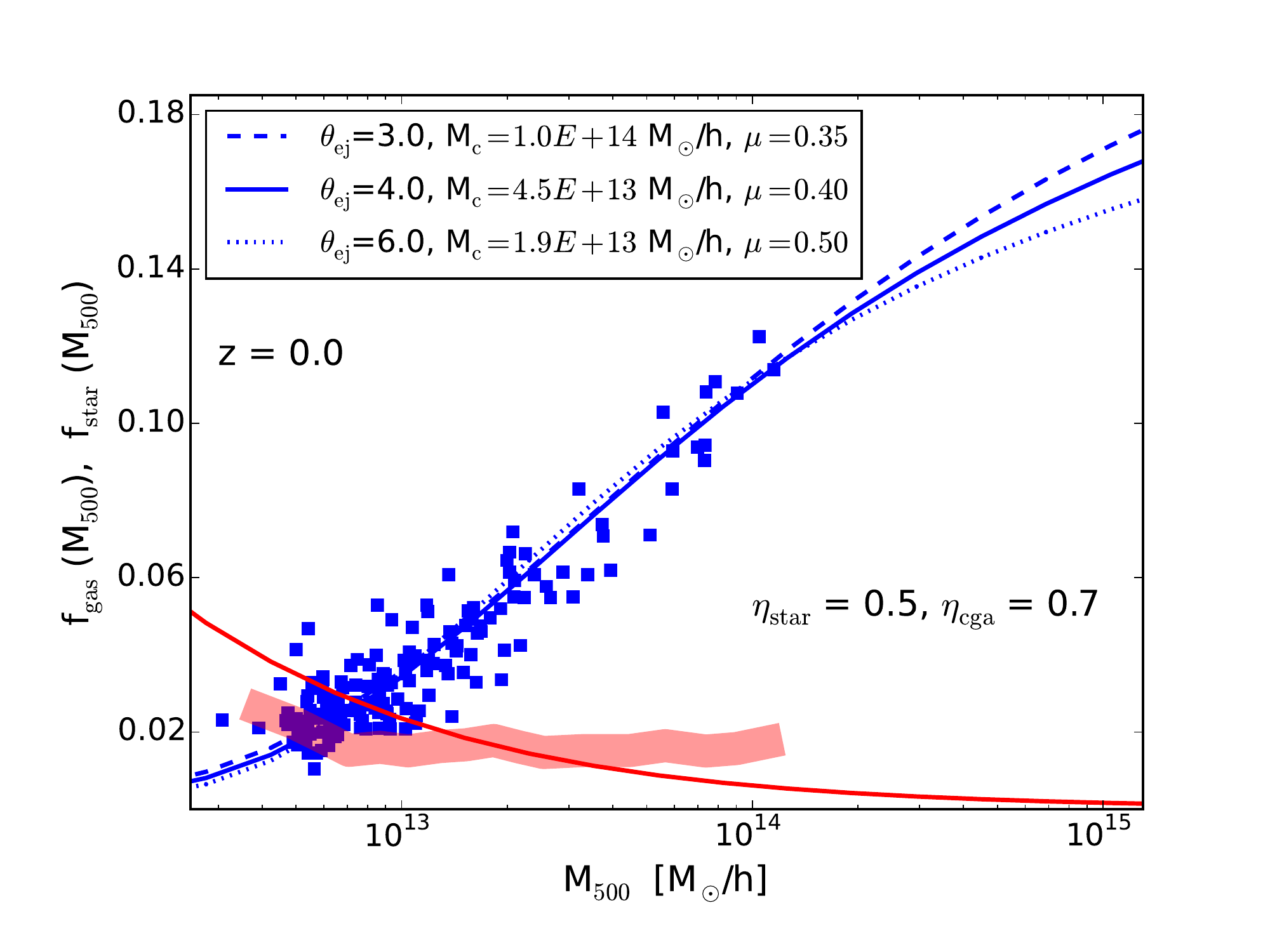}
\includegraphics[width=.49\textwidth,trim=0.9cm 0.8cm 2.0cm 0.9cm,clip]{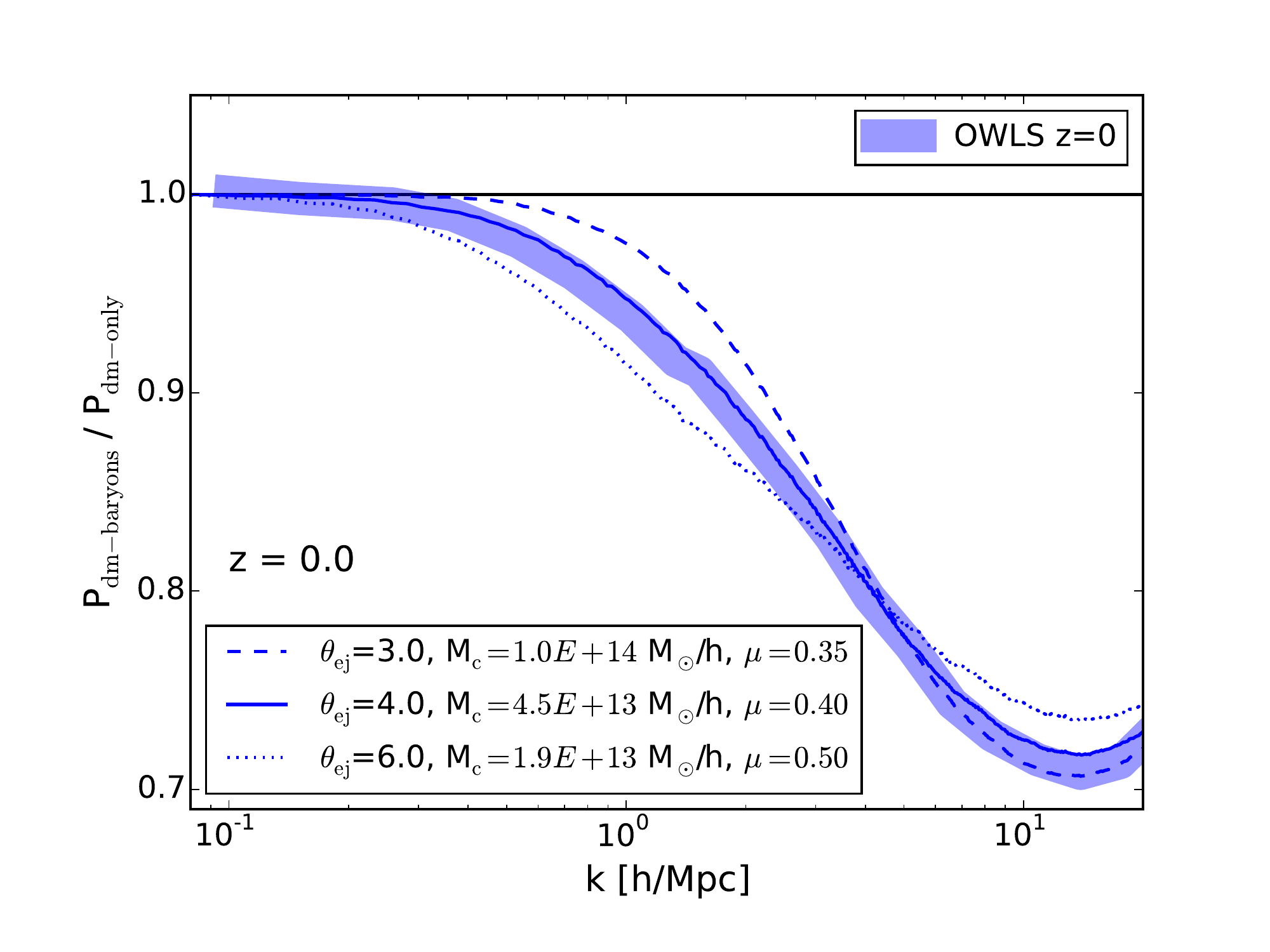}\\
\includegraphics[width=.49\textwidth,trim=0.9cm 0.8cm 2.0cm 0.9cm,clip]{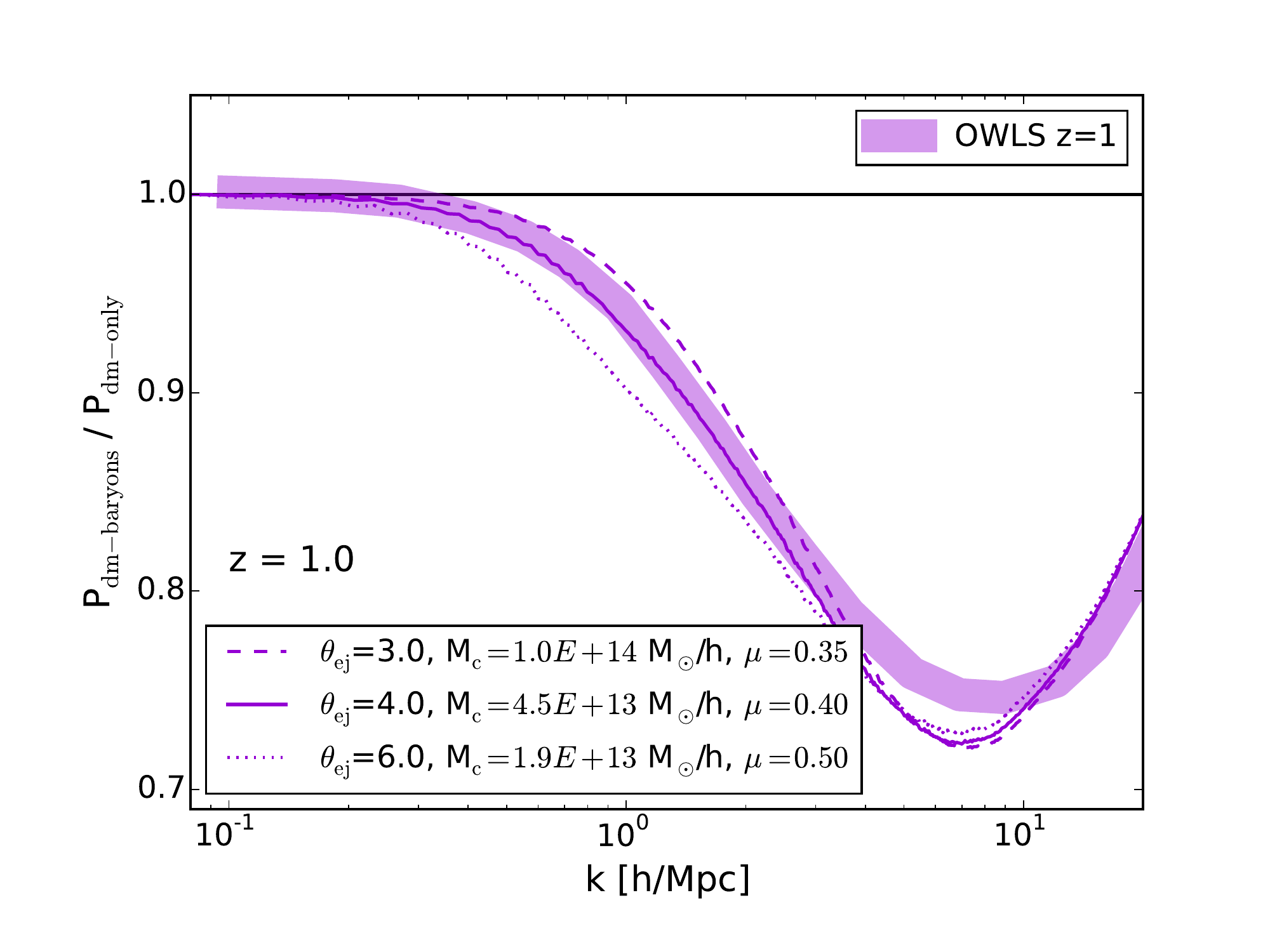}
\includegraphics[width=.49\textwidth,trim=0.9cm 0.8cm 2.0cm 0.9cm,clip]{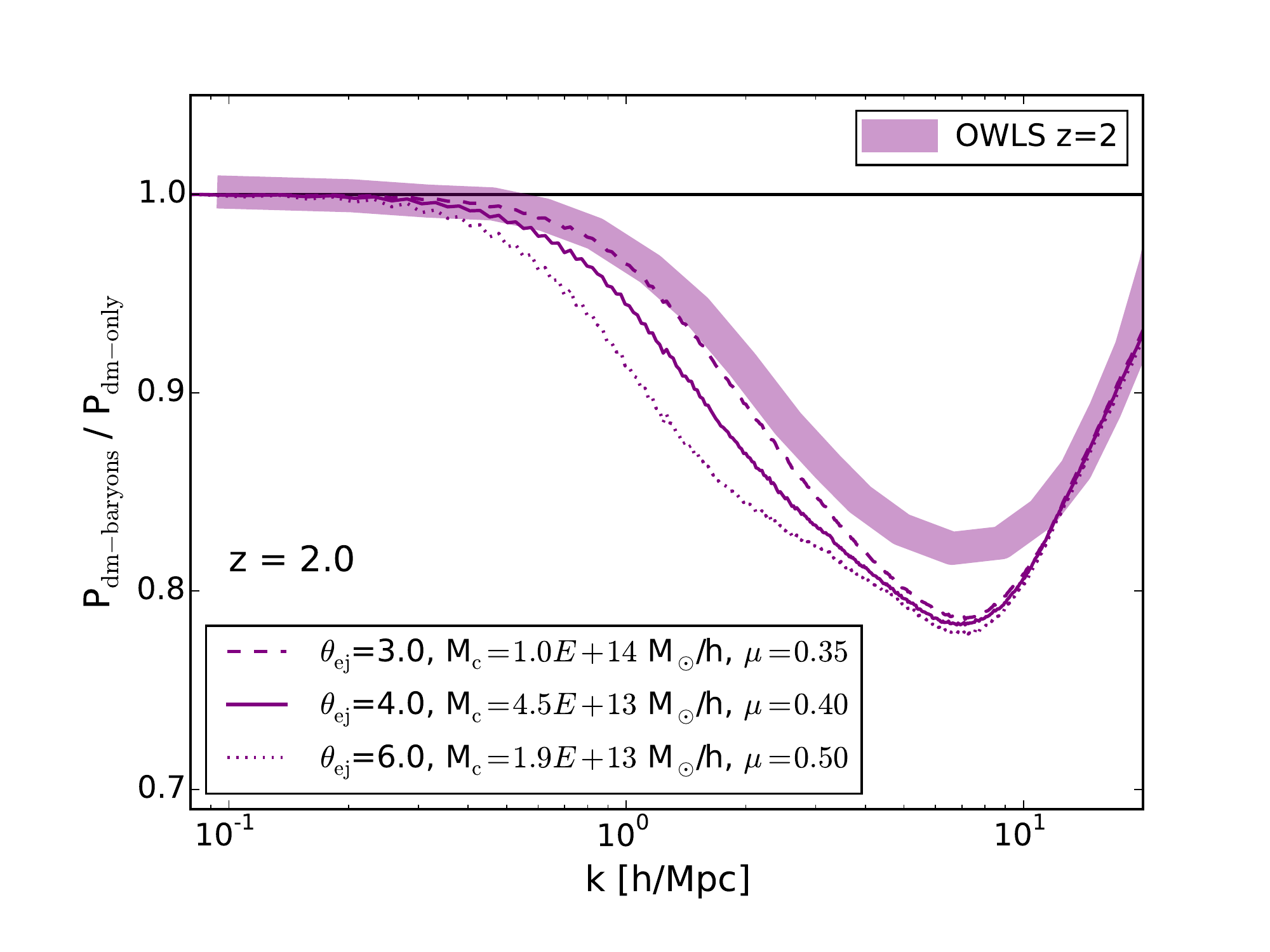}
\caption{\label{fig:OWLS}Comparison between the baryonic correction (BC) model and the OWLS AGN run \citep{Schaye:2010aaa,McCarthy:2010aaa,Semboloni:2011aaa}. The BC parameters $M_c$ and $\mu$ are thereby tuned to match the gas and stellar fraction of OWLS (i.e. the blue dots and red band in the top-left panel). Regarding the gas ejection parameter, we assume three different cases, $\theta_{\rm ej}=3,4,6$. Note that the fits deteriorate substantially for $\theta_{\rm ej}$ below 3 or above 6. The other panels show a comparison of the power spectra predicted by the BC model and measured from OWLS at redshifts 0, 1, and 2.}}
\end{figure}

First of all, it is important to note that the BC model is versatile enough to match the baryonic power suppression of any currently known hydrodynamical simulation to high precision provided all parameters are left free. However, we do not simply want to reproduce results from simulations, but the goal is to check if the BC model is able to \emph{predict} the correct power spectrum if all it knows is the gas and stellar fractions of haloes. We therefore fit the BC model parameters to the gas and stellar fraction of OWLS, before comparing the resulting power spectrum of the BC model with the one measured in OWLS.

The top-left panel of Fig.~\ref{fig:OWLS} shows the gas fraction (blue symbols) and the total stellar fraction (red band) obtained by OWLS at redshift zero \citep[][]{Semboloni:2011aaa,McCarthy:2010aaa}. Regarding the stellar fraction, an acceptable fit is obtained with $\eta_{\rm star}=0.5$ and $\eta_{\rm cga}=0.7$ (solid red line). No better agreement can be obtained without changing the functional form of the stellar fractions in the BC model. This is because stellar fractions from OWLS do not fully agree with results obtained from abundance matching \citep{Moster:2012fv,Behroozi:2012iw} which have been used to setup and motivate the model parametrisation (see Sec.~\ref{stellarparams}). Regarding the gas fraction, the BC model is able to reproduce the data from OWLS at good accuracy for different choices of $\theta_{\rm ej}$ between $\theta_{\rm ej}\sim3-6$. Outside of this range, the fits become significantly worse. In order to account for this degeneracy, we show three cases of the BC model with $\theta_{\rm ej}=3,4,6$ (dashed, solid, and dotted blue lines) bracketing the model space in agreement with the gas fraction of OWLS.

The top-right panel of Fig.~\ref{fig:OWLS} shows a comparison between the power spectrum from OWLS (blue band) and from the BC model (blue lines) at redshift zero. The best agreement is obtained for the $\theta_{\rm ej}=4$ case, where the start of the baryon induced downturn, the amplitude of the maximum suppression, and the functional form are matched very well. The overall difference between the results from OWLS and from the BC model does not exceed one percent below $k\sim 20$ h/Mpc. This is a very good agreement, thoroughly justifying the approach used in this paper.

The bottom panels of Fig.~\ref{fig:OWLS} show a comparison between OWLS and the BC model for the redshifts 1 and 2. The agreement is slightly worse but still better than four percent up to wave modes of $k\sim 20$ h/Mpc. Note that such a good match is only possible because OWLS does not predict any strong redshift evolution of the gas or stellar fraction below redshift 2.

In Appendix \ref{hydrosims}, we compare the baryonic correction model to several other hydrodynamical simulations that feature very different predictions regarding the matter power spectrum. The comparison is performed for all hydrodynamical simulations we could find in the literature that have published gas fractions, stellar fractions, and matter power spectra. As a general trend, the BC model is able to match the redshift zero results very well, while the agreement can degrade somewhat towards higher redshifts depending on the simulation. The main reason for this is that the BC model parameters are always fitted to the simulated gas fractions at redshift zero, and some of the simulations show a non-negligible redshift evolution of these observables.

In summary, the matter power spectra of the \emph{baryonic correction model} and of all hydrodynamical simulations considered in this paper agree to better than two percent at $z=0$, four percent at $z=1$, and ten percent at $z=2$ considering wave modes up to $k=10$ h/Mpc. For a slightly reduced range of interest of $k<5$ h/Mpc, the agreement is better than two percent at $z=0$, three percent at $z=1$, and five percent at $z=2$.


\section{Constraining model parameters with observations}\label{sec:constraints}
After illustrating the effects of the baryonic correction (BC) model on the power spectrum and comparing it to full hydrodynamical simulations, we will now use observations to constrain the BC parameters. Regarding the gas parameters, we rely on X-ray data of individual and stacked galaxy groups and clusters. For the stellar parameters we use abundance matching results from the literature. Furthermore, we specifically discuss the potential systematics of X-ray observations related to the total mass estimate and how it affects the parameter constraints.

\begin{figure}
\center{
\includegraphics[width=.99\textwidth,trim={0.0cm 0.3cm 0.0cm 0.5cm}]{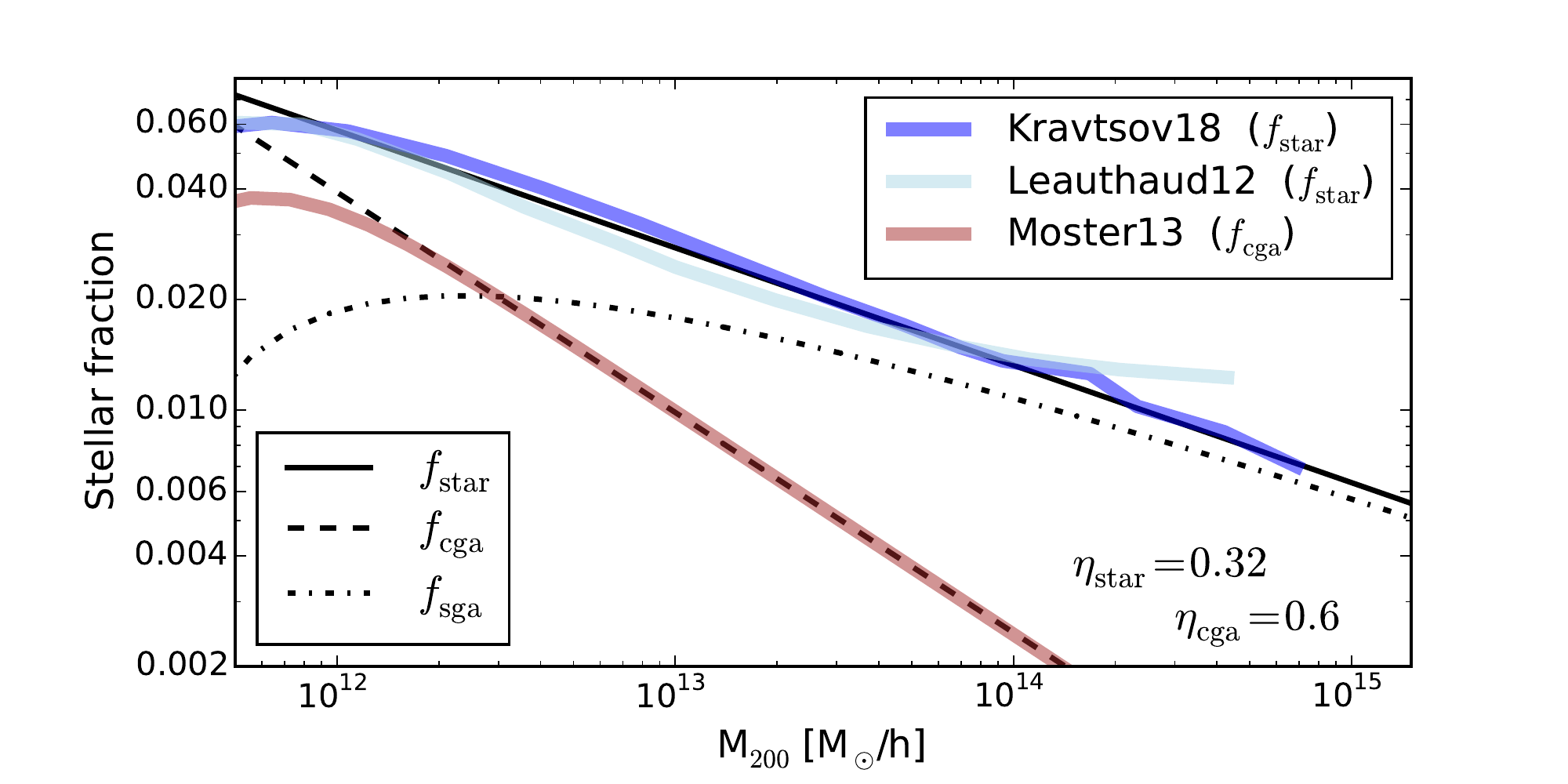}
\caption{\label{fig:stellarfraction}Average stellar fractions of the central galaxies (brown band) and the total stellar component within the virial radius (blue and light-blue band) obtained via abundance matching \citep{Moster:2012fv,Kravtsov:2014sra,Leauthaud:2012aaa}. For the relevant mass scales, both relations are well described by simple power-laws as of Eq.~\ref{stellarfraction}. Optimal fits are found for $\eta_{\rm star}=0.32$ (solid line) and $\eta_{\rm cga}=0.6$ (dashed lines). The dash-dotted line corresponds to the stellar fraction of satellite galaxies and halo stars defined by $f_{\rm sga}=f_{\rm star}-f_{\rm cga}$.}}
\end{figure}

\subsection{Stellar parameters}\label{stellarparams}
The stellar component of the BC model is parametrised by the stellar fractions (Eq.~\ref{stellarfraction}) and the profile of the central galaxy (Eq.,~\ref{rhocga}). Satellite galaxies and stars are assumed to follow the dark matter halo profile. As mentioned before, we do not assume any explicit redshift dependence in the parametrisation. This is likely to be a reasonable approximation for the total stellar component where no significant redshift dependence has been observed for low redshifts (see discussion in Sec.~\ref{BCMredshift}). The stellar fraction of the central galaxy, on the other hand, is subject to a mild redshift dependence which is ignored in our model for simplicity. We have checked that this does not significantly affect our results (see Sec.~\ref{BCMredshift} for a more detailed discussion about the redshift dependence).

Regarding the central galaxy, we rely on the very simple parametrisation of Eq.~(\ref{rhocga}). This is acceptable in our case because the stellar profile only affects very small scales that are not relevant for cosmology (but see Appendix~\ref{systematics} for a more detailed discussion). Note however, that a comparison of the profile to hydrodynamical simulations can be found in Ref.~\citep{Mohammed:2014mba}.

Fig.~\ref{fig:stellarfraction} compares the assumed stellar fractions of the BC model to abundance matching results from the literature (for the central galaxy $f_{\rm cga}$ in brown \citep{Moster:2012fv} and the total stellar fraction $f_{\rm star}$ in blue \citep{Kravtsov:2014sra,Leauthaud:2012aaa}). The best agreement between model and data is obtained for
\be
\eta_{\rm star}=0.32,\hspace{1cm}\eta_{\rm cga}=0.6
\ee
which are the values we adopt for the rest of the paper. The excellent match between model and data for $f_{\rm cga}$ is not surprising, since our model is motivated by the fitting function developed in Ref.~\citep{Moster:2012fv}. In general, Fig.~\ref{fig:stellarfraction} shows that while the stars of the central galaxy dominate the stellar component in haloes hosting Milky-Way type galaxies, they only play a minor role in galaxy clusters.

\subsection{Gas density profiles}\label{Xrayprofiles}
Before constraining the gas parameters of the BC model, it is important to check if the gas profile discussed in Sec.~\ref{sec:gas} provides a good match to observations. We use stacked X-ray observations from {\tt ROSAT/PSPC} \citep[][]{Eckert:2012aaa} and {\tt XMM-Newton} \citep[][]{Eckert:2015rlr} to test the gas profile defined by Eq.~(\ref{rhogas}). These observations are very suitable because they extend out to the virial radius and they cover the most important mass scales from galaxy groups to clusters.

\begin{figure}[tbp]
\center{
\includegraphics[width=.44\textwidth,trim=0.4cm 0.1cm 1.2cm 0.8cm,clip]{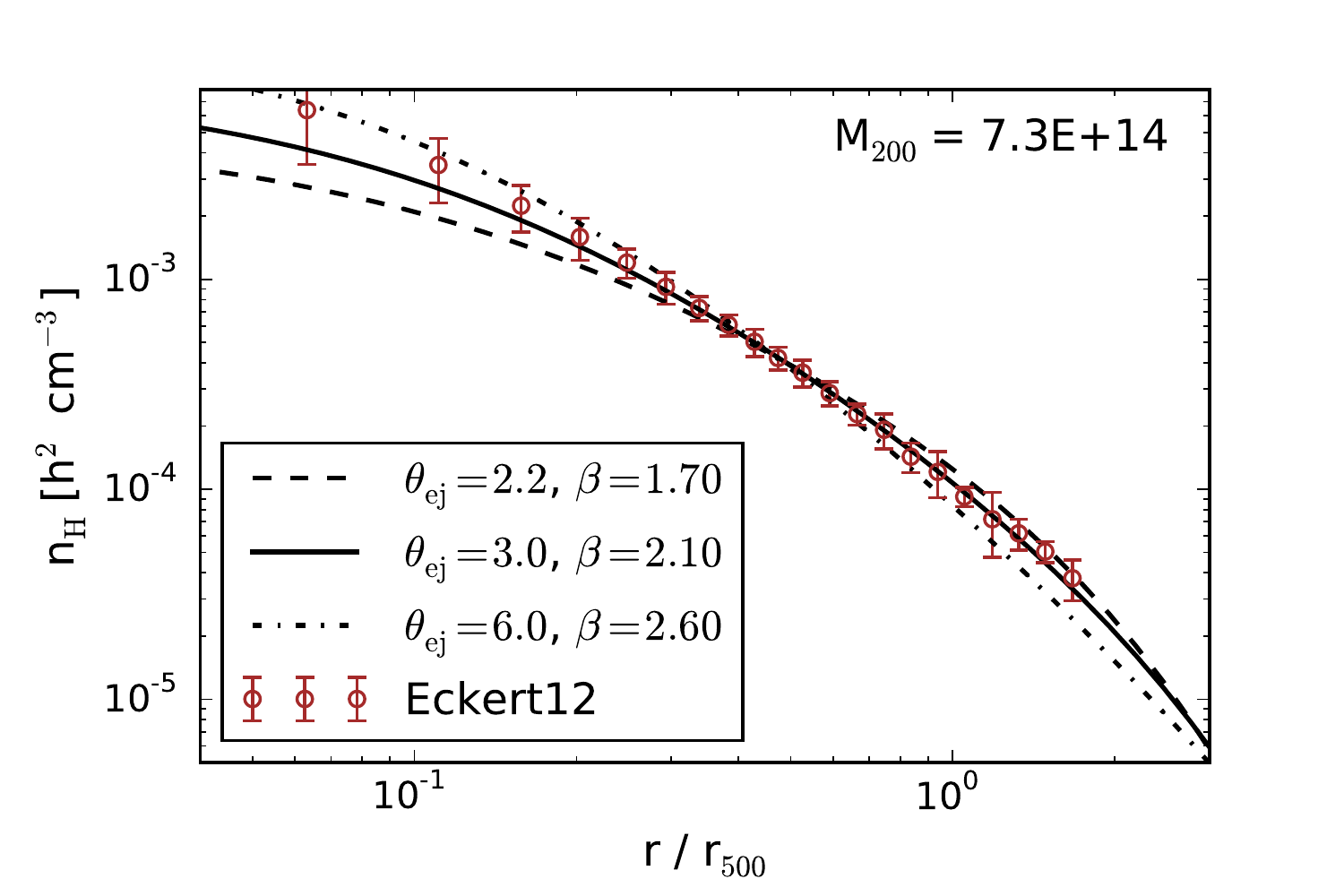}
\includegraphics[width=.44\textwidth,trim=0.4cm 0.1cm 1.2cm 0.8cm,clip]{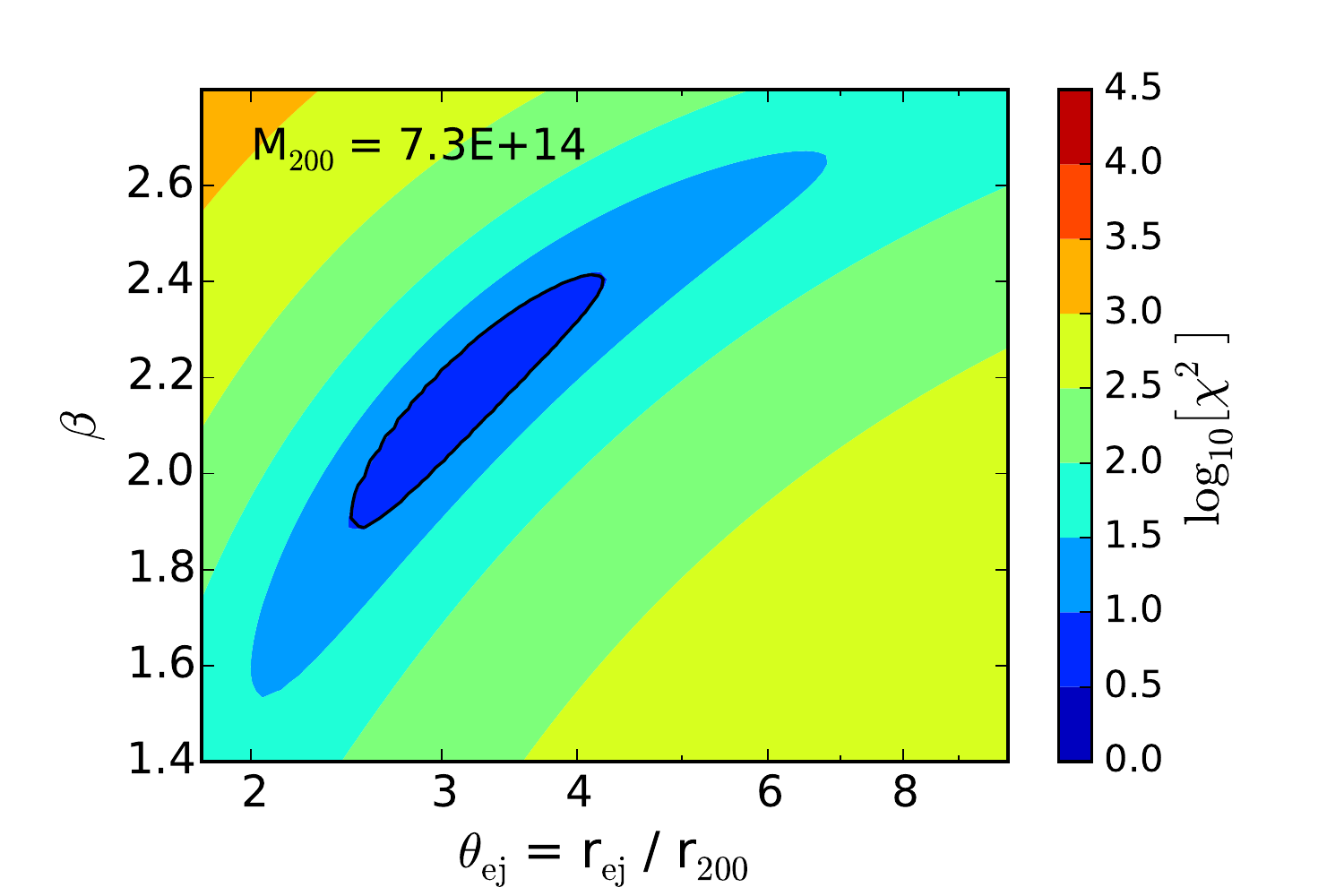}
\caption{\label{fig:profiles1}\emph{Left:} Stacked gas density profile from {\tt ROSAT/PSPC} X-ray observations of \citet{Eckert:2012aaa} (brown data points) compared to the gas profile of Eq.~(\ref{rhogas}) with selected values for $\theta_{\rm ej}$ and $\beta$ (black lines). \emph{Right:} Colour map indicating the chi-square ($\chi^2$) values for any combination of the profile parameters $\beta$ and $\theta_{\rm ej}$. The solid black line shows the 2-$\sigma$ likelihood contours.}}
\end{figure}

The left panel of Fig.~\ref{fig:profiles1} shows the profile of stacked galaxy clusters based on {\tt ROSAT/PSPC} data from \citet[][]{Eckert:2012aaa}. The average halo mass $M_{\rm 200}=7.3\times 10^{14}$ M$_{\odot}$/h is obtained with the standard assumption of hydrostatic equilibrium. Three example profiles from our model are added as black lines. The plot shows that, depending on the exact values of $\beta$ and $\theta_{\rm ej}$, the BC profiles are able to provide a good match to the data.

The right panel of Fig.~\ref{fig:profiles1} illustrates the chi-square ($\chi^2$) values of a least-square regression analysis over the full $\beta-\theta_{\rm ej}$ parameter space. The 2-$\sigma$ likelihood contours are shown as black line. Acceptable agreement between the model and the data is found for parameter values $\theta_{\rm ej}\sim 2-6$ and $\beta\sim 1.6-2.6$. Beyond this range the fits deteriorate substantially. The two parameters are degenerate in the sense that larger values $\theta_{\rm ej}$ require larger values of $\beta$.

\begin{figure}[tbp]
\center{
\includegraphics[width=.44\textwidth,trim=0.4cm 0.1cm 1.2cm 0.8cm,clip]{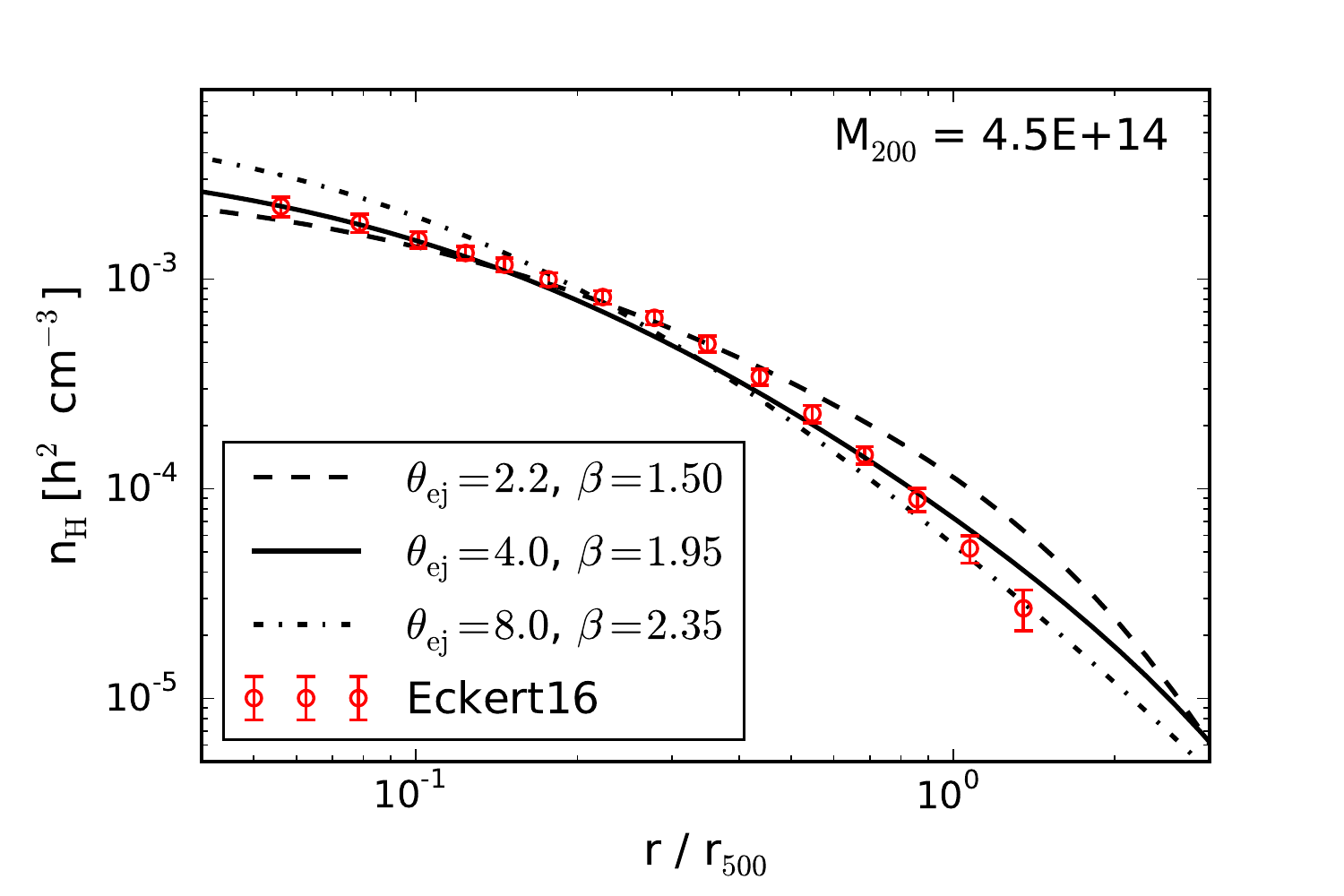}
\includegraphics[width=.44\textwidth,trim=0.4cm 0.1cm 1.2cm 0.8cm,clip]{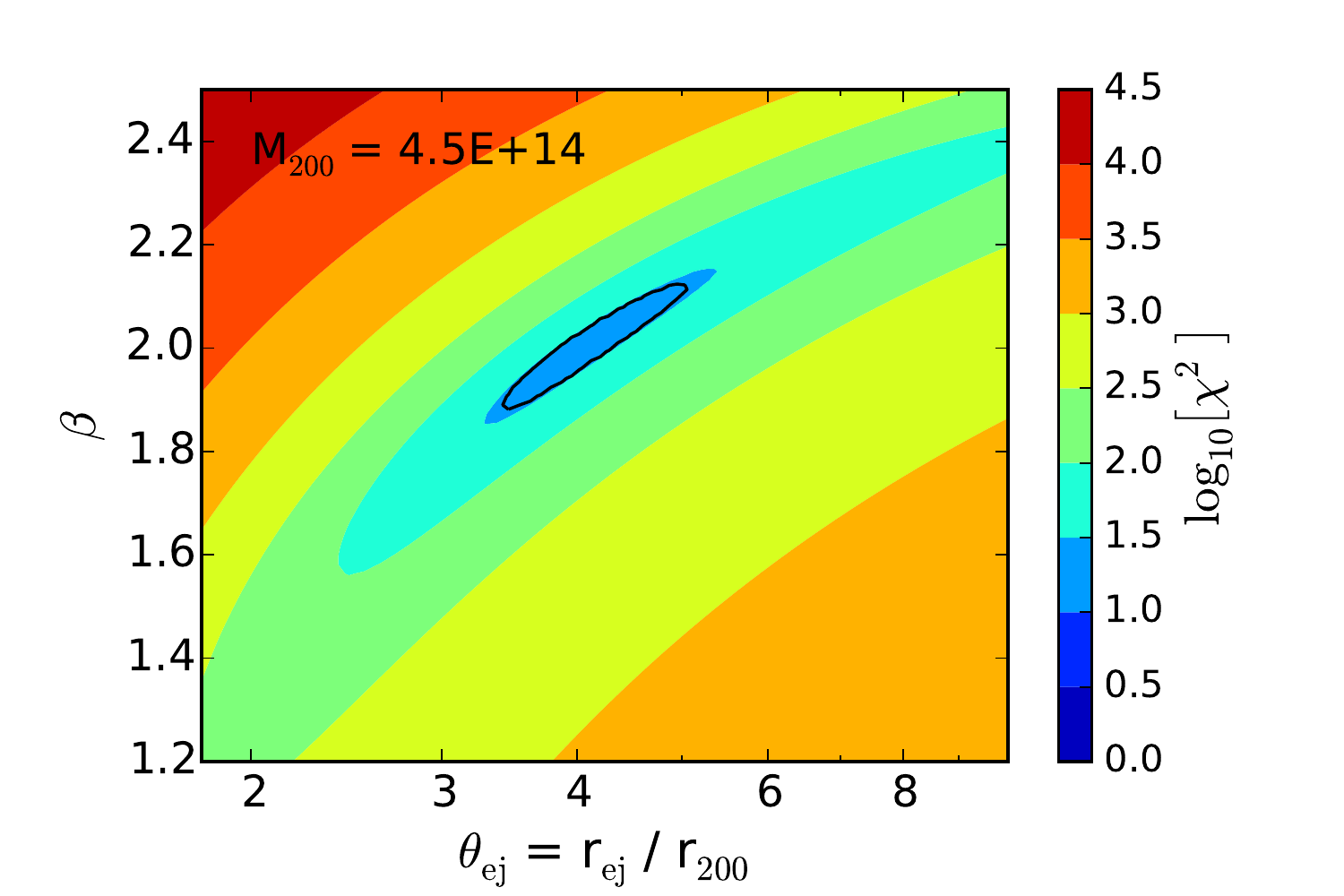}\\
\includegraphics[width=.44\textwidth,trim=0.4cm 0.1cm 1.2cm 0.8cm,clip]{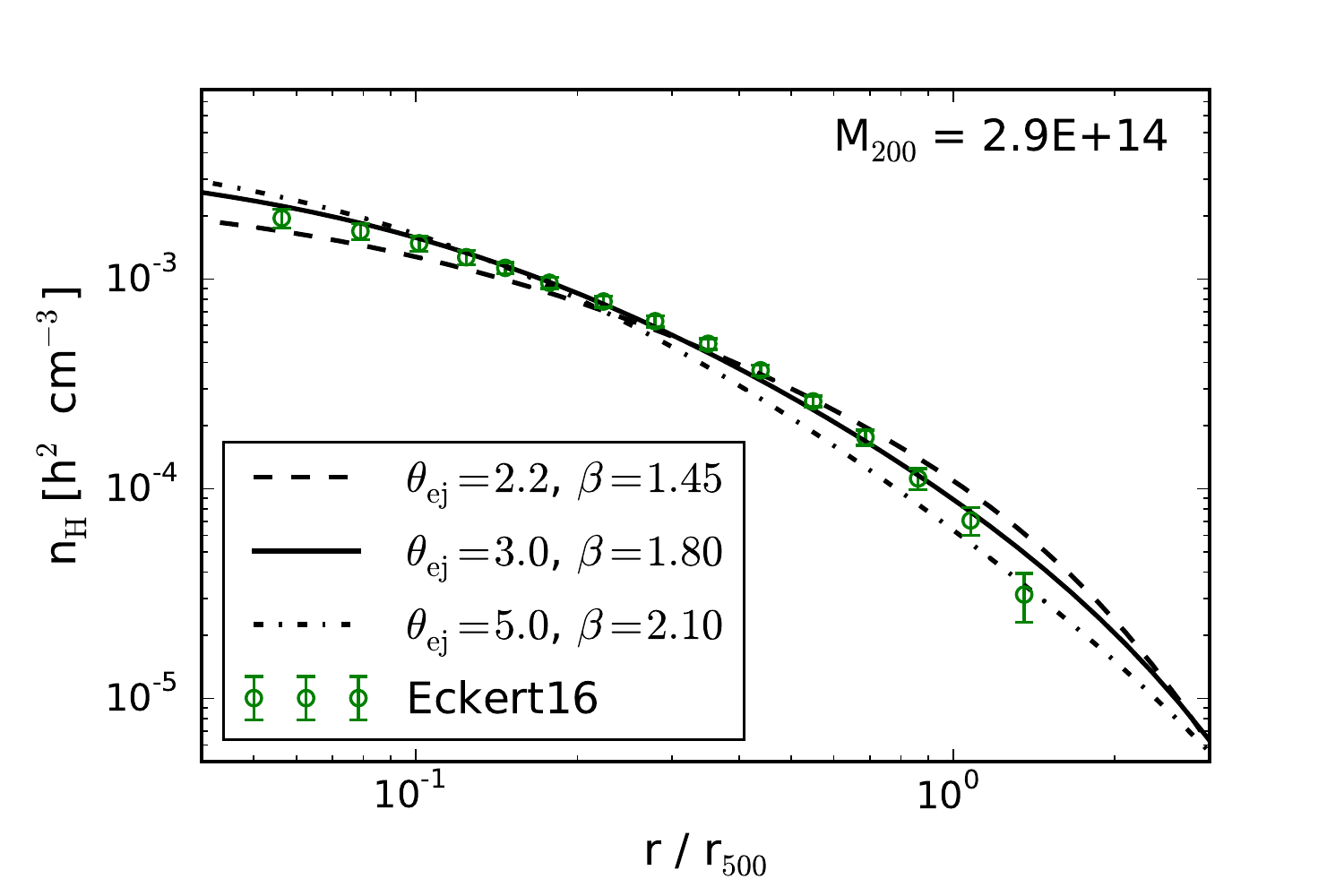}
\includegraphics[width=.44\textwidth,trim=0.4cm 0.1cm 1.2cm 0.8cm,clip]{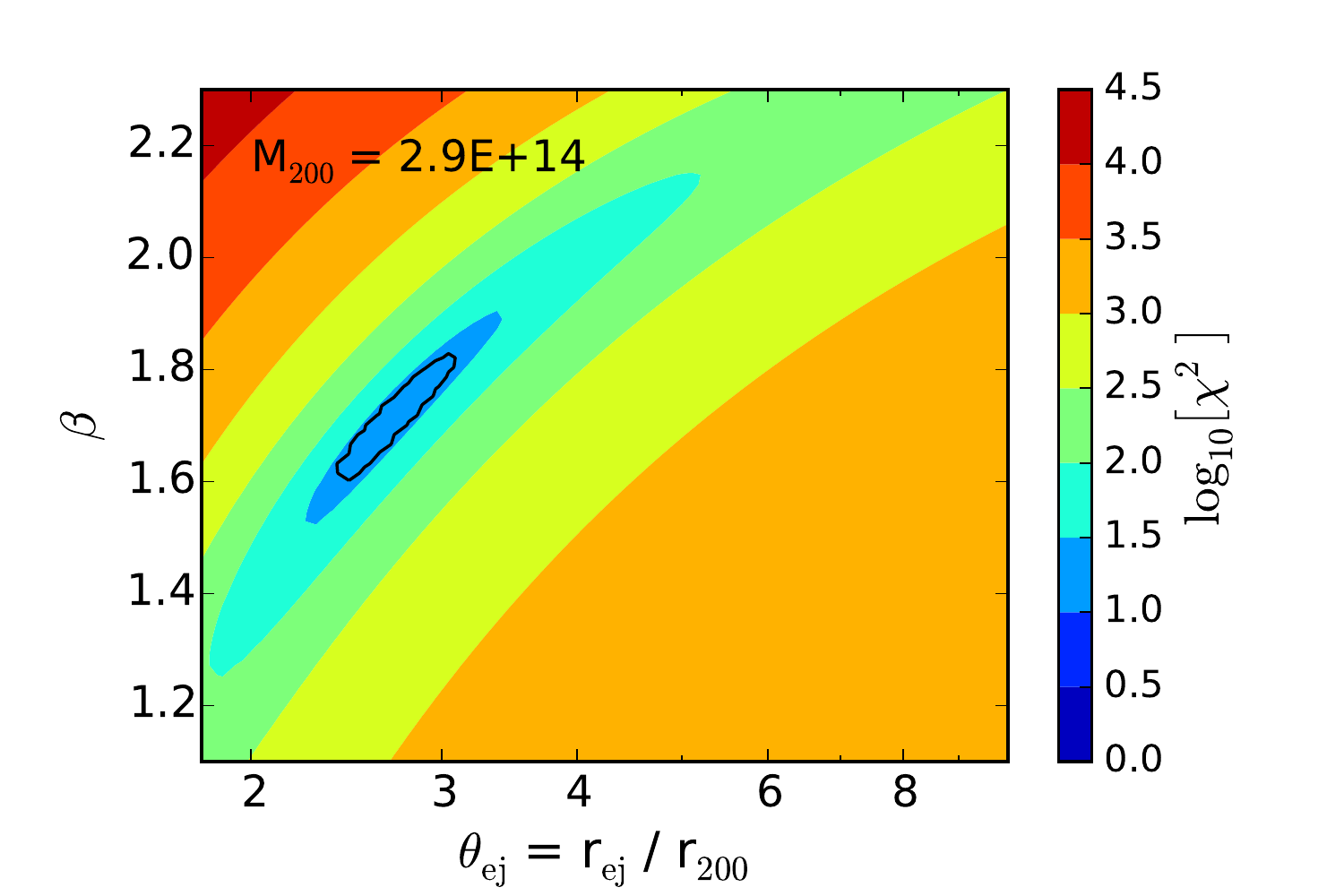}\\
\includegraphics[width=.44\textwidth,trim=0.4cm 0.1cm 1.2cm 0.8cm,clip]{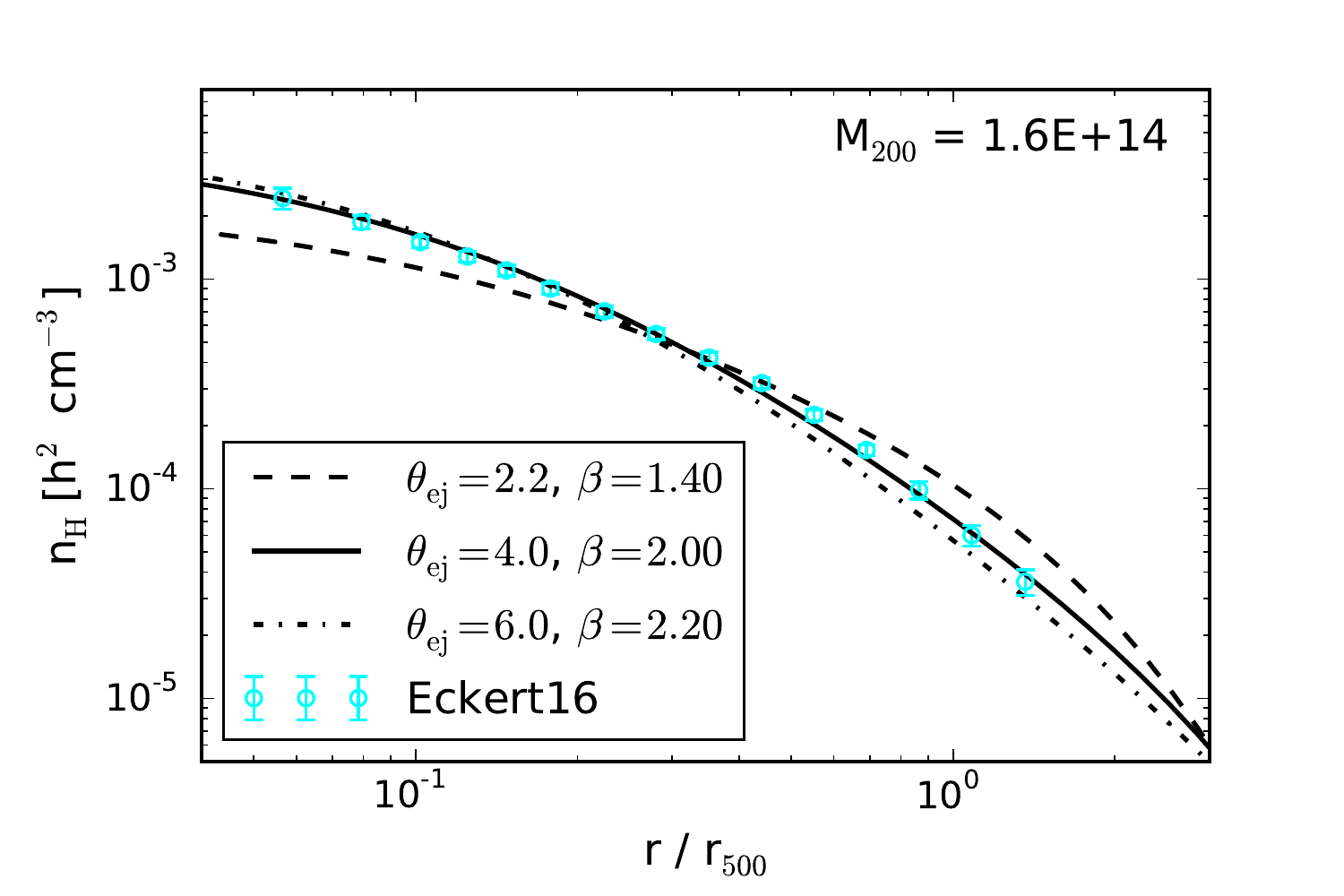}
\includegraphics[width=.44\textwidth,trim=0.4cm 0.1cm 1.2cm 0.8cm,clip]{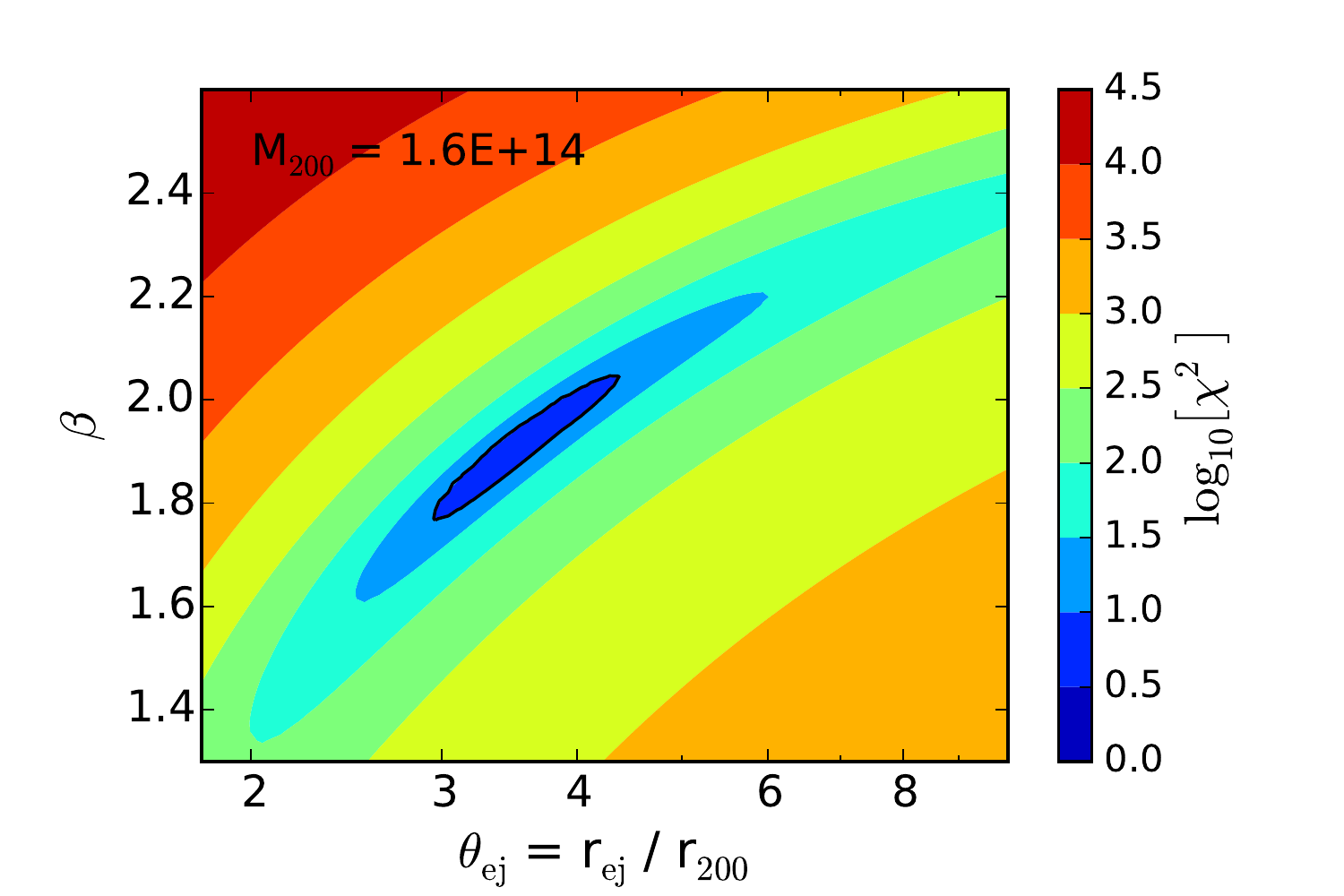}\\
\includegraphics[width=.44\textwidth,trim=0.4cm 0.1cm 1.2cm 0.8cm,clip]{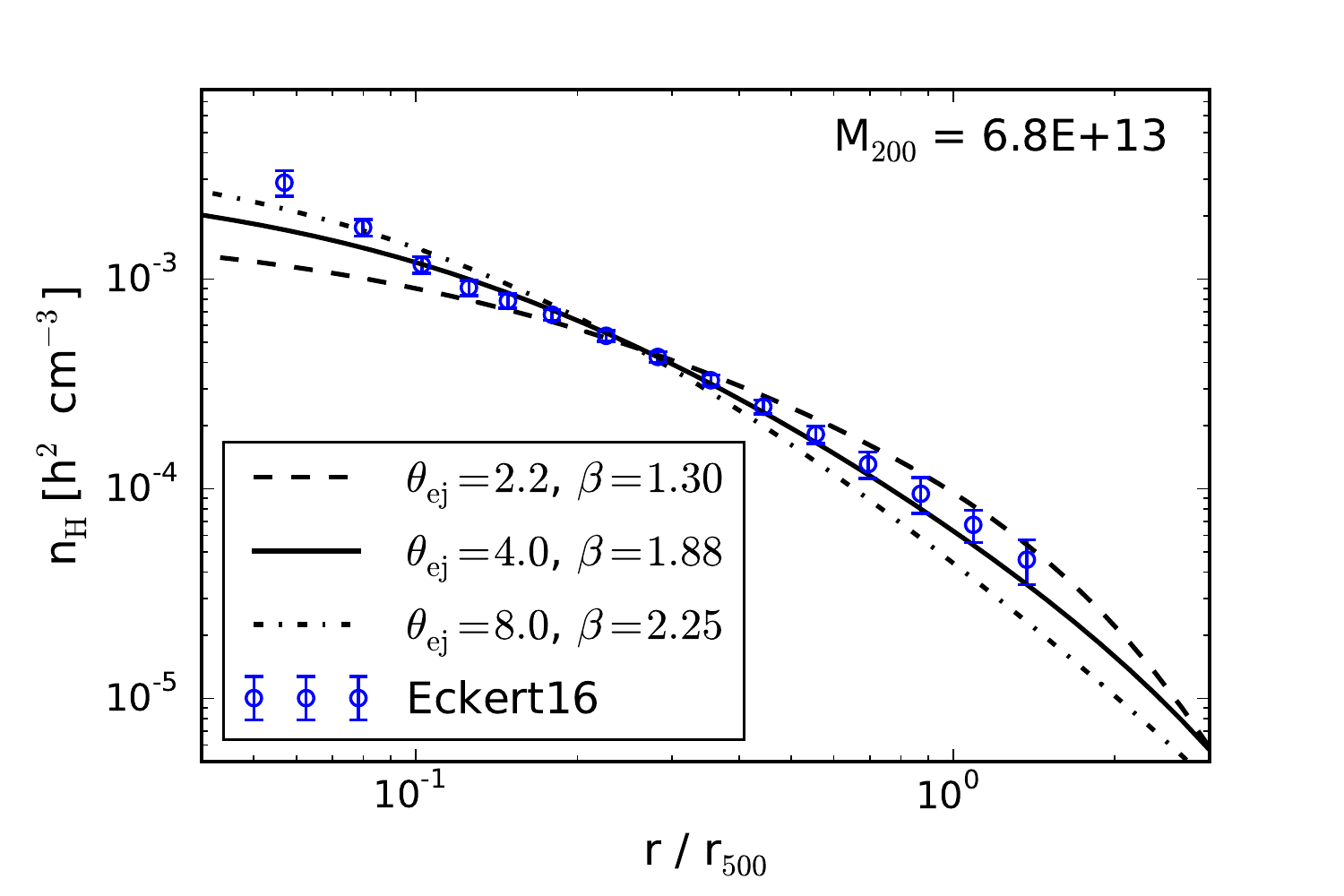}
\includegraphics[width=.44\textwidth,trim=0.4cm 0.1cm 1.2cm 0.8cm,clip]{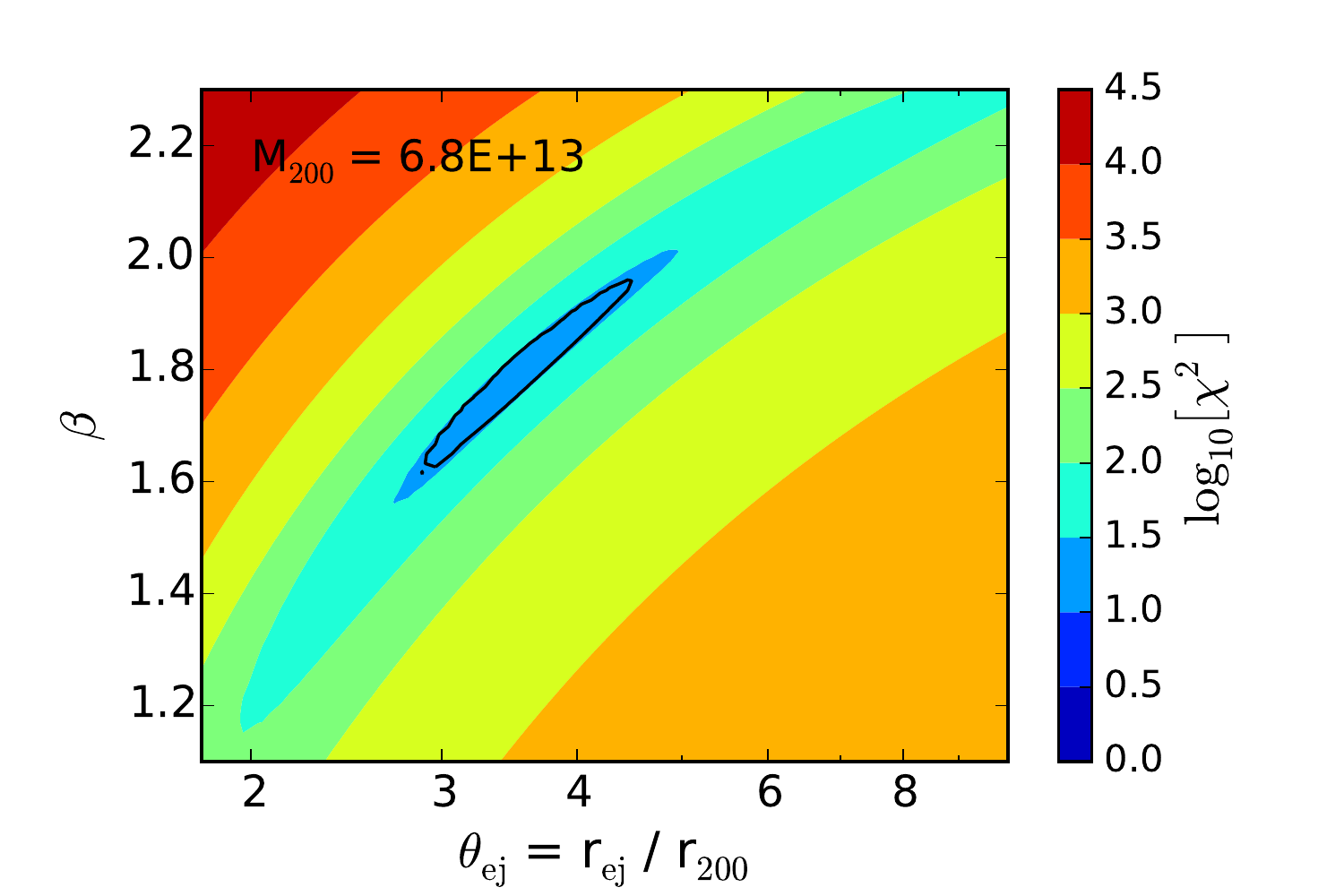}\\
\caption{\label{fig:profiles2}Same than Fig.~\ref{fig:profiles1} but for XXL survey data from {\tt XMM-Newton} which extends to smaller halo masses \citep[see Ref.][]{Eckert:2015rlr}. Four different mass bins from large clusters to galaxy groups are illustrated (from top to bottom).}}
\end{figure}

Fig.~\ref{fig:profiles2} shows the same analysis for stacked profiles of the XXL survey \citep[obtained with the {\tt XMM-Newton} satellite]{Pierre:2015cqe} split in four different temperature bins from clusters down to galaxy groups \citep[][]{Eckert:2015rlr}. The corresponding halo masses are derived based on the hydrostatic equilibrium assumption \citep[following Ref.][]{Arnaud:2005ur}. They range from $6.8\times10^{13}$ M$_{\odot}$/h to $4.5\times 10^{14}$ M$_{\odot}$/h.

The panels on the left-hand-side of Fig.~\ref{fig:profiles2} show the stacked X-ray data (coloured symbols) together with some selected BC profiles (black lines). Good agreement between model and data can be obtained for all mass bins assuming a sensible choice of $\beta$ and $\theta_{\rm ej}$. A notable exception is the smallest mass bin (bottom panel), where the data seems to suggest a central cusp in disagreement with the cored profiles assumed by the BC model. However, this has a negligible effect on the larger physical scales we are primarily interested in.

The right-hand-side panels of Fig.~\ref{fig:profiles2} show contour maps of the $\beta-\theta_{\rm ej}$ plane with the chi-square ($\chi^2$) values from a least-square regression analysis. The 2-$\sigma$ contour-lines of the likelihood distribution are shown as black lines. Regarding the ejection radius (parametrised by $\theta_{\rm ej}$), values of $\theta_{\rm ej}\sim3-6$ seem to be favoured by the data with no clear trend with total halo mass. The $\beta$-parameter, on the other hand, shows a mild mass dependence with decreasing values towards smaller halo masses.

Based on the results shown in Fig.~\ref{fig:profiles1} and \ref{fig:profiles2} we allow the gas ejection parameter ($\theta_{\rm ej}$) to vary within the very conservative range of
\be\label{thejrange}
\theta_{\rm ej}\in[2,8].
\ee
Outside of this range, the gas profile of Eq.~(\ref{rhogas}) fails to match the X-ray observations for all mass bins. The $\beta$-parameter is assumed to stay below $\beta<3$ with a potential dependence on halo mass as parametrised in Eq.~(\ref{beta}). In the following section, we will see that a decreasing value of $\beta$ with decreasing halo mass is indeed required to match the observed gas fractions of galaxy groups and clusters.

Note that in this section we have used estimates of the total halo mass based on the assumption of hydrostatic equilibrium \citep[obtained via the temperature-mass relation from Ref.][]{Arnaud:2005ur}. If we use weak-lensing data instead, the halo mass estimates are about 40 percent larger. However, we have checked that in terms of the profile fitting and the chi-square analysis, the choice of the mass estimator has very little influence on the BC parameters.


\subsection{Gas parameters}\label{sec:gasparams}
So far we have convinced ourselves that the BC model yields a good fit to the observed gas profiles of galaxy groups and clusters and we have set prior ranges on the parameters $\beta$ and $\theta_{\rm ej}$. As a next step, we now constrain the parameters $M_c$ and $\mu$ describing the slope of the gas density profile as a function of halo mass via Eq.~(\ref{beta}). 

Instead of stacked X-ray profiles, we constrain the BC model using observed fractions of gas inside of haloes (more precisely within $r_{\rm 500}$). The advantage of X-ray gas fractions is that they consist of integrated quantities that can be measured down to smaller halo masses without relying on halo stacking. The downside is that quantifying a gas fraction requires knowledge of the total halo mass inside of $r_{500}$. Traditionally, the halo mass is computed directly from the X-ray measurement assuming the gas to be in hydrostatic equilibrium. This assumption has, however, been questioned due to the potential influence of non-thermal pressure contributions significantly biasing the mass estimate. X-ray studies of large clusters have found that the hydrostatic equilibrium assumption leads to an underestimation of the total halo mass of less than ten percent \citep{Eckert:2018mlz,Ettori:2018tus}. Studies involving hydrodynamical simulations, on the other hand, point towards a stronger effect of the order of 10-30 percent with little or no dependence on halo mass \citep[e.g. Refs.][]{Nagai:2006sz,Brun:2013yva}. Finally, direct mass estimates from weak-lensing measurements yield up to 40 percent larger halo masses compared to hydrostatic estimates \citep{Lieu:2015pit}. It should be noted, however, that, even though mass estimates based on weak lensing are independent of the dynamical state of a system and therefore in principle superior, they are subject to large uncertainties \citep{Sereno:2014pfa}.

\begin{figure}[tbp]
\center{
\includegraphics[height=.28\textwidth,trim=0.0cm 0.3cm 0.9cm 0.1cm,clip]{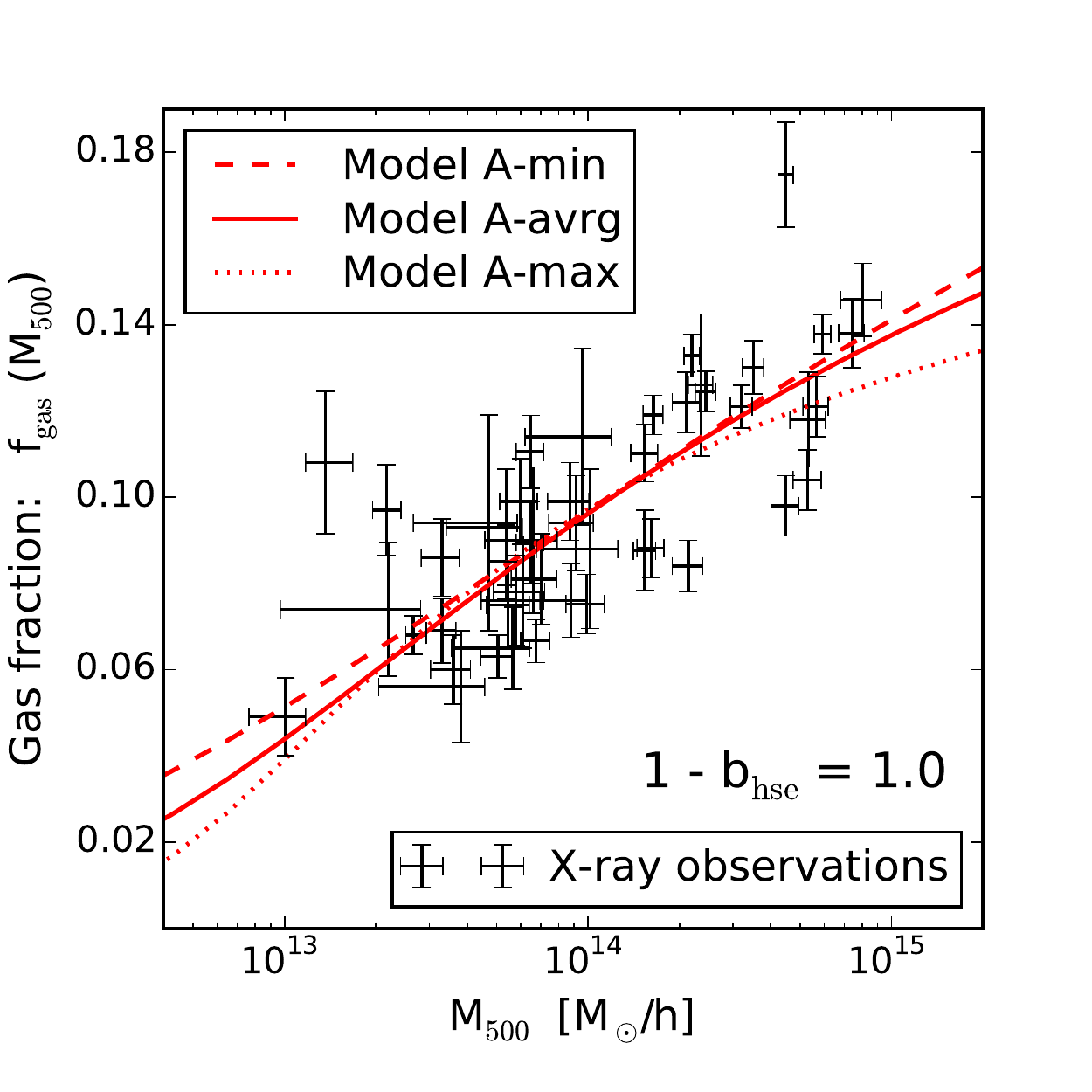}
\includegraphics[height=.28\textwidth,trim=0.5cm 0.3cm 2.8cm 0.1cm,clip]{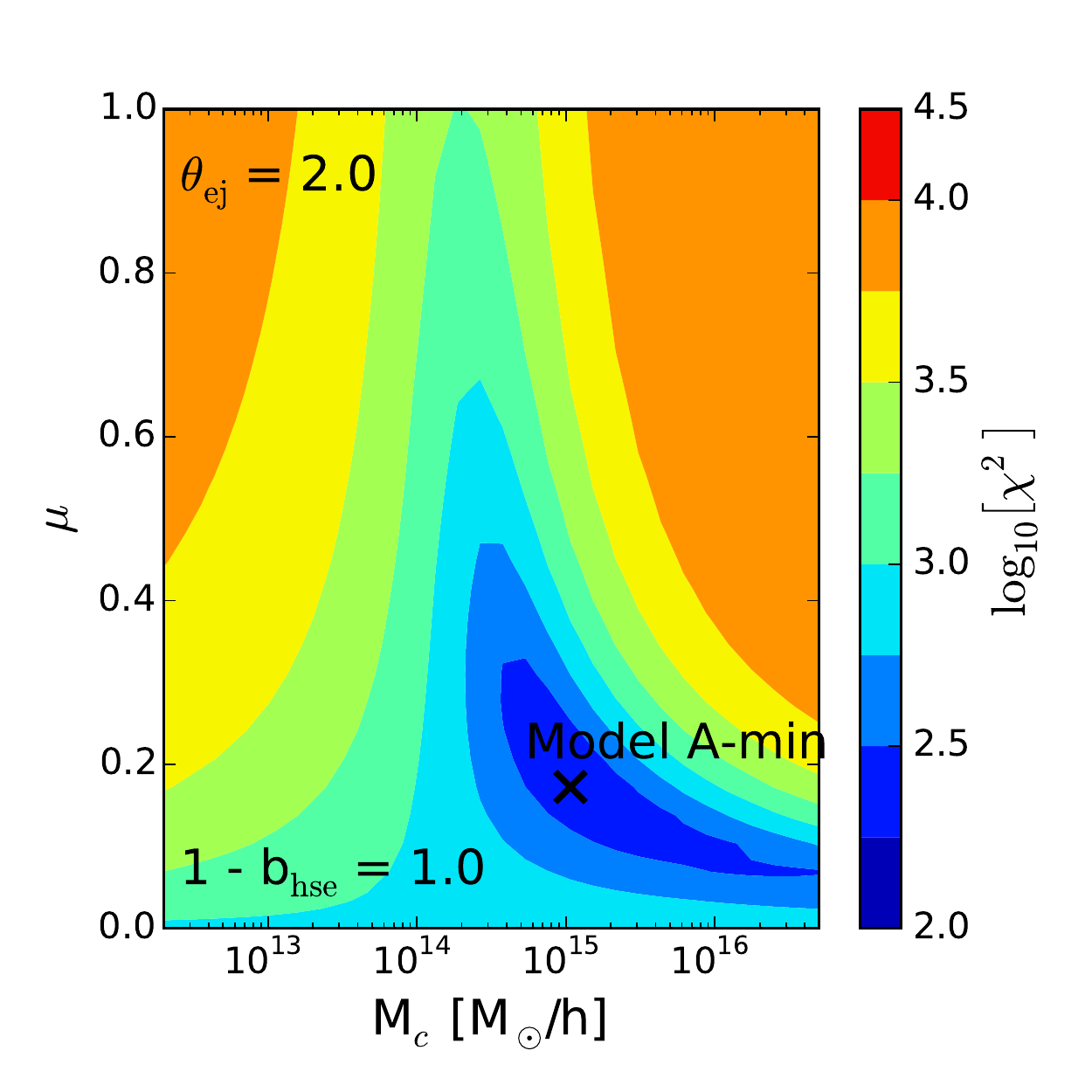}
\includegraphics[height=.28\textwidth,trim=0.5cm 0.3cm 2.8cm 0.1cm,clip]{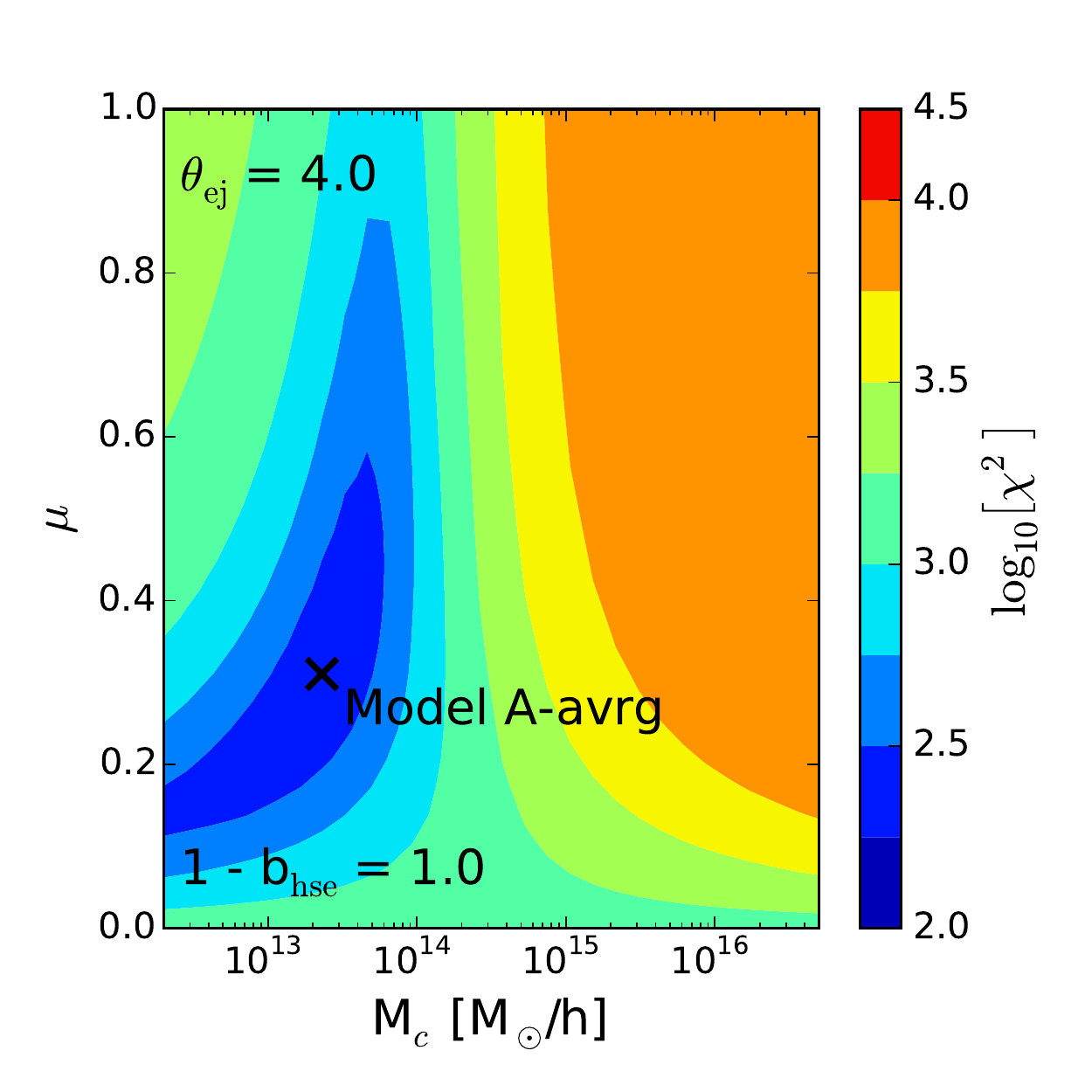}
\includegraphics[height=.28\textwidth,trim=0.5cm 0.3cm 0.5cm 0.1cm,clip]{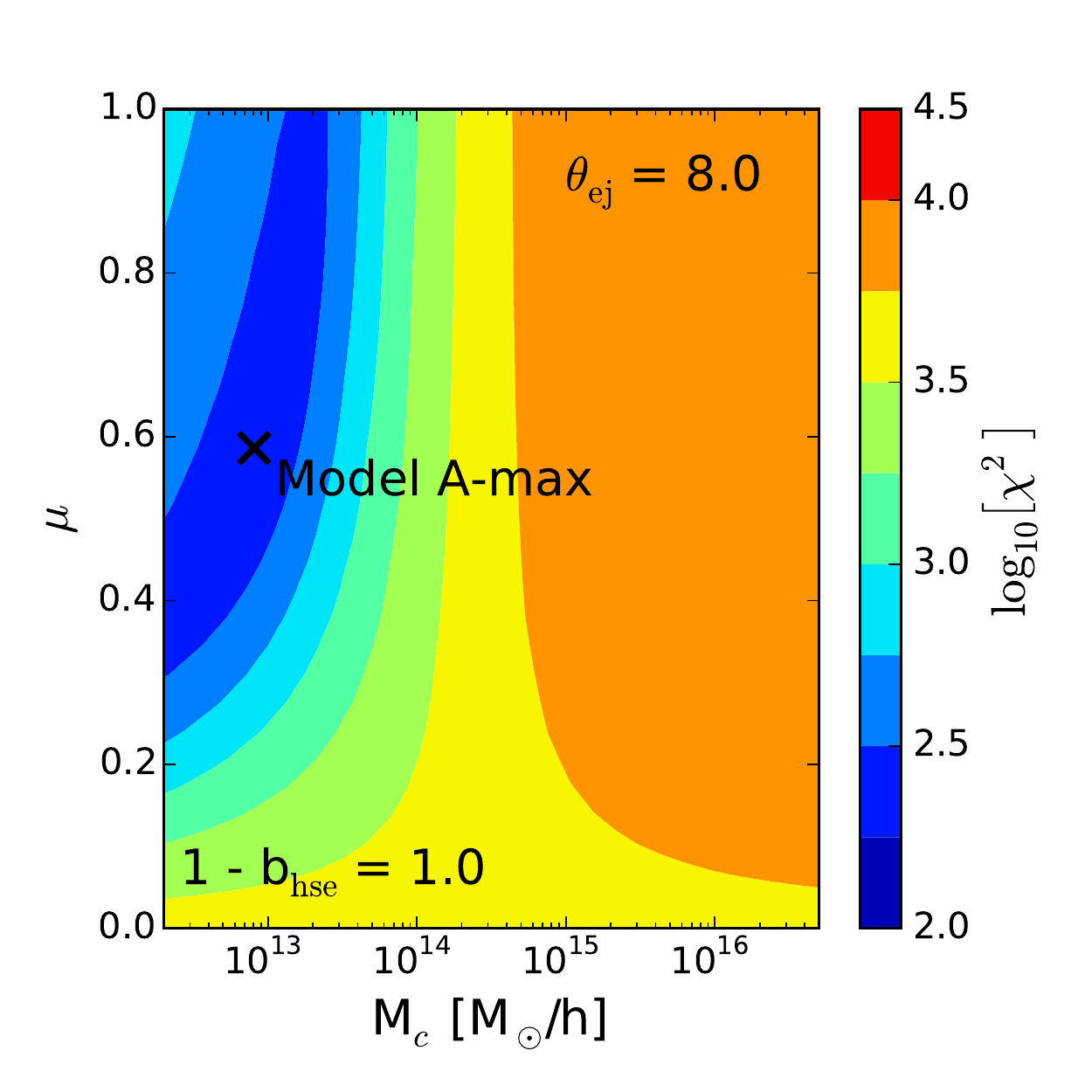}\\
\includegraphics[height=.28\textwidth,trim=0.0cm 0.3cm 0.9cm 0.1cm,clip]{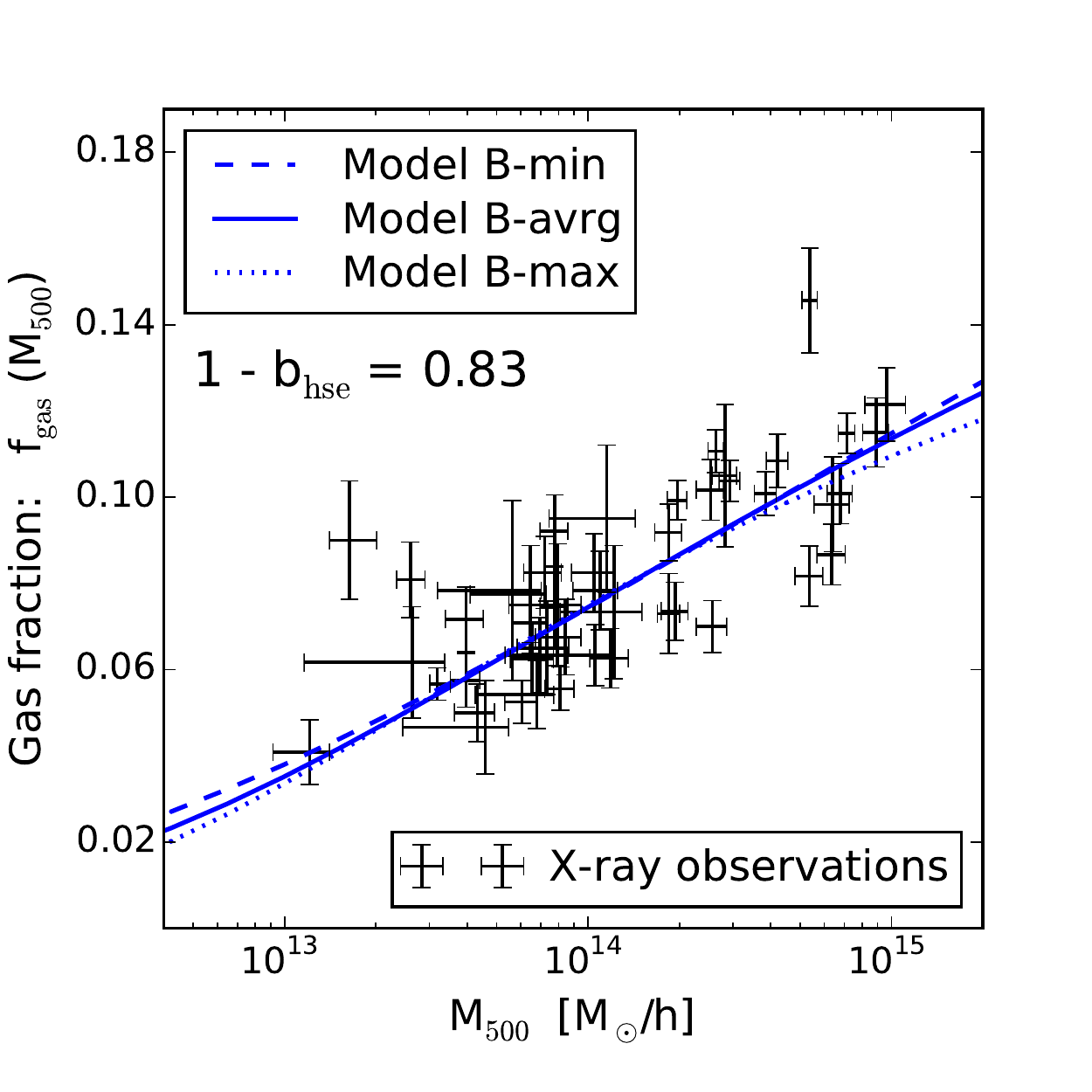}
\includegraphics[height=.28\textwidth,trim=0.5cm 0.3cm 2.8cm 0.1cm,clip]{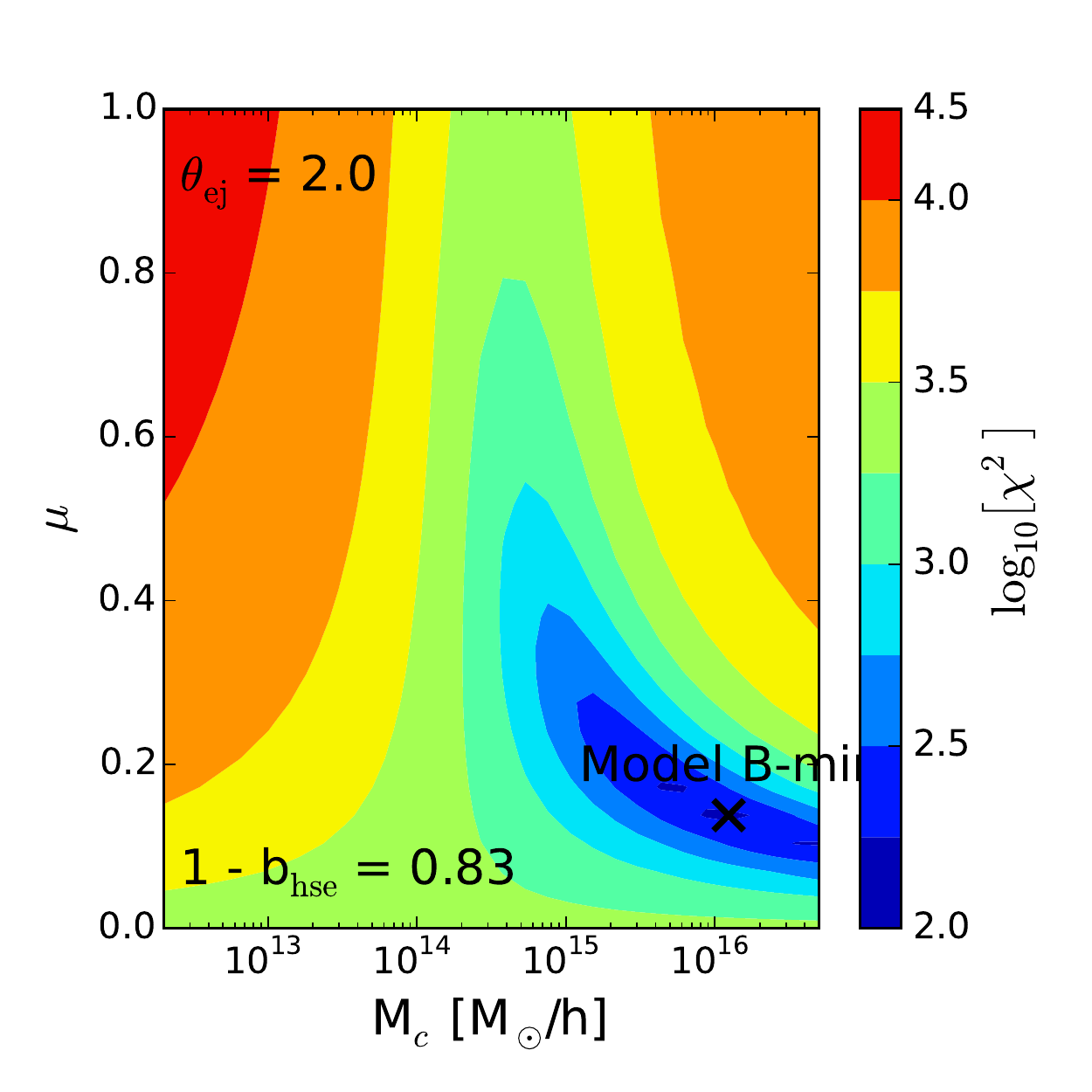}
\includegraphics[height=.28\textwidth,trim=0.5cm 0.3cm 2.8cm 0.1cm,clip]{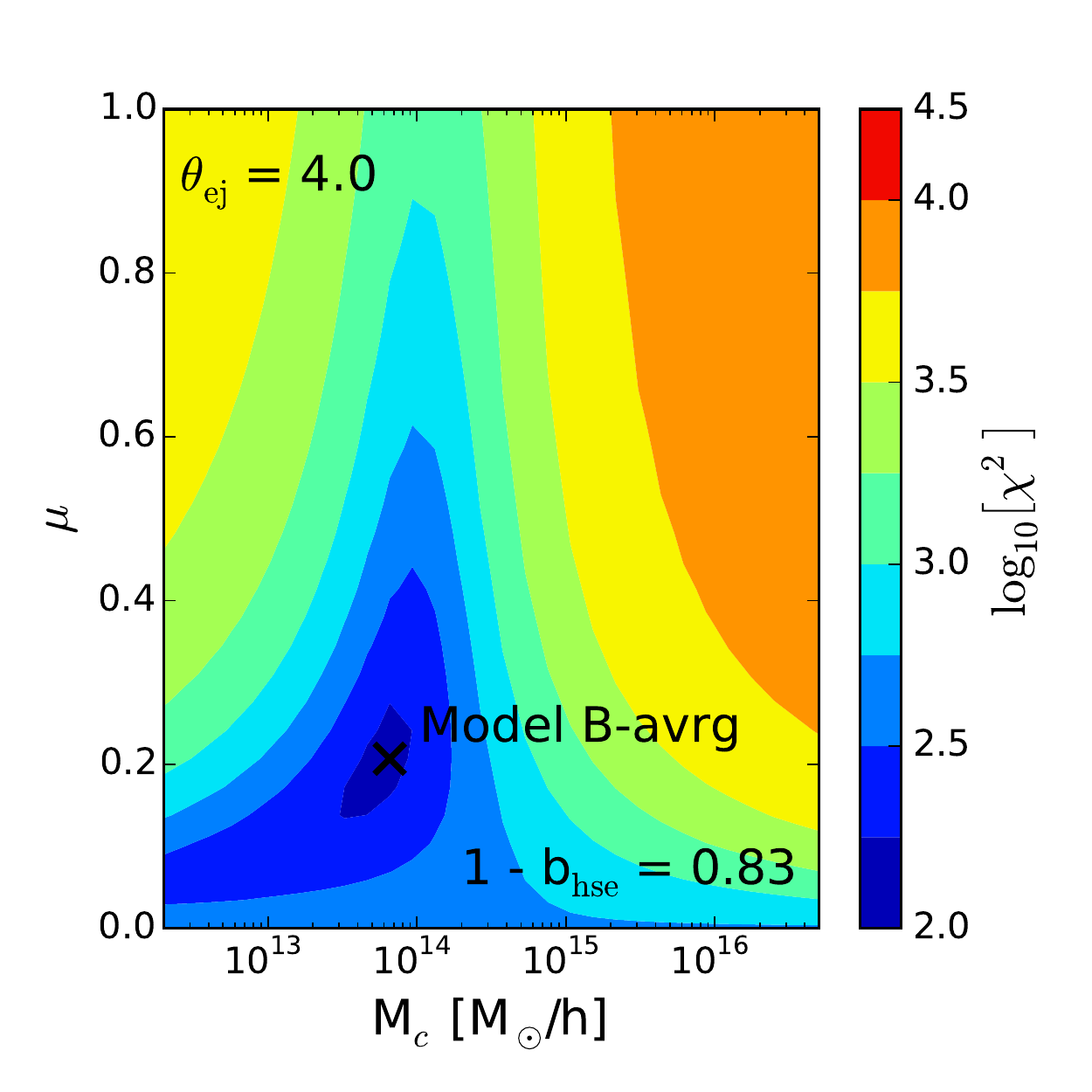}
\includegraphics[height=.28\textwidth,trim=0.5cm 0.3cm 0.5cm 0.1cm,clip]{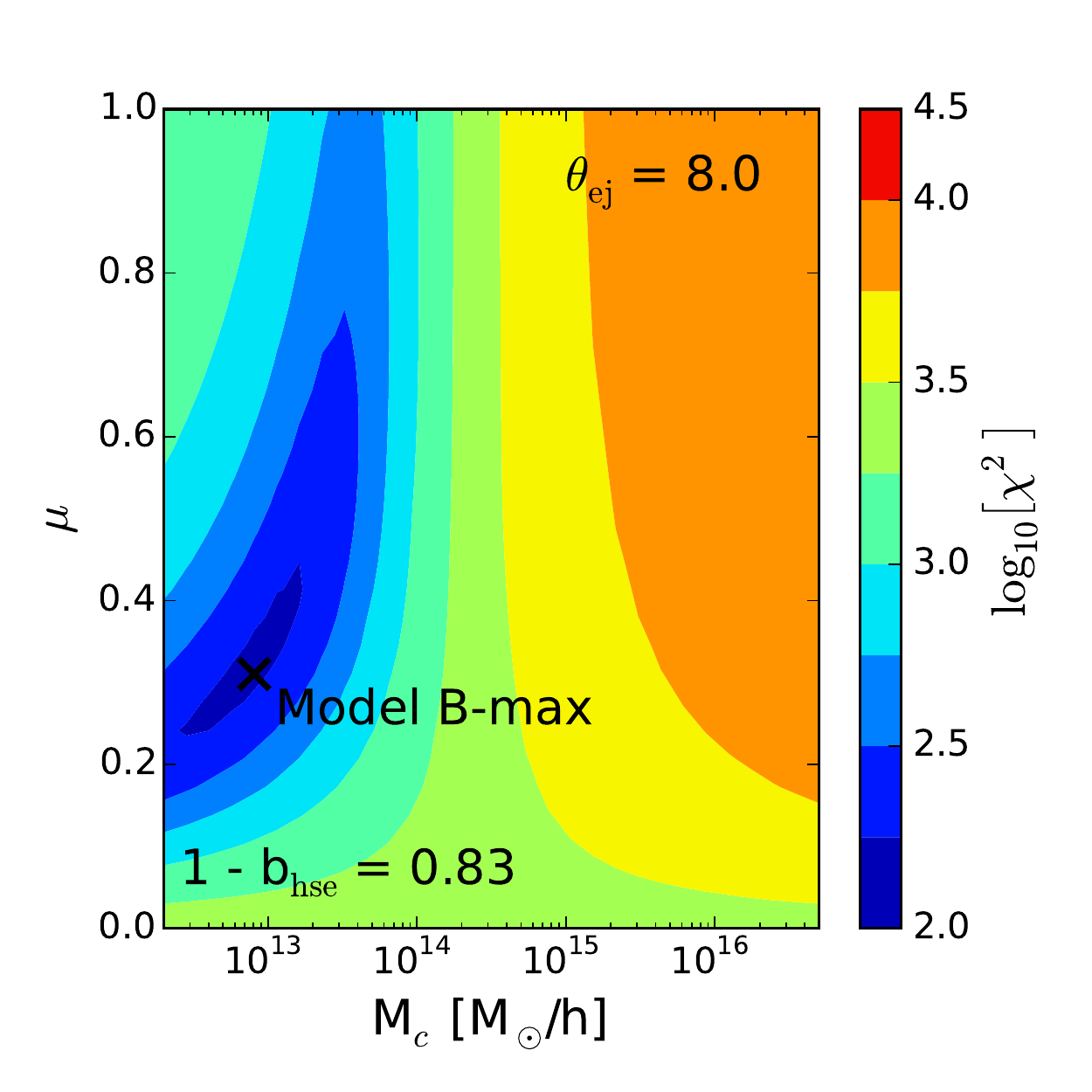}\\
\includegraphics[height=.28\textwidth,trim=0.0cm 0.3cm 0.9cm 0.1cm,clip]{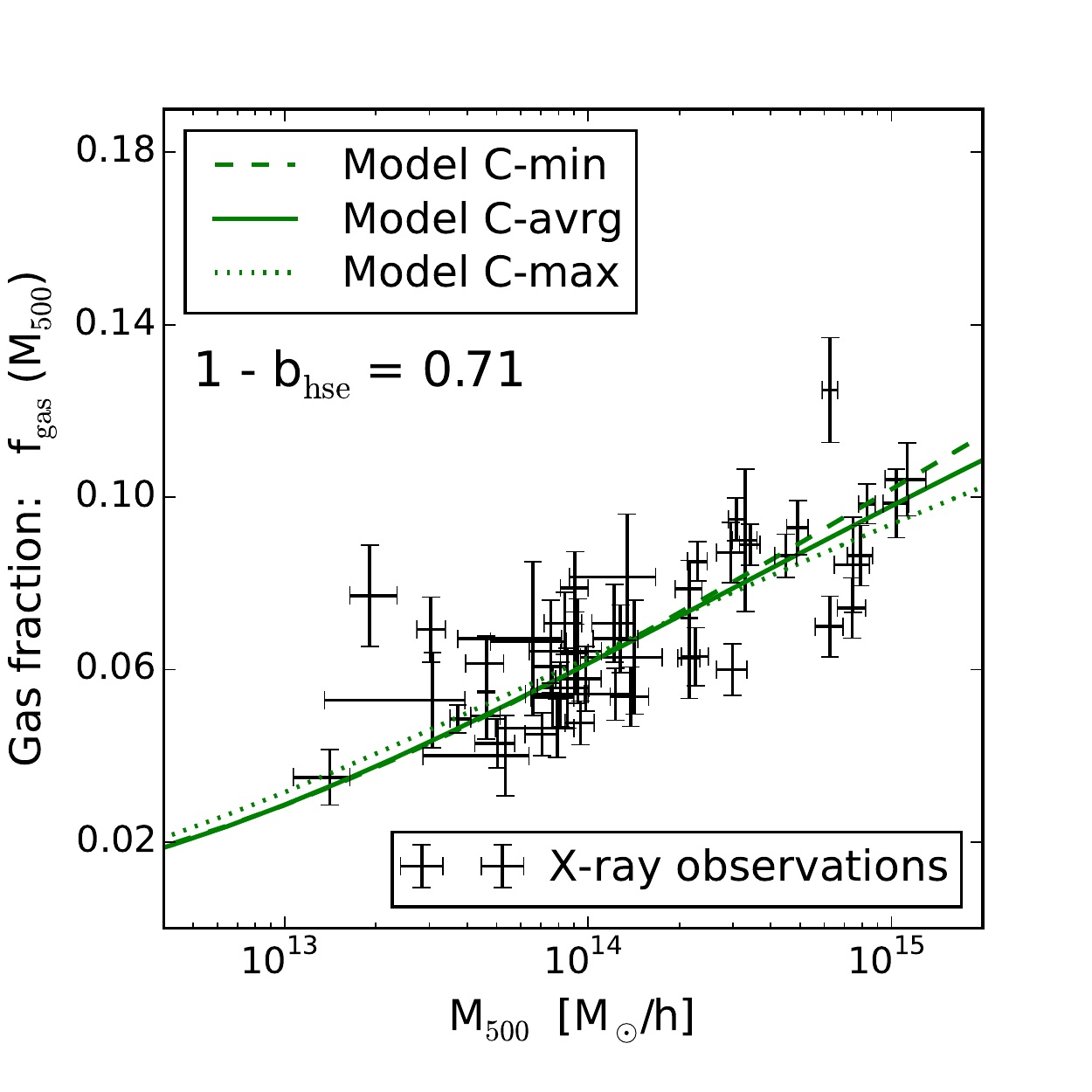}
\includegraphics[height=.28\textwidth,trim=0.5cm 0.3cm 2.8cm 0.1cm,clip]{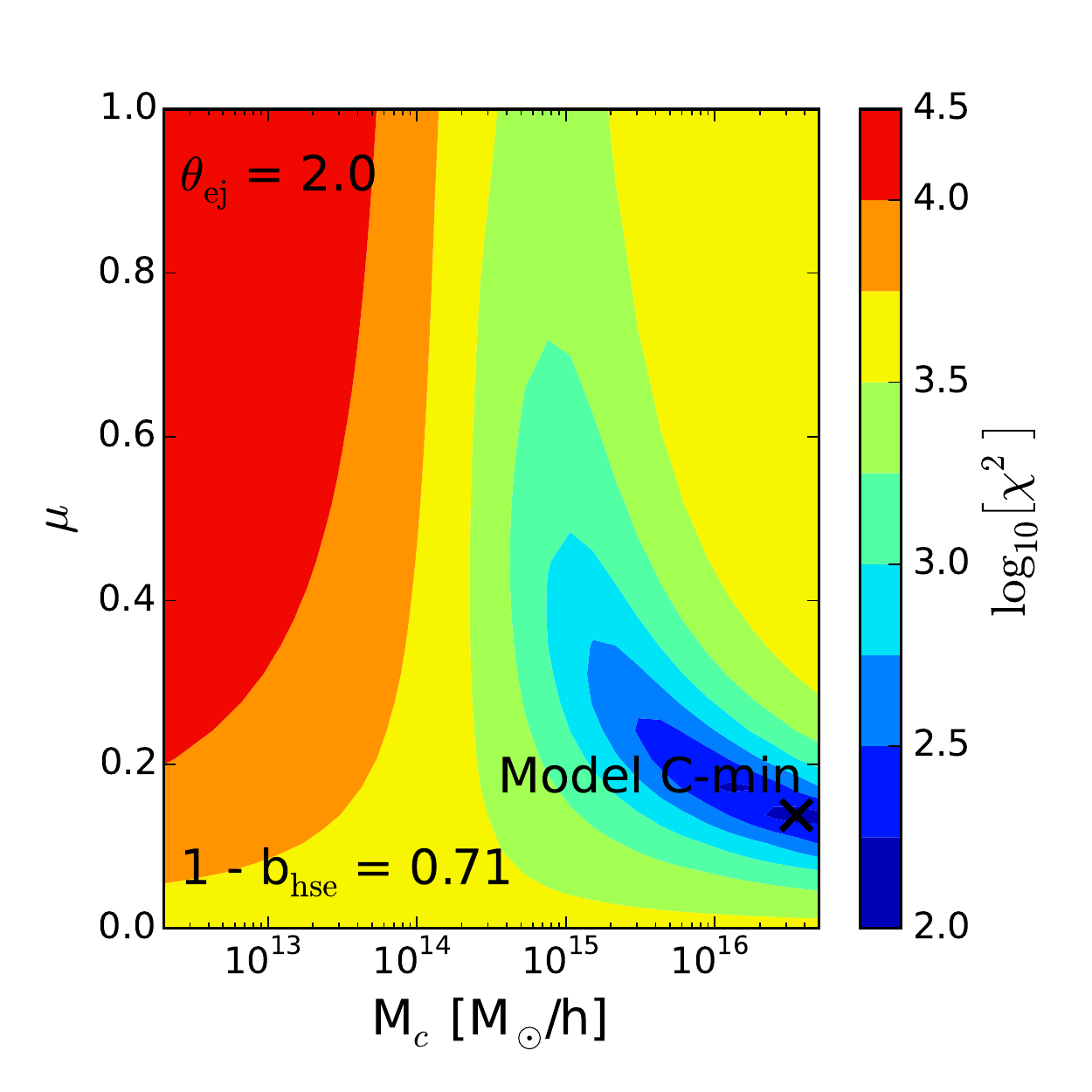}
\includegraphics[height=.28\textwidth,trim=0.5cm 0.3cm 2.8cm 0.1cm,clip]{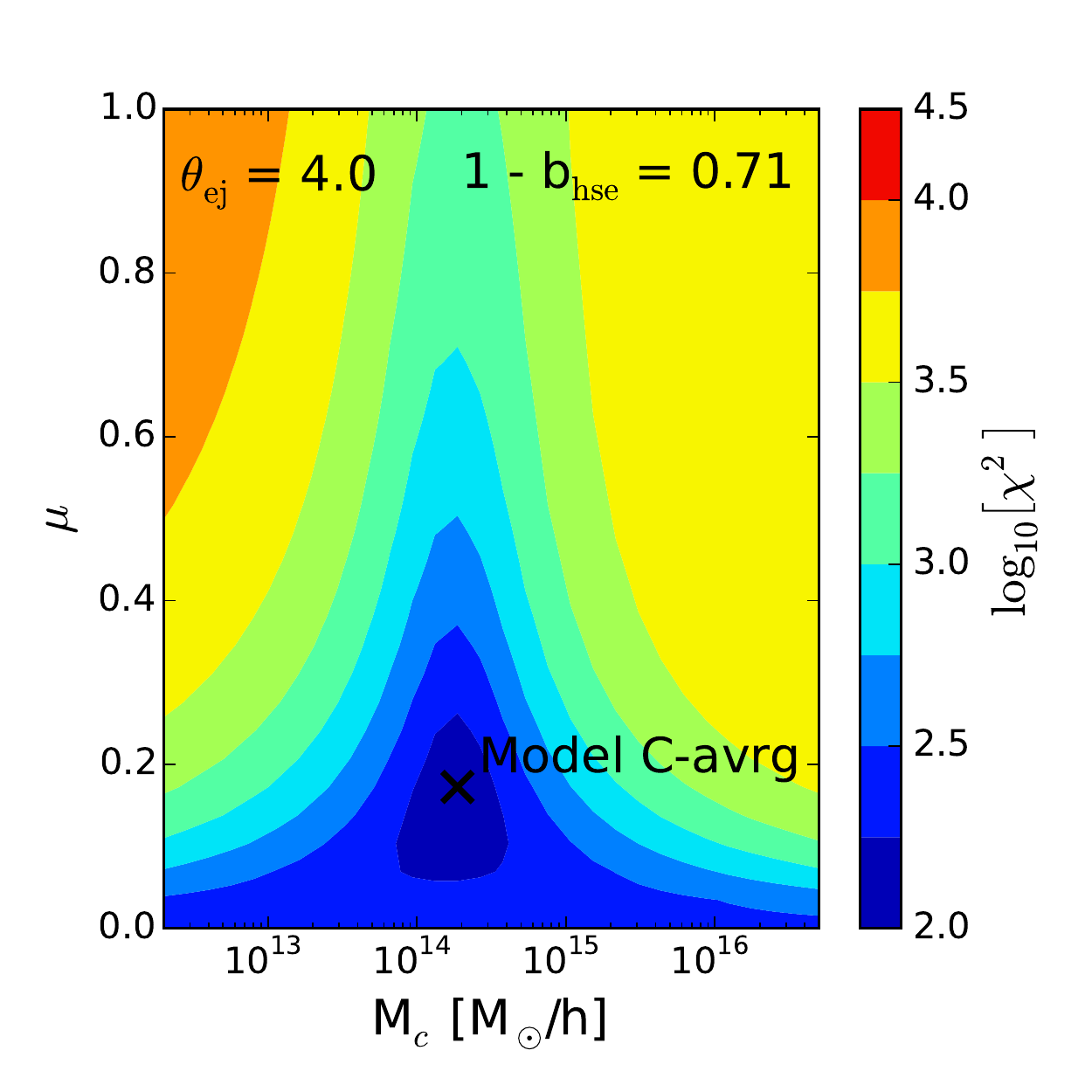}
\includegraphics[height=.28\textwidth,trim=0.5cm 0.3cm 0.5cm 0.1cm,clip]{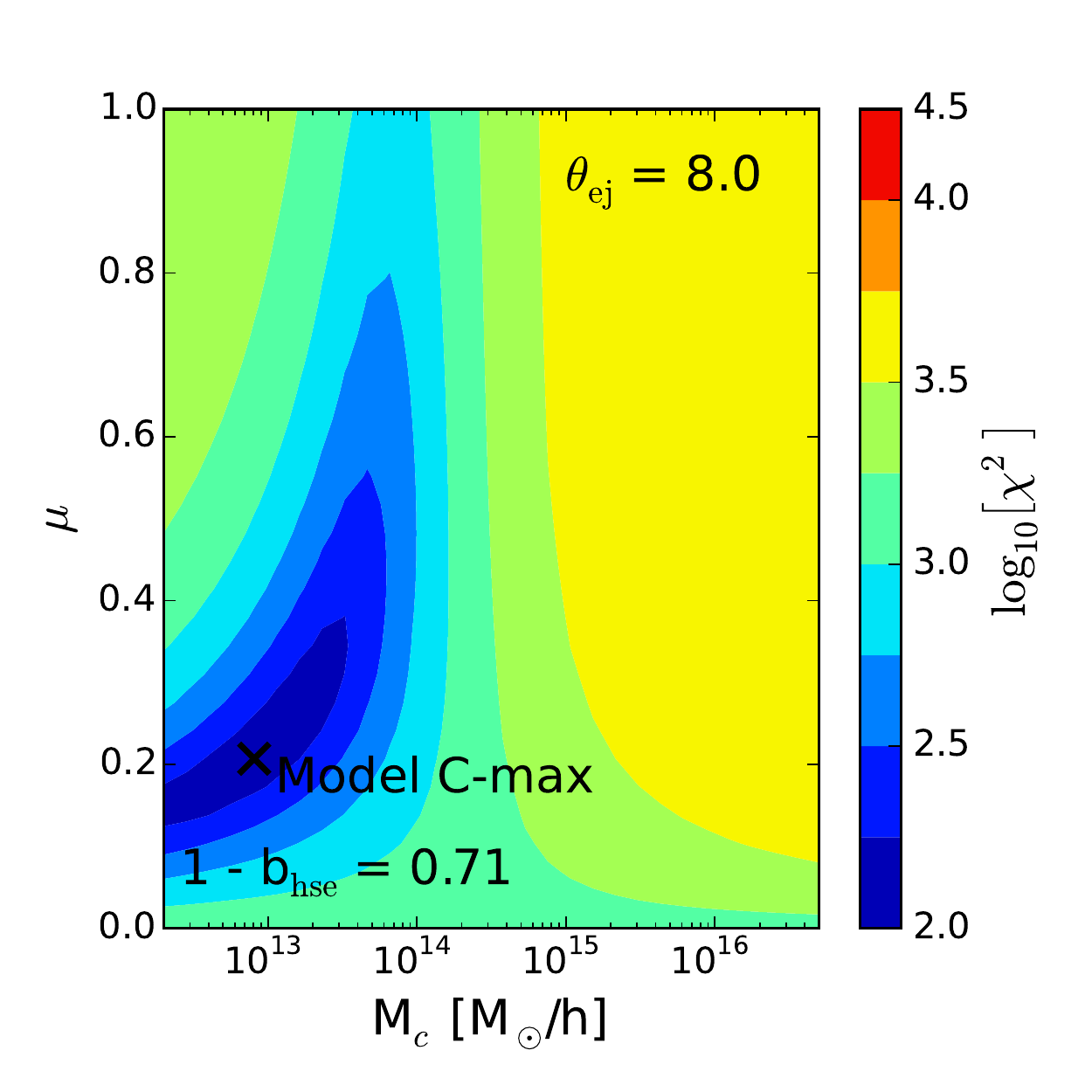}
\caption{\label{fig:fraction}Constraints on the gas parameters ($M_c$, $\mu$) for $\theta_{\rm ej}=2,4,8$ corresponding to a minimum (min), best-guess (avrg), and maximum (max) gas ejection radius. The leftmost panels show the gas fraction obtained from X-ray observations of Refs.~\citep{Sun:2008eh,Vikhlinin:2008cd,Gonzalez:2013awy} and from the BC model (coloured lines). The other panels show the chi-square ($\chi^2$) values of a least-square regression analysis, where the best-fitting models are highlighted with a black cross. The top, middle and bottom rows assume a different hydrostatic mass bias ($b_{\rm hse}$) of the X-ray observations, corresponding to Model A, B, and C in the text. While Model A assumes no bias, Model B and C assume a bias motivated by results from hydrodynamical simulations and weak-lensing mass estimates, respectively.}}
\end{figure}

In order to quantify the uncertainty related to the assumption of hydrostatic equilibrium, it is convenient to introduce a hydrostatic mass bias ($b_{\rm hse}$) via the definition
\be\label{hsebias}
1-b_{\rm hse} =  \frac{M_{\rm 500,hse}}{M_{500}},
\ee
where $M_{\rm 500,hse}$ is the mass obtained with the approximation of hydrostatic equilibrium and $M_{\rm 500}$  is the true halo mass at $r_{500}$. In the following, we will consider three models with different values for the hydrostatic mass bias, i.e.,
\be
\text{Model A:}\hspace{0.5cm}1-b_{\rm hse}&=&1.00,\nonumber\\
\text{Model B:}\hspace{0.5cm}1-b_{\rm hse}&=&0.83,\\
\text{Model C:}\hspace{0.5cm}1-b_{\rm hse}&=&0.71.\nonumber
\ee
Model A implies the hydrostatic mass to be correct, while Model B and C assume $M_{\rm hse}$ to be 20 and 40 percent below the true halo mass $M_{500}$. We assume the true answer to lie somewhere between Model A and C.

In the top-left panel of Fig.~\ref{fig:fraction} we show the gas fractions of different X-ray measurements between $M_{500}\sim10^{13}$ and $10^{15}$ M$_{\odot}$/h \citep[obtained from Refs.][]{Sun:2008eh,Vikhlinin:2008cd,Gonzalez:2013awy} based on the hydrostatic mass approximation and therefore following the assumptions of Model A. The centre-left and bottom-left panels, on the other hand, show the same observations but this time corrected to account for a hydrostatic bias of $(1-b_{\rm hse})=0.83$ and $0.71$ corresponding to Model B and C. The correction has been performed by recalculating the total mass using Eq.~(\ref{hsebias}). This leads to a shift of the data points both downwards and to the right\footnote{The additional change of the gas mass due to the recalibration of $r_{500}$ has been shown to be very small \citep[see e.g. Ref.][]{Eckert:2015rlr} and is ignored for simplicity.}.

Next to the X-ray data shown in the left-most panels of Fig.~\ref{fig:fraction}, we plot the predictions of the BC model, i.e.,
\be
f_{\rm gas}(r_{500})=M_{\rm gas}(r_{500})/M_{\rm dmb}(r_{500}),
\ee
for the best best-fitting values of the gas parameters $\mu$ and $M_{\rm c}$ (coloured lines). We consider three cases $\theta_{\rm ej}=2,4,8$ corresponding to a minimum (min), best-guess (average, denoted avrg), and maximum (max) allowed gas ejection radius as defined in Eq.~(\ref{thejrange}).

The results of a least-square regression analysis over the full $M_c$-$\mu$ parameter space is shown at the right-hand-side of Fig.~\ref{fig:fraction}. The colour maps indicate the chi-square ($\chi^2$) values of the fits to the X-ray gas fractions shown in the corresponding panel on the left. The best-fitting values are highlighted with a black cross. They correspond to the benchmark models for a given hydrostatic mass bias and gas ejection radius.

A summary of these benchmark models is provided in Table~\ref{tab:paramvalues}. Next to the assumed hydrostatic mass bias ($b_{\rm hse}$) and gas ejection parameter ($\theta_{\rm ej}$), we list the best fitting values for $M_c$ and $\mu$. Most notably, we find $M_c>0$ and $\mu>0$ for all models, indicating a clear mass dependence of the gas density slope with shallower densities for galaxy groups compared to clusters. This is a direct result of the observed decrease of X-ray gas fractions towards smaller halo masses shown in Fig.~\ref{fig:fraction}. The fact that gas profiles are shallower compared to the NFW profile of dark matter is a direct consequence of AGN feedback pushing gas away from the halo centres. This mechanism is more effective for galaxy groups compared to clusters (as the former have shallower gravitational potentials) leading to the observed mass dependence.

\begin{table}
\caption{Benchmark models with best-fitting parameters ($M_c$, $\mu$) for different assumptions regarding the hydrostatic mass bias ($b_{\rm hse}$) and the gas ejection radius ($\theta_{\rm ej}$). The stellar parameters are fixed at $\eta_{\rm star}=0.32$ and $\eta_{\rm cga}=0.6$.\label{tab:paramvalues}}
\begin{center}
\small
 \setcellgapes{1pt}\makegapedcells
\begin{tabular}{p{3.5cm}p{3.0cm}p{2.0cm}p{4.0cm}p{0.6cm}}
\hline
 Name  & $1-b_{\rm hse}$ & $\theta_{\rm ej}$ & $M_{\rm c}$ [M$_{\odot}$/h] & $\mu$  \\
 \hline
 \hline
 Model A-min & 1.0 & 2 & $1.1\times10^{15}$ & 0.17\\
 Model A-avrg & 1.0 & 4 &$2.3\times10^{13}$ & 0.31\\
 Model A-max & 1.0 & 8 & $8.1\times10^{12}$ & 0.59\\
 \hline
 Model B-min & 0.833 & 2 & $1.2\times10^{16}$ & 0.14\\
 Model B-avrg & 0.833 & 4 & $6.6\times10^{13}$ & 0.21\\
 Model B-max & 0.833 & 8 & $8.1\times10^{12}$ & 0.31\\
 \hline
 Model C-min & 0.714 & 2 &$3.5\times10^{16}$ & 0.14\\
 Model C-avrg & 0.714 & 4 &$1.9\times10^{14}$ & 0.17\\
 Model C-max & 0.714 & 8 &$8.1\times10^{12}$ & 0.21\\
 \hline
\end{tabular}
\end{center}
\end{table}


\section{Predictions for the matter clustering signal}\label{sec:predictions}
In the last two sections we have established a simple model for the total density distribution and constrained the model parameters for the gas and stellar components with observations. In order to account for potential systematics of the X-ray gas fractions, we have investigated three different values for the hydrostatic mass bias motivated by direct X-ray estimates, hydrodynamical simulations, and weak lensing mass measurements. In the present section, we build upon these best-fitting models to provide predictions for the matter power spectrum and the weak lensing shear correlation.

\begin{figure}[tbp]
\center{
\includegraphics[width=.8\textwidth,trim=0.2cm 1.25cm 1.2cm 0.6cm,clip]{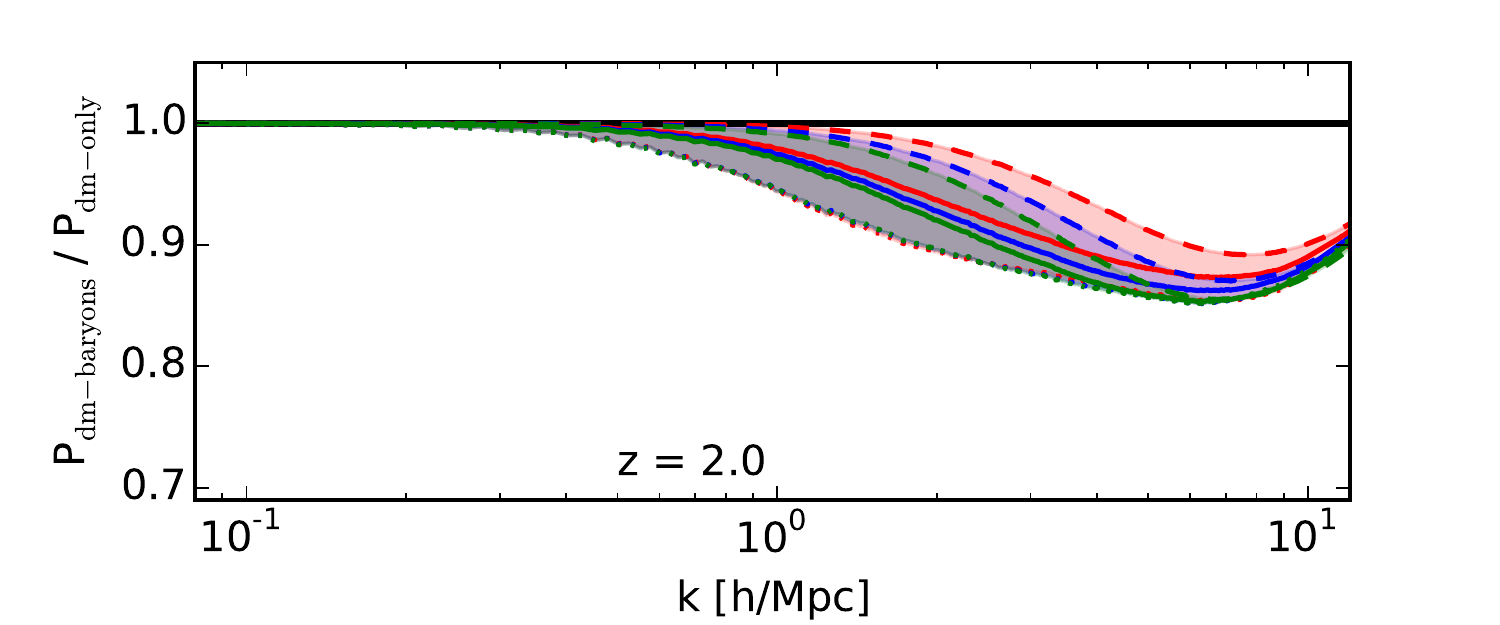}\\
\includegraphics[width=.8\textwidth,trim=0.2cm 1.25cm 1.2cm 0.6cm,clip]{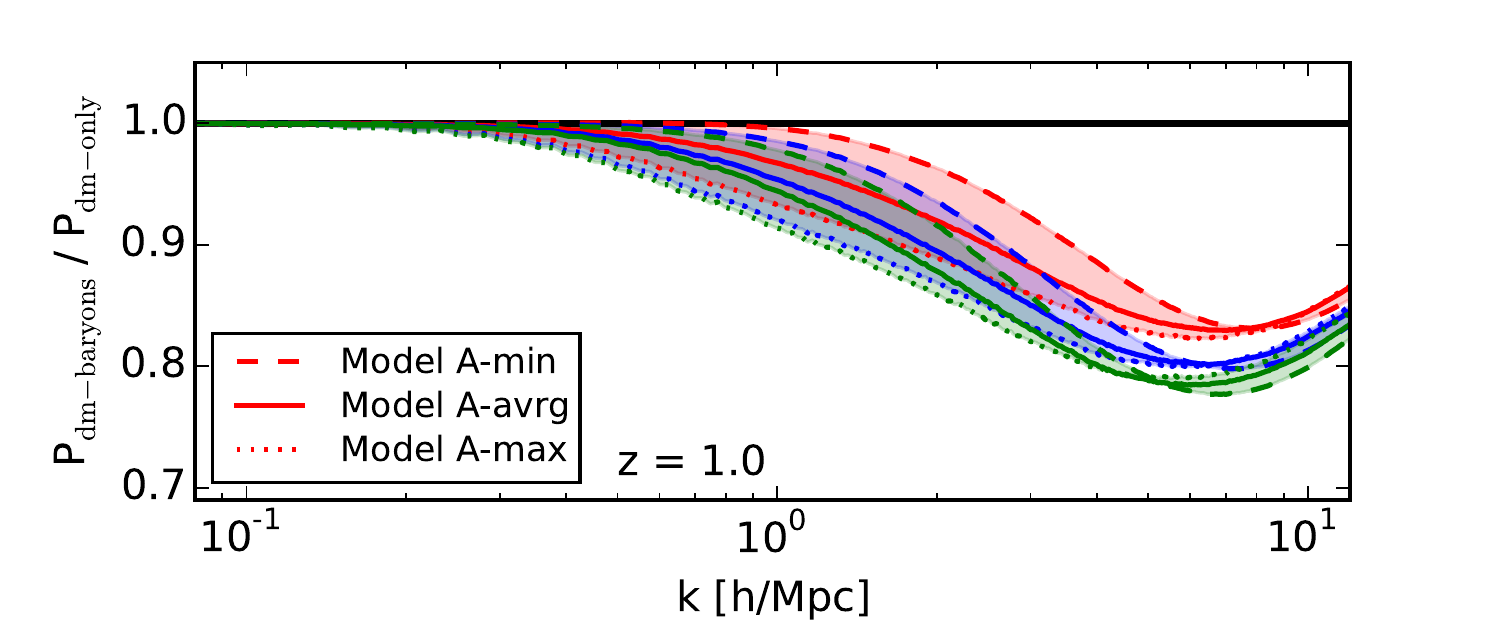}\\
\includegraphics[width=.8\textwidth,trim=0.2cm 1.25cm 1.2cm 0.6cm,clip]{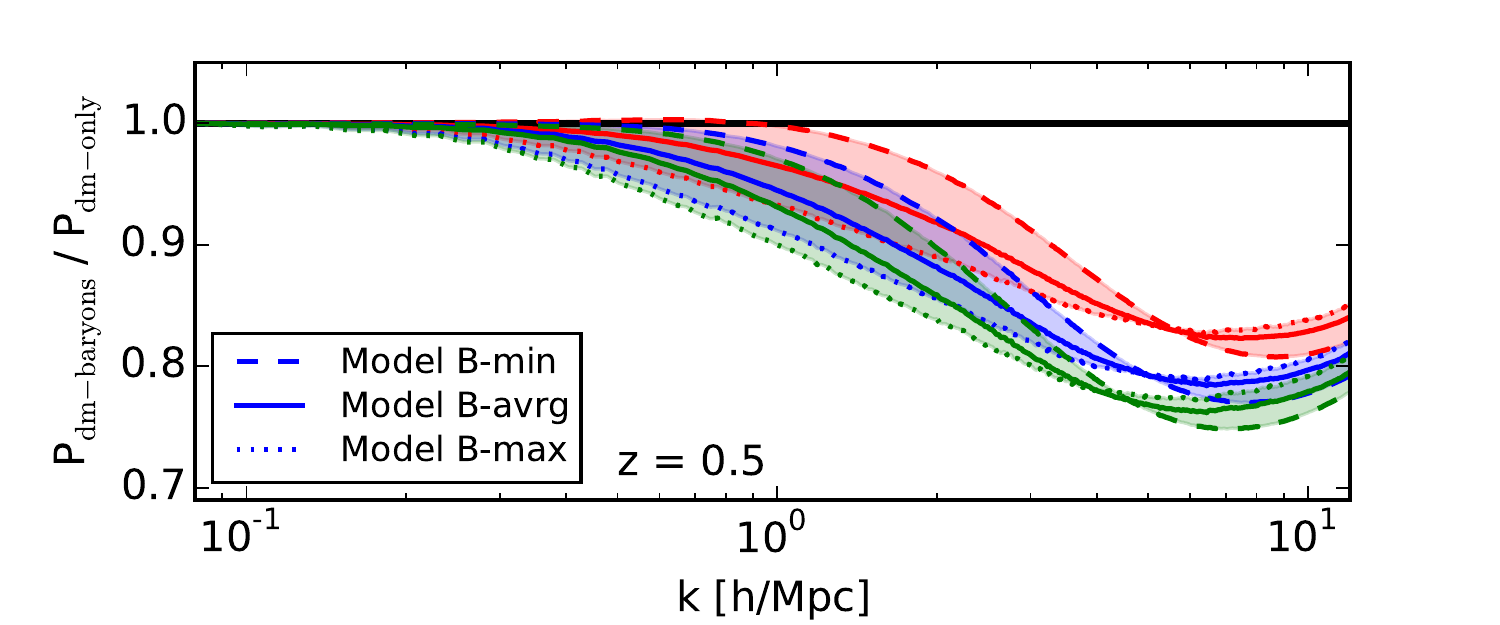}\\
\includegraphics[width=.8\textwidth,trim=0.2cm 0.1cm 1.2cm 0.6cm,clip]{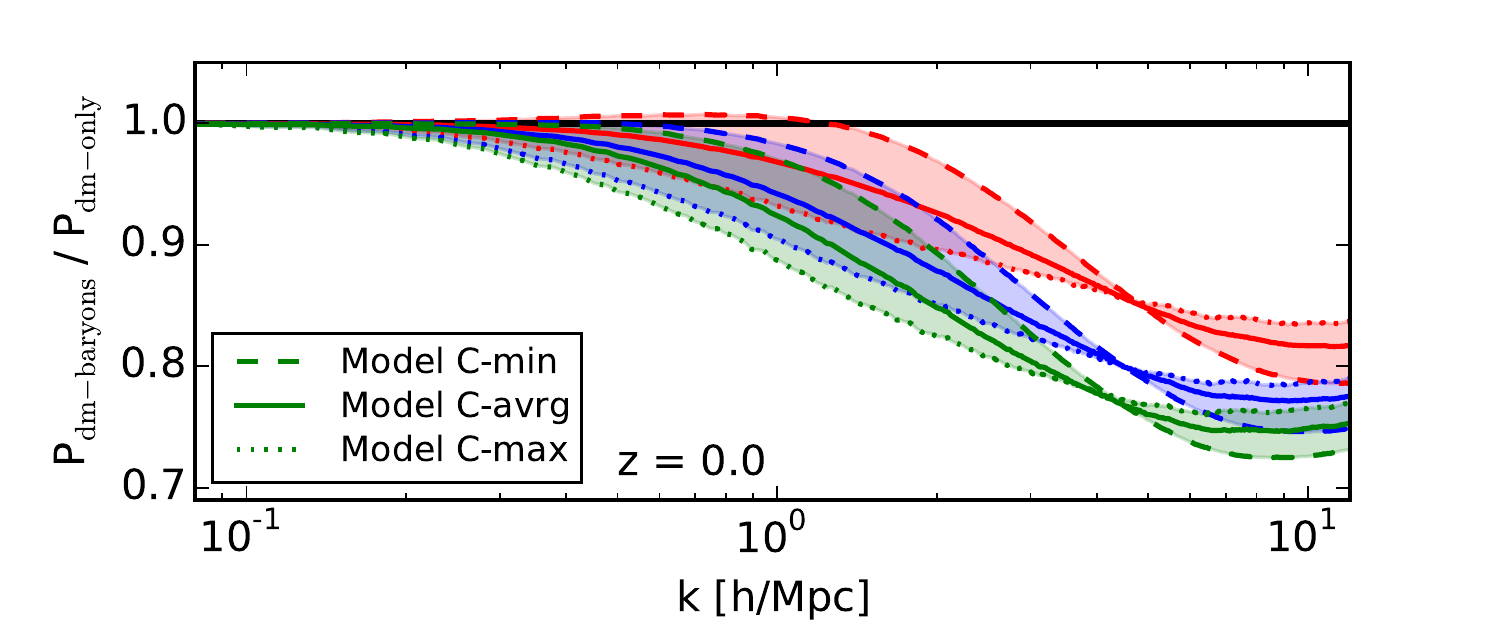}
\caption{\label{fig:PS}Ratio between the dark-matter-baryon and the dark-matter-only power spectra from the \emph{baryonic correction model} with parameters calibrated to X-ray observations. Different colours correspond to different assumptions regarding the X-ray hydrostatic mass bias. We furthermore assume $\theta_{\rm ej}=2,\,4,\,8$ for the minimum (min), best-guess (avrg), and maximum (max) allowed gas ejection radius (dashed, solid, and dotted lines). Each panel refers to a different redshift. All benchmark models are summarised in Table~\ref{tab:paramvalues}.}}
\end{figure}

\subsection{Matter power spectrum}
It is well known that baryon effects lead to a power suppression at medium and a power enhancement at very small cosmological scales compared to results from gravity-only $N$-body simulations. The suppression of power is due to feedback effects pushing gas out of haloes, while the enhancement is a result of star formation and subsequent dark matter contraction at the halo centres. Current hydrodynamical simulations reproduce this general trend \citep{vanDaalen:2011xb,Vogelsberger:2014dza,Hellwing:2016ucy,Mummery:2017lcn,Springel:2017tpz,Chisari:2018prw} but their predictions do not agree at a quantitative level. The latter is not surprising because different feedback recipes are at work depending on the simulation. Note, however, that only few current simulations reproduce the observed X-ray baryon fraction of galaxy groups and clusters (shown in Fig.~\ref{fig:fraction}).

The benchmark models listed in Table~\ref{tab:paramvalues} are calibrated to the X-ray observations assuming different values for the hydrostatic mass bias and the gas ejection radius within current uncertainties. Based on this, the BC method can then be used to predict the matter power spectrum.

In Fig.~\ref{fig:PS} we plot the power spectrum of the BC model relative to the dark-matter-only case. The different benchmark models A, B, and C (each of them with a minimum, a best-guess, and a maximum allowed gas ejection radius) are illustrated with coloured lines. Any realistic model is expected to lie within this range (which is further highlighted by the shaded areas).

The different panels of Fig.~\ref{fig:PS} refer to redshift bins from $z=2$ (top) to $z=0$ (bottom). While the baryon suppression is of order 10 percent at $k\gtrsim 1$ h/Mpc for $z\sim2$, it deepens and extents to smaller $k$-values towards lower redshifts. At redshift zero, the BC model predicts wave modes above $k\sim0.2-1$ h/Mpc to be affected by baryons with a maximum suppression of 15-25 percent at $k\sim 10$ h/Mpc.

Note that there is no explicit redshift dependence in the parametrisation of the BC model (see discussion in Sec.~\ref{BCMredshift}). The redshift evolution visible in Fig.~\ref{fig:PS} comes from the fact that at different redshifts the signal of the power spectrum is dominated by different mass scales. At redshift zero the power spectrum is strongly influenced by haloes hosting galaxy groups and small clusters, while Milky-Way sized haloes do not affect the signal below $k\sim10$ h/Mpc \citep{vanDaalen:2015msa,Schneider:2011yu}. This is very different for higher redshifts where cluster haloes have not formed yet. The redshift evolution of the BC model is in good agreement with results from hydrodynamical simulations, at least for $z\lesssim2$ (see Sec.~\ref{comparingtohydro}). This is very encouraging since any explicit redshift dependence would make the current parametrisation of the BC model considerably more complicated.

\begin{figure}[tbp]
\center{
\includegraphics[width=.99\textwidth,trim=0.2cm 0.6cm 1.2cm 0.6cm,clip]{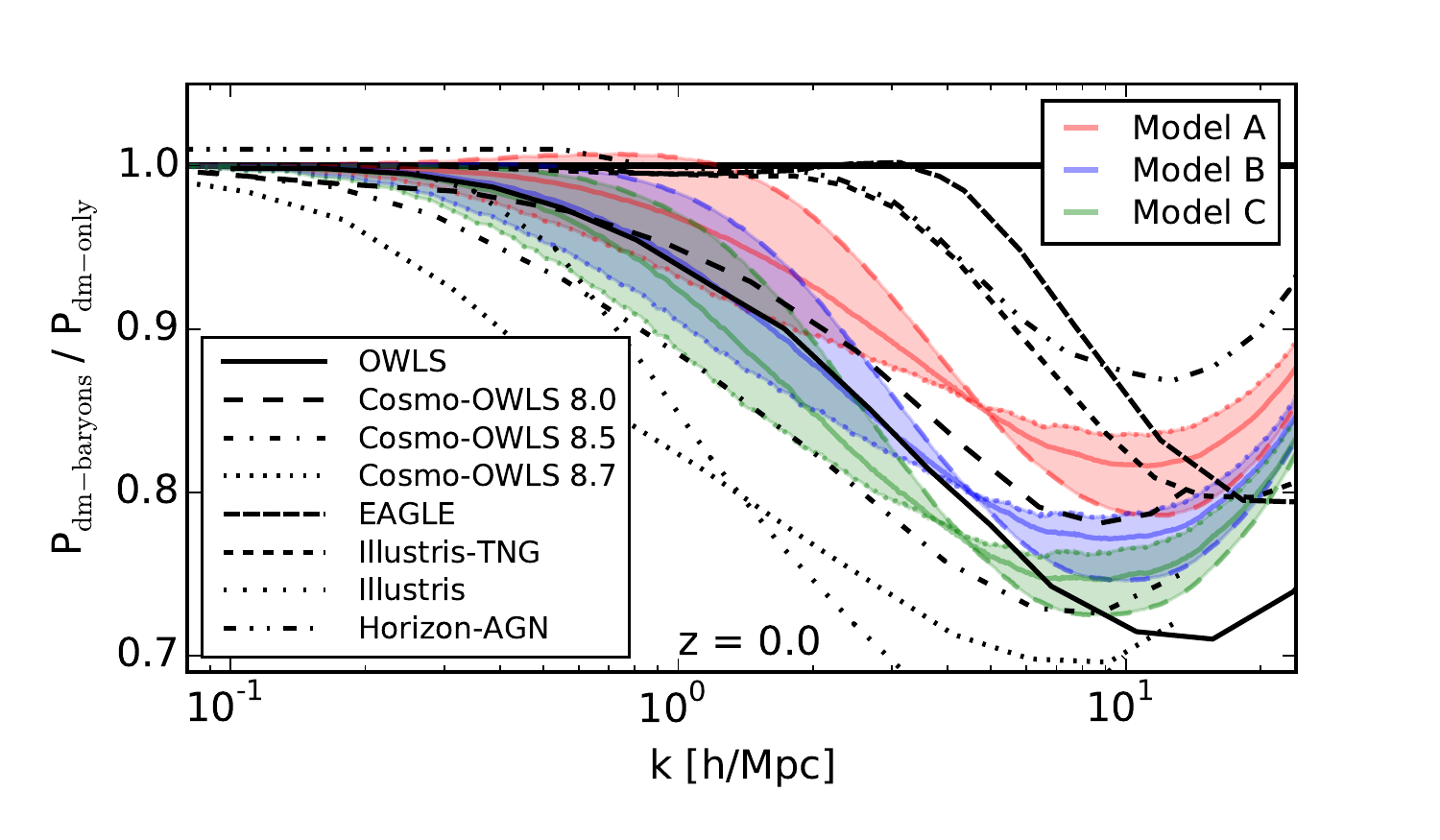}
\caption{\label{fig:PShydro}Comparison between the benchmark models A, B, and C (same as Fig.~\ref{fig:PS}) and results from hydrodynamical simulations with AGN feedback \citep{vanDaalen:2011xb,Mummery:2017lcn,Hellwing:2016ucy,Springel:2017tpz,Vogelsberger:2014dza,Chisari:2018prw}. Note that the majority of these simulations do not match the observed X-ray baryon fraction. A thorough comparison between the BC model and hydrodynamical simulations is provided in Sec.~\ref{comparingtohydro} and Appendix~\ref{hydrosims}.}}
\end{figure}

In Fig.~\ref{fig:PShydro} we again plot the benchmark power spectra of the BC model, but this time we compare them to predictions of various hydrodynamical simulations from the literature. The first thing to notice is the large differences between different simulations. While some of the simulations predict a beyond percent suppression for $k\gtrsim 0.1$ h/Mpc, others show no effect until $k\sim 4$ h/Mpc, i.e. at about 40 times smaller scales. The maximum amplitude also varies substantially ranging from about 10 to 40 percent. At face value, this large spread in both scale and amplitude between different hydrodynamical simulations is very worrying, as it means that we cannot predict the relevant scales of future weak lensing and galaxy clustering surveys. However, it is important to notice that many of these simulations do not reproduce the observed fraction of gas in galaxy groups and clusters, which has been shown to strongly correlate with the baryon suppression of the matter power spectrum (see ST15). The BC model is an ideal tool to make use of this connection in order to obtain better predictions for the matter power spectrum.

In summary, the colour shaded bands in Fig.~\ref{fig:PShydro} provide a measure for the uncertainty of the baryon suppression effect on the matter power spectrum based on the X-ray gas fractions from Refs~\citep{Sun:2008eh,Vikhlinin:2008cd,Gonzalez:2013awy} (see Fig.~\ref{fig:fraction}). By far the strongest uncertainty comes from the hydrostatic mass bias which is still poorly known. However, we have allowed for an up to forty percent bias, which is a very conservative assumption. Other potential systematics of the baryonic correction model are considerable smaller and therefore not included here (see Appendix \ref{systematics} for a detailed discussion). As a result, we conclude that the true baryon power suppression lies somewhere between the uppermost red and the lowermost green line.

\subsection{Weak lensing shear spectrum}
While the matter power spectrum provides a useful measure of the clustering process, it is not a direct observable. In this section we therefore derive the projected angular power spectrum of the weak lensing shear. Using the Limber approximation, the shear power spectrum becomes
\be
C(\ell)=\int_0^{\chi_{H}}g^2(\chi)P\left(\frac{\ell}{\chi},z(\chi)\right)d\chi,
\ee
where $\chi$ the comoving distance (with $\chi_H$ being the distance to the horizon), $P(k,z)$ is the matter power spectrum at redshift $z=z(\chi)$, and $g(\chi)$ is the lensing weight. The latter is given by
\be
g(\chi)=\frac{3\Omega_m}{2}\left(\frac{H_0}{c}\right)^2\left(1+z(\chi)\right)\int_{\chi}^{\chi_H}n_s\left(z(\chi')\right)\frac{(\chi'-\chi)}{\chi'}\frac{dz}{d\chi'}d\chi',
\ee
The galaxy redshift distribution $n_s(z)$ can be parametrised as follows
\be\label{redshiftdistr}
n_s(z)= \frac{1}{2z_0^3}\times z^2\exp\left(-\frac{z}{z_0}\right)
\ee
where $z_0=0.24$ provides a reasonable fit to the galaxy distribution for a typical stage-II lensing survey like {\tt KiDS} or a stage-IV survey like {\tt Euclid}.

\begin{figure}[tbp]
\center{
\includegraphics[width=.95\textwidth,trim=0.2cm 0.6cm 1.2cm 0.6cm,clip]{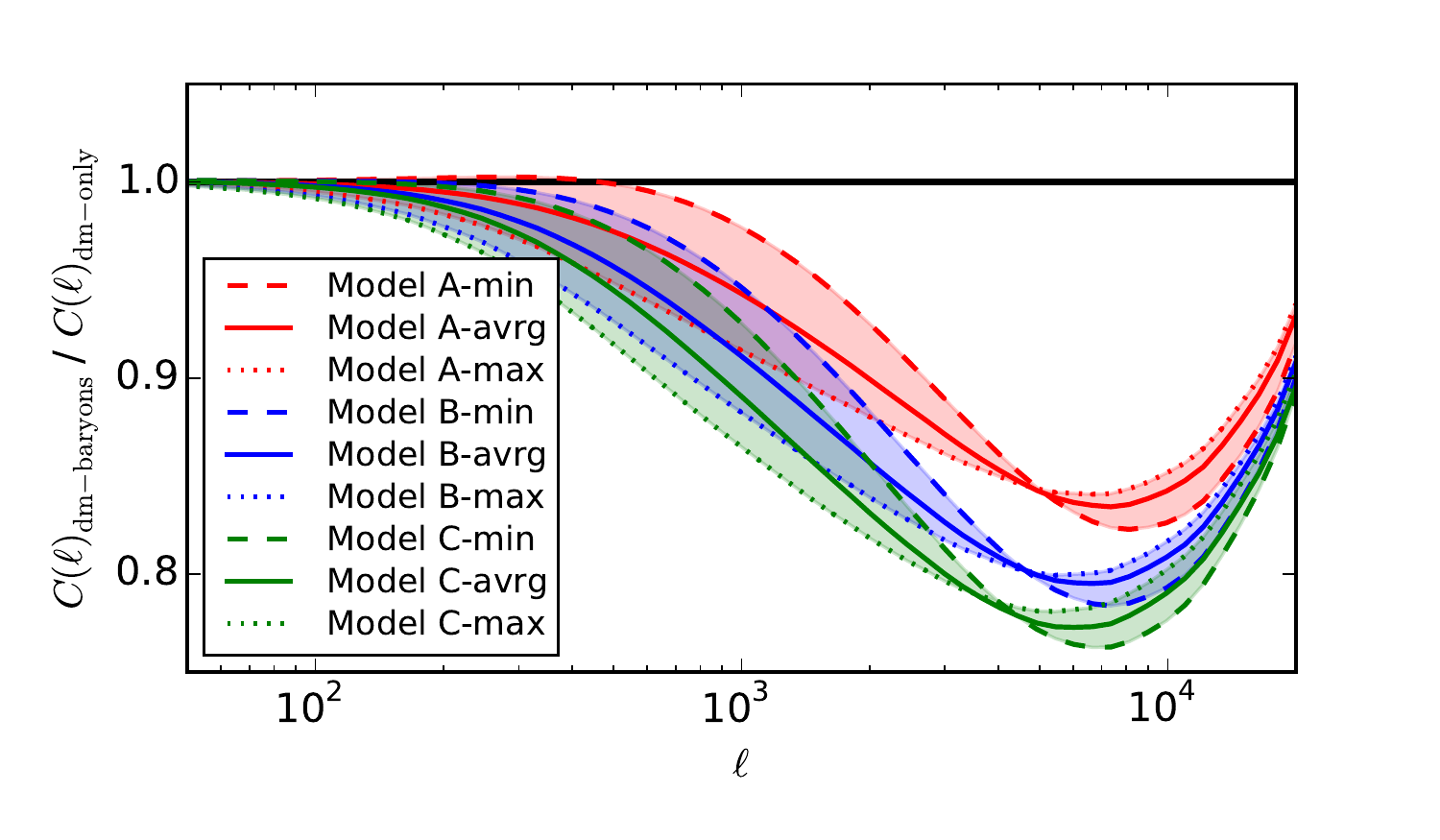}
\caption{\label{fig:Cofl}Relative effect of baryons on the weak lensing angular shear power spectrum for the benchmark models A, B, and C , assuming a minimum (min), best-guess (avrg), and maximum (max) allowed value for the gas ejection radius (same as Fig.~\ref{fig:PS}).}}
\end{figure}

The baryon effect on the angular shear power spectrum is illustrated in Fig.~\ref{fig:Cofl}. We again plot the three benchmark models A, B, and C for each assuming a minimum (min), best-guess (avrg), and maximum (max) gas ejection radius ($\theta_{\rm ej}=2,\,4,\,8$). Beyond this range, the gas profiles are in disagreement with X-ray observations of stacked galaxy groups and clusters (see Sec.~\ref{sec:constraints}). The BC model therefore predicts the baryon suppression effect of the shear power spectrum to lie within the range covered by these three models. Hence, we expect a maximum suppression of 15 - 25 percent with a beyond percent effect above $\ell\sim100-600$. These scales are relevant for current and future weak lensing surveys.

\subsection{Cosmic shear correlation}
As a final result of this paper, we derive the expected baryon effect on the cosmic shear correlation and compare it to the measured shear correlation from the lensing survey {\tt CFHTLenS} \citep{Kilbinger:2012qz,Heymans:2013fya}. The angular shear power can be converted into the shear correlation via the transformation \citep{Kaiser:1992aaa}
\be
\xi_{+}(\theta)= \frac{1}{2\pi}\int_0^{\infty}d\ell \ell J_0(\ell\theta) C(\ell),\hspace{1cm}
\xi_{-}(\theta)= \frac{1}{2\pi}\int_0^{\infty}d\ell \ell J_4(\ell\theta) C(\ell),
\ee
where $J_0$ and $J_4$ are Bessel functions of the first kind of order 0 and 4, respectively. This makes it straight-forward to calculate the weak-lensing shear correlation from the angular power spectrum derived in the previous section\footnote{In order to allow for a direct comparison with {\tt CFHTLenS}, we use a tabulated galaxy redshift distribution from Ref.~\citep{Kilbinger:2012qz} instead of Eq.~(\ref{redshiftdistr}).}.

\begin{figure}[tbp]
\center{
\includegraphics[width=.49\textwidth,trim=0.2cm 0.6cm 1.2cm 0.6cm,clip]{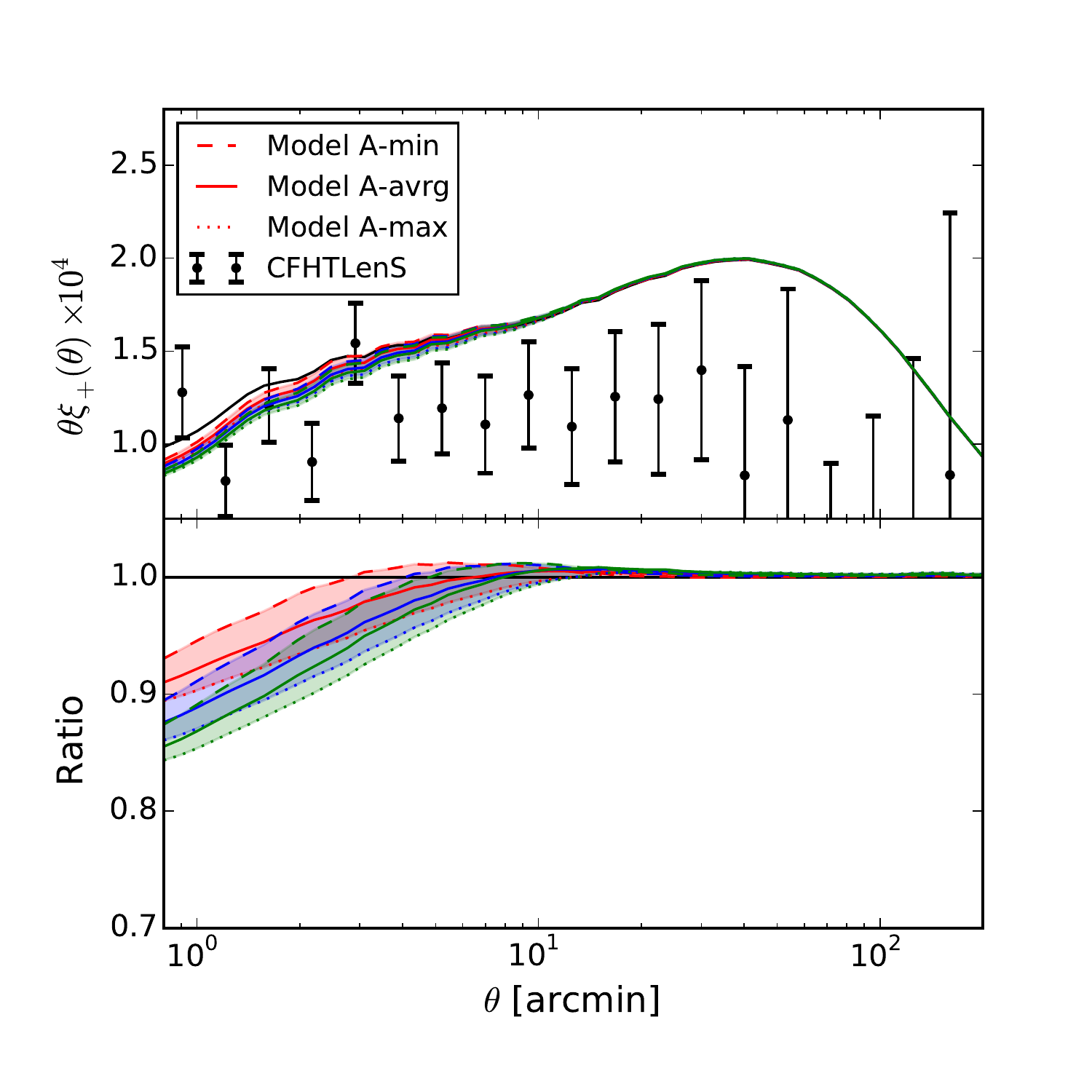}
\includegraphics[width=.49\textwidth,trim=0.2cm 0.6cm 1.2cm 0.6cm,clip]{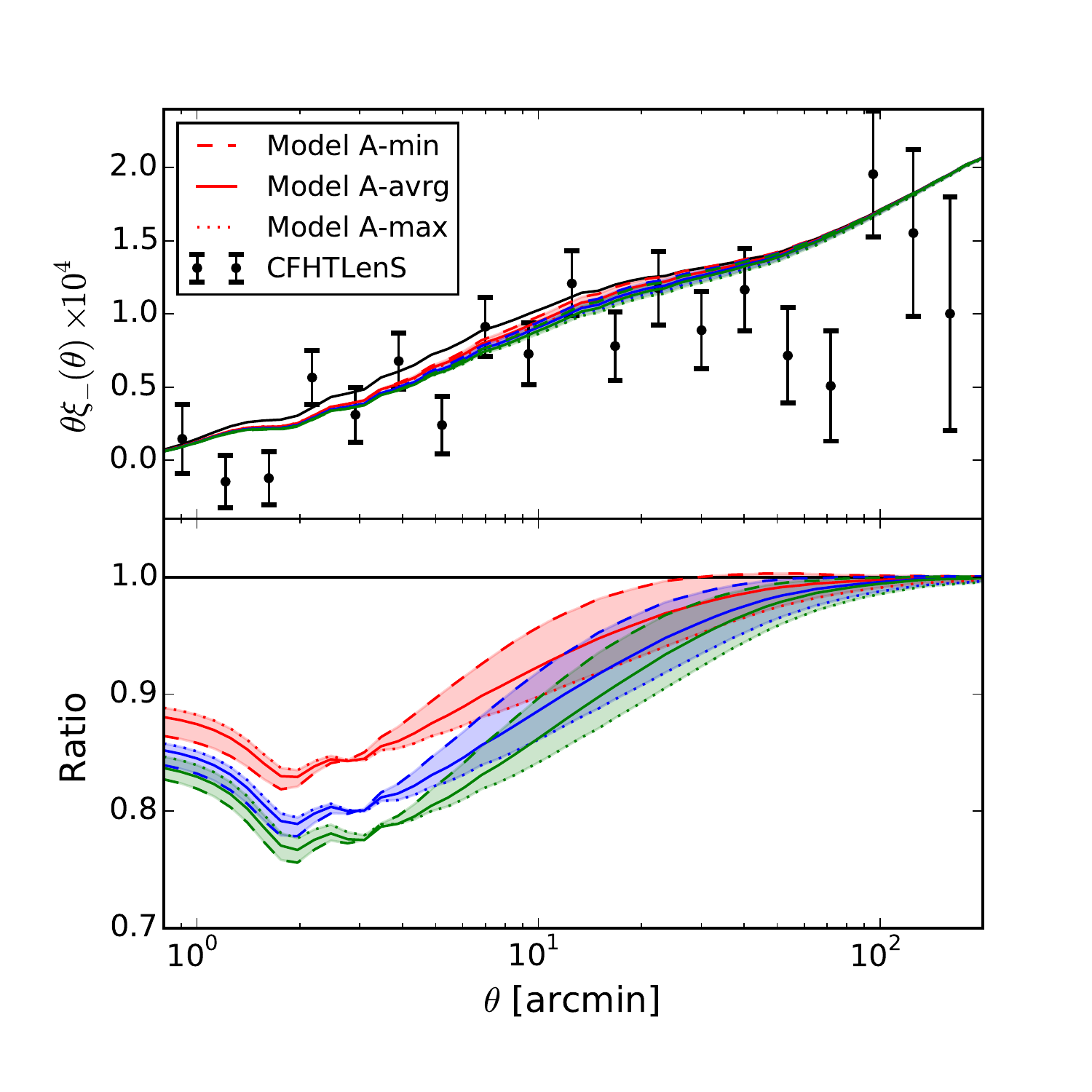}
\caption{\label{fig:shearcorrelation}Weak lensing shear correlations $\xi+$ (left) and $\xi_-$ (right) for the benchmark models A, B, and C with minimum (min), best guess (avrg) and maximum (max) allowed ejected gas radius based on the {\tt Planck} cosmology \citep{Ade:2015xua}. Observations from {\tt CFHTLenS} \citep{Heymans:2013fya} are shown for comparison. Note that at small angles, the baryon suppression effects predicted by the BC model are only slightly smaller than the error-bars of {\tt CFHTLenS}.}}
\end{figure}

The top panels of Fig.~\ref{fig:shearcorrelation} show the weak-lensing shear correlations $\xi_+$ (left) and $\xi_-$ (right).  Assuming a {\tt Planck} cosmology \citep{Ade:2015xua}, we plot predictions of a dark-matter-only $N$-body simulation (black line) together with the BC benchmark models A, B, and C. The colour-shaded areas again illustrate the acceptable range regarding the gas ejection radius. The observed shear correlations from {\tt CFHTLenS} are added as black symbols \cite[see Ref.][]{Heymans:2013fya}. Compared to the theory, they reveal a slight offset, which can be attributed to the fact that {\tt CFHTLenS} favours a cosmology with reduced clustering compared to {\tt Planck} \citep[see e.g. Ref.][]{Hildebrandt:2016iqg}. Our results are in agreement with previous calculations from Refs.~\citep{Mead:2015yca,McCarthy:2017csu}.

The bottom-panels of Fig.~\ref{fig:shearcorrelation} illustrate the relative effect of baryons on the shear correlation. Regarding $\xi_{+}$ (left), the baryon suppression stays below 5-15 percent at scales $0.8\lesssim\theta\lesssim5$ arc-minutes. For $\xi_-$ (right), baryon effects become visible at ten times larger angular scales (below $\theta\sim50$ arc-minutes) and reach a maximum amplitude of 15-25 percent around $\theta\sim 2$ arc-minutes. While these values provide reasonable estimates, the exact suppression signal depends on the galaxy redshift distribution of a given weak-lensing survey.

It is important to note that at the small angular scales shown in Fig.~\ref{fig:shearcorrelation}, the size of the maximum allowed suppression from baryons is of the order of the observational error-bars \citep[see also Ref.][]{Joudaki:2016mvz}. This conclusion is in agreement with other findings on the importance of baryon effects from {\tt KiDS} \citep{Hildebrandt:2016iqg,Joudaki:2016kym} and the {\tt HSC survey} \citep{Hikage:2018qbn}. Compared to current weak lensing measurements, stage IV surveys such as {\tt Euclid} and {\tt LSST} will both extend to smaller angular scales and have significantly reduced errors. This means that baryon effects will become a dominating systematics in the future. Hence, more accurate model predictions will be crucial to achieve the ambitious goals of {\tt Euclid} and {\tt LSST}.


\section{Conclusions}\label{sec:conclusions}
In this paper we present an updated version of the \emph{baryonic correction} (BC) model originally introduced in ST15 \citep{Schneider:2015wta}. The  model is based on a description of how baryons affect the halo profiles and it uses this information to displace particles in outputs of gravity-only $N$-body simulations. This approach has the advantage of being computationally inexpensive and to allow for a simple, physically motivated parametrisation of baryon effects on the total density field of the universe.

The BC model describes halo profiles based on the three components gas, stars, and dark matter. Compared to ST15, we consider a simplified parametrisation which we show to be in good agreement with both observations and hydrodynamical simulations. The gas profile is assumed to follow a power-law with an additional central core and a truncation radius determined by the maximum gas ejection (see Eq.~\ref{rhogas}). The stellar profile follows a power law with an exponential cutoff at the half light radius (see Eq.~\ref{rhocga}). The dark matter component is described by a truncated NFW profile (Eq.~\ref{rhoNFW}), which is allowed to contract and expand adopting to the presence of stars and gas (see Eq.~\ref{ACmodel}).

In a first step, we compare the BC model to results from hydrodynamical simulations. We show that the method is not only able to match the various shapes of the reported baryon power suppression from the literature, but it also reproduces the power spectra from hydrodynamical simulations up to a few percent once the BC parameters are fitted to the gas fraction of the same simulation. This is a strong indication that the method of displacing particles in $N$-body simulations accurately mimics the effects of baryons on the large-scale structure.

In a next step, we use observations to further test the parametrisation of the BC model. We show that the assumed gas profile is in good agreement with stacked X-ray density profiles of galaxy groups and clusters. The same observations are also used to show that gas profiles become flatter for haloes with smaller mass and to constrain the gas ejection parameter to the range $\theta_{\rm ej}=2-8$.

One of the main advantages of the BC model is that it makes it easy to connect observations of gas and stars with statistical probes of the large scale structure. In this paper, we rely on X-ray gas fractions from individual galaxy groups and clusters to predict (i) the baryon effects on the matter power spectrum, (ii) the weak-lensing angular power spectrum, and (iii) the real-space correlations. The main results are listed below:
\begin{itemize}
\item[(i)] At redshift zero, the matter power spectrum is predicted to show beyond percent effects above $k=0.2-1.0$ h/Mpc with a maximum suppression of $15-25$ percent around $k\sim10$ h/Mpc. The suppression signal becomes smaller towards higher redshifts, decreasing to values below 20 percent at $z\gtrsim 1$.
\item[(ii)] For the weak-lensing angular spectrum, we predict a suppression of 10-20 percent at scales beyond $\ell\sim200-1000$. This estimate is based on the galaxy redshift distribution of a typical stage-II lensing survey.
\item[(iii)] The shear correlations $\xi_+$ and $\xi_-$ are affected at the 10-25 percent level for scales below 5 and 50 arc-minutes. The amplitude of the baryon suppression effect is comparable to the observational error bars of past weak-lensing surveys such as {\tt CFHTLenS}.
\end{itemize}
Note that it is currently impossible to provide more accurate predictions than the ones listed above. This is because the X-ray gas fractions are subject to large uncertainties due to the poorly known value of the hydrostatic mass bias. In this paper we have conservatively assumed a bias between zero and forty percent.

We conclude that baryonic effects will become a leading systematical uncertainty for upcoming weak-lensing surveys like {\tt Euclid} or {\tt LSST}. A potential way to address this problem is to cross-correlate the weak-lensing signal with observations of the gas distribution in the universe. Next to X-ray data used in this paper, thermal and kinetic Sunyaev-Zel'dovich measurements are the most promising probes to achieve this goal \citep[see e.g. Refs.][]{Battaglia:2017neq,Ade:2018sbj}. Only by combining different observables will it be possible to simultaneously quantify the effects of cosmology and baryonic feedback. The \emph{baryonic correction model} provides a fast and accurate tool to achieve this goal.

\section*{Acknowledgements}
AS is supported by the Swiss National Science Foundation (project number PZ00P2\_161363). NEC acknowledges support from a Royal Astronomical Society Research Fellowship. AMCLB was supported by the European Research Council under the European Union's Seventh Framework Programme (FP7/2007-2013) / ERC grant agreement number 340519. We thank Julien Devriendt, Yohan Dubois, Christophe Pichon and the Horizon-AGN team for making their results available to us.


\bibliographystyle{unsrtnat}
\bibliography{ASbib}

\begin{thebibliography}{81}
\providecommand{\natexlab}[1]{#1}
\providecommand{\url}[1]{\texttt{#1}}
\expandafter\ifx\csname urlstyle\endcsname\relax
  \providecommand{\doi}[1]{doi: #1}\else
  \providecommand{\doi}{doi: \begingroup \urlstyle{rm}\Url}\fi

\bibitem[Schneider and Teyssier(2015)]{Schneider:2015wta}
Aurel Schneider and Romain Teyssier.
\newblock {A new method to quantify the effects of baryons on the matter power
  spectrum}.
\newblock \emph{JCAP}, 1512\penalty0 (12):\penalty0 049, 2015.
\newblock \doi{10.1088/1475-7516/2015/12/049}.

\bibitem[Heitmann et~al.(2008)]{Heitmann:2007hr}
Katrin Heitmann et~al.
\newblock {The Cosmic Code Comparison Project}.
\newblock \emph{Comput. Sci. Dis.}, 1:\penalty0 015003, 2008.
\newblock \doi{10.1088/1749-4699/1/1/015003}.

\bibitem[Schneider et~al.(2016)Schneider, Teyssier, Potter, Stadel, Onions,
  Reed, Smith, Springel, Pearce, and Scoccimarro]{Schneider:2015yka}
Aurel Schneider, Romain Teyssier, Doug Potter, Joachim Stadel, Julian Onions,
  Darren~S. Reed, Robert~E. Smith, Volker Springel, Frazer~R. Pearce, and Roman
  Scoccimarro.
\newblock {Matter power spectrum and the challenge of percent accuracy}.
\newblock \emph{JCAP}, 1604\penalty0 (04):\penalty0 047, 2016.
\newblock \doi{10.1088/1475-7516/2016/04/047}.

\bibitem[Knabenhans et~al.(2018)]{Knabenhans:2018cng}
Mischa Knabenhans et~al.
\newblock {Euclid preparation: II. The EuclidEmulator -- A tool to compute the
  cosmology dependence of the nonlinear matter power spectrum}.
\newblock 2018.

\bibitem[Smith and Angulo(2018)]{Smith:2018zcj}
Robert~E. Smith and Raul~E. Angulo.
\newblock {Precision modelling of the matter power spectrum in a Planck-like
  Universe}.
\newblock 2018.

\bibitem[van Daalen et~al.(2011)van Daalen, Schaye, Booth, and
  Vecchia]{vanDaalen:2011xb}
Marcel~P. van Daalen, Joop Schaye, C.~M. Booth, and Claudio~Dalla Vecchia.
\newblock {The effects of galaxy formation on the matter power spectrum: A
  challenge for precision cosmology}.
\newblock \emph{Mon. Not. Roy. Astron. Soc.}, 415:\penalty0 3649--3665, 2011.
\newblock \doi{10.1111/j.1365-2966.2011.18981.x}.

\bibitem[{Semboloni} et~al.(2011){Semboloni}, {Hoekstra}, {Schaye}, {van
  Daalen}, and {McCarthy}]{Semboloni:2011aaa}
E.~{Semboloni}, H.~{Hoekstra}, J.~{Schaye}, M.~P. {van Daalen}, and I.~G.
  {McCarthy}.
\newblock {Quantifying the effect of baryon physics on weak lensing
  tomography}.
\newblock \emph{Mon. Not. Roy. Astron. Soc.}, 417:\penalty0 2020--2035,
  November 2011.
\newblock \doi{10.1111/j.1365-2966.2011.19385.x}.

\bibitem[Huang et~al.(2018)Huang, Eifler, Mandelbaum, and
  Dodelson]{Huang:2018wpy}
Hung-Jin Huang, Tim Eifler, Rachel Mandelbaum, and Scott Dodelson.
\newblock {Modeling baryonic physics in future weak lensing surveys}.
\newblock 2018.

\bibitem[Parimbelli et~al.(2018)Parimbelli, Viel, and
  Sefusatti]{Parimbelli:2018yzv}
Gabriele Parimbelli, Matteo Viel, and Emiliano Sefusatti.
\newblock {On the degeneracy between baryon feedback and massive neutrinos as
  probed by matter clustering and weak lensing}.
\newblock 2018.

\bibitem[Mummery et~al.(2017)Mummery, McCarthy, Bird, and
  Schaye]{Mummery:2017lcn}
Benjamin~O. Mummery, Ian~G. McCarthy, Simeon Bird, and Joop Schaye.
\newblock {The separate and combined effects of baryon physics and neutrino
  free-streaming on large-scale structure}.
\newblock \emph{Mon. Not. Roy. Astron. Soc.}, 471\penalty0 (1):\penalty0
  227--242, 2017.
\newblock \doi{10.1093/mnras/stx1469}.

\bibitem[Hellwing et~al.(2016)Hellwing, Schaller, Frenk, Theuns, Schaye, Bower,
  and Crain]{Hellwing:2016ucy}
Wojciech~A. Hellwing, Matthieu Schaller, Carlos~S. Frenk, Tom Theuns, Joop
  Schaye, Richard~G. Bower, and Robert~A. Crain.
\newblock {The effect of baryons on redshift space distortions and cosmic
  density and velocity fields in the EAGLE simulation}.
\newblock \emph{Mon. Not. Roy. Astron. Soc.}, 461\penalty0 (1):\penalty0
  L11--L15, 2016.
\newblock \doi{10.1093/mnrasl/slw081}.

\bibitem[Springel et~al.(2017)]{Springel:2017tpz}
Volker Springel et~al.
\newblock {First results from the IllustrisTNG simulations: matter and galaxy
  clustering}.
\newblock 2017.

\bibitem[Chisari et~al.(2018)Chisari, Richardson, Devriendt, Dubois, Schneider,
  Brun, Beckmann, Peirani, Slyz, and Pichon]{Chisari:2018prw}
Nora~Elisa Chisari, Mark L.~A. Richardson, Julien Devriendt, Yohan Dubois,
  Aurel Schneider, M.~C. Brun, Amandine~Le, Ricarda~S. Beckmann, Sebastien
  Peirani, Adrianne Slyz, and Christophe Pichon.
\newblock {The impact of baryons on the matter power spectrum from the
  Horizon-AGN cosmological hydrodynamical simulation}.
\newblock 2018.

\bibitem[Mohammed and Seljak(2014)]{Mohammed:2014lja}
Irshad Mohammed and Uros Seljak.
\newblock {Analytic model for the matter power spectrum, its covariance matrix,
  and baryonic effects}.
\newblock \emph{Mon. Not. Roy. Astron. Soc.}, 445\penalty0 (4):\penalty0
  3382--3400, 2014.
\newblock \doi{10.1093/mnras/stu1972}.

\bibitem[Dai et~al.(2018)Dai, Feng, and Seljak]{Dai:2018vvv}
Biwei Dai, Yu~Feng, and Uros Seljak.
\newblock {A gradient based method for modeling baryons and matter in halos of
  fast simulations}.
\newblock 2018.

\bibitem[Fedeli et~al.(2014)Fedeli, Semboloni, Velliscig, Van~Daalen, Schaye,
  and Hoekstra]{Fedeli:2014gja}
C.~Fedeli, E.~Semboloni, M.~Velliscig, M.~Van~Daalen, J.~Schaye, and
  H.~Hoekstra.
\newblock {The clustering of baryonic matter. II: halo model and hydrodynamic
  simulations}.
\newblock \emph{JCAP}, 1408:\penalty0 028, 2014.
\newblock \doi{10.1088/1475-7516/2014/08/028}.

\bibitem[Mohammed et~al.(2014)Mohammed, Martizzi, Teyssier, and
  Amara]{Mohammed:2014mba}
Irshad Mohammed, Davide Martizzi, Romain Teyssier, and Adam Amara.
\newblock {Baryonic effects on weak-lensing two-point statistics and its
  cosmological implications}.
\newblock 2014.

\bibitem[Mead et~al.(2015)Mead, Peacock, Heymans, Joudaki, and
  Heavens]{Mead:2015yca}
Alexander Mead, John Peacock, Catherine Heymans, Shahab Joudaki, and Alan
  Heavens.
\newblock {An accurate halo model for fitting non-linear cosmological power
  spectra and baryonic feedback models}.
\newblock \emph{Mon. Not. Roy. Astron. Soc.}, 454\penalty0 (2):\penalty0
  1958--1975, 2015.
\newblock \doi{10.1093/mnras/stv2036}.

\bibitem[{Schaye} et~al.(2010){Schaye}, {Dalla Vecchia}, {Booth}, {Wiersma},
  {Theuns}, {Haas}, {Bertone}, {Duffy}, {McCarthy}, and {van de
  Voort}]{Schaye:2010aaa}
J.~{Schaye}, C.~{Dalla Vecchia}, C.~M. {Booth}, R.~P.~C. {Wiersma},
  T.~{Theuns}, M.~R. {Haas}, S.~{Bertone}, A.~R. {Duffy}, I.~G. {McCarthy}, and
  F.~{van de Voort}.
\newblock {The physics driving the cosmic star formation history}.
\newblock \emph{Mon. Not. Roy. Astron. Soc.}, 402:\penalty0 1536--1560, March
  2010.
\newblock \doi{10.1111/j.1365-2966.2009.16029.x}.

\bibitem[Navarro et~al.(1996)Navarro, Frenk, and White]{Navarro:1995iw}
Julio~F. Navarro, Carlos~S. Frenk, and Simon D.~M. White.
\newblock {The Structure of cold dark matter halos}.
\newblock \emph{Astrophys. J.}, 462:\penalty0 563--575, 1996.
\newblock \doi{10.1086/177173}.

\bibitem[Baltz et~al.(2009)Baltz, Marshall, and Oguri]{Baltz:2007vq}
Edward~A. Baltz, Phil Marshall, and Masamune Oguri.
\newblock {Analytic models of plausible gravitational lens potentials}.
\newblock \emph{JCAP}, 0901:\penalty0 015, 2009.
\newblock \doi{10.1088/1475-7516/2009/01/015}.

\bibitem[Diemer and Kravtsov(2015)]{Diemer:2014gba}
Benedikt Diemer and Andrey~V. Kravtsov.
\newblock {A universal model for halo concentrations}.
\newblock \emph{Astrophys. J.}, 799\penalty0 (1):\penalty0 108, 2015.
\newblock \doi{10.1088/0004-637X/799/1/108}.

\bibitem[{Oguri} and {Hamana}(2011)]{Oguri:2011aaa}
Masamune {Oguri} and Takashi {Hamana}.
\newblock {Detailed cluster lensing profiles at large radii and the impact on
  cluster weak lensing studies}.
\newblock \emph{Mon. Not. Roy. Astron. Soc.}, 414:\penalty0 1851--1861, July
  2011.
\newblock \doi{10.1111/j.1365-2966.2011.18481.x}.

\bibitem[Sheth and Tormen(1999)]{Sheth:1999mn}
Ravi~K. Sheth and Giuseppe Tormen.
\newblock {Large scale bias and the peak background split}.
\newblock \emph{Mon. Not. Roy. Astron. Soc.}, 308:\penalty0 119, 1999.
\newblock \doi{10.1046/j.1365-8711.1999.02692.x}.

\bibitem[Hayashi and White(2008)]{Hayashi:2007uk}
E.~Hayashi and S.~D.~M. White.
\newblock {Understanding the shape of the halo-mass and galaxy-mass
  cross-correlation functions}.
\newblock \emph{Mon. Not. Roy. Astron. Soc.}, 388:\penalty0 2, 2008.
\newblock \doi{10.1111/j.1365-2966.2008.13371.x}.

\bibitem[Kravtsov et~al.(2018)Kravtsov, Vikhlinin, and
  Meshscheryakov]{Kravtsov:2014sra}
Andrey Kravtsov, Alexey Vikhlinin, and Alexander Meshscheryakov.
\newblock {Stellar mass -- halo mass relation and star formation efficiency in
  high-mass halos}.
\newblock \emph{Astron. Lett.}, 44\penalty0 (1):\penalty0 8--34, 2018.
\newblock \doi{10.1134/S1063773717120015}.

\bibitem[Moster et~al.(2013)Moster, Naab, and White]{Moster:2012fv}
Benjamin~P. Moster, Thorsten Naab, and Simon D.~M. White.
\newblock {Galactic star formation and accretion histories from matching
  galaxies to dark matter haloes}.
\newblock \emph{Mon. Not. Roy. Astron. Soc.}, 428:\penalty0 3121, 2013.
\newblock \doi{10.1093/mnras/sts261}.

\bibitem[Croston et~al.(2008)Croston, Pratt, Boehringer, Arnaud, Pointecouteau,
  Ponman, Sanderson, Temple, Bower, and Donahue]{Croston:2008yr}
J.~H. Croston, G.~W. Pratt, H.~Boehringer, M.~Arnaud, E.~Pointecouteau, T.~J.
  Ponman, A.~J.~R. Sanderson, R.~F. Temple, R.~G. Bower, and M.~Donahue.
\newblock {Galaxy-cluster gas-density distributions of the Representative
  XMM-Newton Cluster Structure Survey (REXCESS)}.
\newblock \emph{Astron. Astrophys.}, 487:\penalty0 431, 2008.
\newblock \doi{10.1051/0004-6361:20079154}.

\bibitem[Sanders et~al.(2018)Sanders, Fabian, Russell, and
  Walker]{Sanders:2017lce}
J.~S. Sanders, A.~C. Fabian, H.~R. Russell, and S.~A. Walker.
\newblock {Hydrostatic Chandra X-ray analysis of SPT-selected galaxy clusters ?
  I. Evolution of profiles and core properties}.
\newblock \emph{Mon. Not. Roy. Astron. Soc.}, 474\penalty0 (1):\penalty0
  1065--1098, 2018.
\newblock \doi{10.1093/mnras/stx2796}.

\bibitem[Eckert et~al.(2016)]{Eckert:2015rlr}
D.~Eckert et~al.
\newblock {The XXL Survey. XIII. Baryon content of the bright cluster sample}.
\newblock \emph{Astron. Astrophys.}, 592:\penalty0 A12, 2016.
\newblock \doi{10.1051/0004-6361/201527293}.

\bibitem[{Barnes} and {White}(1984)]{Barnes:1984aaa}
J.~{Barnes} and S.~D.~M. {White}.
\newblock {The response of a spheroid to a disc field or were bulges ever
  ellipticals?}
\newblock \emph{Mon. Not. Roy. Astron. Soc.}, 211:\penalty0 753--765, December
  1984.
\newblock \doi{10.1093/mnras/211.4.753}.

\bibitem[Blumenthal et~al.(1986)Blumenthal, Faber, Flores, and
  Primack]{Blumenthal:1985qy}
George~R. Blumenthal, S.~M. Faber, Ricardo Flores, and Joel~R. Primack.
\newblock {Contraction of Dark Matter Galactic Halos Due to Baryonic Infall}.
\newblock \emph{Astrophys. J.}, 301:\penalty0 27, 1986.
\newblock \doi{10.1086/163867}.

\bibitem[Abadi et~al.(2010)Abadi, Navarro, Fardal, Babul, and
  Steinmetz]{Abadi:2009ve}
Mario~G. Abadi, Julio~F. Navarro, Mark Fardal, Arif Babul, and Matthias
  Steinmetz.
\newblock {Galaxy-Induced Transformation of Dark Matter Halos}.
\newblock \emph{Mon. Not. Roy. Astron. Soc.}, 407:\penalty0 435--446, 2010.
\newblock \doi{10.1111/j.1365-2966.2010.16912.x}.

\bibitem[{Teyssier} et~al.(2011){Teyssier}, {Moore}, {Martizzi}, {Dubois}, and
  {Mayer}]{Teyssier:2011aaa}
R.~{Teyssier}, B.~{Moore}, D.~{Martizzi}, Y.~{Dubois}, and L.~{Mayer}.
\newblock {Mass distribution in galaxy clusters: the role of Active Galactic
  Nuclei feedback}.
\newblock \emph{Mon. Not. Roy. Astron. Soc.}, 414:\penalty0 195--208, June
  2011.
\newblock \doi{10.1111/j.1365-2966.2011.18399.x}.

\bibitem[{Stadel}(2001)]{Stadel:2001aaa}
J.~G. {Stadel}.
\newblock \emph{{Cosmological N-body simulations and their analysis}}.
\newblock PhD thesis, UNIVERSITY OF WASHINGTON, 2001.

\bibitem[Potter et~al.(2016)Potter, Stadel, and Teyssier]{Potter:2016ttn}
Douglas Potter, Joachim Stadel, and Romain Teyssier.
\newblock {PKDGRAV3: Beyond Trillion Particle Cosmological Simulations for the
  Next Era of Galaxy Surveys}.
\newblock 2016.

\bibitem[Ade et~al.(2015)]{Planck:2015xua}
P.~A.~R. Ade et~al.
\newblock {Planck 2015 results. XIII. Cosmological parameters}.
\newblock 2015.

\bibitem[{Knollmann} and {Knebe}(2009)]{Knollmann:2009aaa}
S.~R. {Knollmann} and A.~{Knebe}.
\newblock {AHF: Amiga's Halo Finder}.
\newblock \emph{Astrophys. J. Supp.}, 182:\penalty0 608--624, June 2009.
\newblock \doi{10.1088/0067-0049/182/2/608}.

\bibitem[LaRoque et~al.(2006)LaRoque, Bonamente, Carlstrom, Joy, Nagai, Reese,
  and Dawson]{LaRoque:2006te}
Samuel LaRoque, M.~Bonamente, J.~Carlstrom, M.~Joy, D.~Nagai, E.~Reese, and
  K.~Dawson.
\newblock {X-ray and Sunyaev-Zel'dovich Effect Measurements of the Gas Mass
  Fraction in Galaxy Clusters}.
\newblock \emph{Astrophys. J.}, 652:\penalty0 917--936, 2006.
\newblock \doi{10.1086/508139}.

\bibitem[Morandi et~al.(2015)Morandi, Sun, Forman, and Jones]{Morandi:2015pra}
Andrea Morandi, Ming Sun, William Forman, and Christine Jones.
\newblock {The galaxy cluster outskirts probed by Chandra}.
\newblock \emph{Mon. Not. Roy. Astron. Soc.}, 450\penalty0 (3):\penalty0
  2261--2278, 2015.
\newblock \doi{10.1093/mnras/stv660}.

\bibitem[Chiu et~al.(2018)]{Chiu:2017nwm}
I.~Chiu et~al.
\newblock {Baryon Content in a Sample of 91 Galaxy Clusters Selected by the
  South Pole Telescope at 0.2 < z < 1.25}.
\newblock \emph{Mon. Not. Roy. Astron. Soc.}, 478\penalty0 (3):\penalty0
  3072--3099, 2018.
\newblock \doi{10.1093/mnras/sty1284}.

\bibitem[{Leauthaud} et~al.(2012){Leauthaud}, {George}, {Behroozi}, {Bundy},
  {Tinker}, {Wechsler}, {Conroy}, {Finoguenov}, and
  {Tanaka}]{Leauthaud:2012aaa}
A.~{Leauthaud}, M.~R. {George}, P.~S. {Behroozi}, K.~{Bundy}, J.~{Tinker},
  R.~H. {Wechsler}, C.~{Conroy}, A.~{Finoguenov}, and M.~{Tanaka}.
\newblock {The Integrated Stellar Content of Dark Matter Halos}.
\newblock \emph{Ap.J.}, 746:\penalty0 95, February 2012.
\newblock \doi{10.1088/0004-637X/746/1/95}.

\bibitem[Behroozi et~al.(2013)Behroozi, Wechsler, and Conroy]{Behroozi:2012iw}
Peter~S. Behroozi, Risa~H. Wechsler, and Charlie Conroy.
\newblock {The Average Star Formation Histories of Galaxies in Dark Matter
  Halos from $z=$0-8}.
\newblock \emph{Astrophys. J.}, 770:\penalty0 57, 2013.
\newblock \doi{10.1088/0004-637X/770/1/57}.

\bibitem[Giodini et~al.(2009)]{Giodini:2009qf}
S.~Giodini et~al.
\newblock {Stellar and total baryon mass fractions in groups and clusters since
  redshift 1}.
\newblock \emph{Astrophys. J.}, 703:\penalty0 982--993, 2009.
\newblock \doi{10.1088/0004-637X/703/1/982}.

\bibitem[{McCarthy} et~al.(2010){McCarthy}, {Schaye}, {Ponman}, {Bower},
  {Booth}, {Dalla Vecchia}, {Crain}, {Springel}, {Theuns}, and
  {Wiersma}]{McCarthy:2010aaa}
I.~G. {McCarthy}, J.~{Schaye}, T.~J. {Ponman}, R.~G. {Bower}, C.~M. {Booth},
  C.~{Dalla Vecchia}, R.~A. {Crain}, V.~{Springel}, T.~{Theuns}, and R.~P.~C.
  {Wiersma}.
\newblock {The case for AGN feedback in galaxy groups}.
\newblock \emph{Mon. Not. Roy. Astron. Soc.}, 406:\penalty0 822--839, August
  2010.
\newblock \doi{10.1111/j.1365-2966.2010.16750.x}.

\bibitem[Springel(2005)]{Springel:2005mi}
Volker Springel.
\newblock {The Cosmological simulation code GADGET-2}.
\newblock \emph{Mon. Not. Roy. Astron. Soc.}, 364:\penalty0 1105--1134, 2005.
\newblock \doi{10.1111/j.1365-2966.2005.09655.x}.

\bibitem[Spergel et~al.(2007)]{Spergel:2006hy}
D.~N. Spergel et~al.
\newblock {Wilkinson Microwave Anisotropy Probe (WMAP) three year results:
  implications for cosmology}.
\newblock \emph{Astrophys. J. Suppl.}, 170:\penalty0 377, 2007.
\newblock \doi{10.1086/513700}.

\bibitem[{Eckert} et~al.(2012){Eckert}, {Vazza}, {Ettori}, {Molendi}, {Nagai},
  {Lau}, {Roncarelli}, {Rossetti}, {Snowden}, and
  {Gastaldello}]{Eckert:2012aaa}
D.~{Eckert}, F.~{Vazza}, S.~{Ettori}, S.~{Molendi}, D.~{Nagai}, E.~T. {Lau},
  M.~{Roncarelli}, M.~{Rossetti}, S.~L. {Snowden}, and F.~{Gastaldello}.
\newblock {The gas distribution in the outer regions of galaxy clusters}.
\newblock \emph{A\&A}, 541:\penalty0 A57, May 2012.
\newblock \doi{10.1051/0004-6361/201118281}.

\bibitem[Pierre et~al.(2016)]{Pierre:2015cqe}
M.~Pierre et~al.
\newblock {The XXL Survey - I. Scientific motivations ? XMM-Newton observing
  plan ? Follow-up observations and simulation programme}.
\newblock \emph{Astron. Astrophys.}, 592:\penalty0 A1, 2016.
\newblock \doi{10.1051/0004-6361/201526766}.

\bibitem[Arnaud et~al.(2005)Arnaud, Pointecouteau, and Pratt]{Arnaud:2005ur}
Monique Arnaud, E.~Pointecouteau, and G.~W. Pratt.
\newblock {The Structural and scaling properties of nearby galaxy clusters. 2.
  The M-T relation}.
\newblock \emph{Astron. Astrophys.}, 441:\penalty0 893--903, 2005.
\newblock \doi{10.1051/0004-6361:20052856}.

\bibitem[Eckert et~al.(2018)]{Eckert:2018mlz}
D.~Eckert et~al.
\newblock {Non-thermal pressure support in X-COP galaxy clusters}.
\newblock 2018.

\bibitem[Ettori et~al.(2018)Ettori, Ghirardini, Eckert, Pointecouteau,
  Gastaldello, Sereno, Gaspari, Ghizzardi, Roncarelli, and
  Rossetti]{Ettori:2018tus}
S.~Ettori, V.~Ghirardini, D.~Eckert, E.~Pointecouteau, F.~Gastaldello,
  M.~Sereno, M.~Gaspari, S.~Ghizzardi, M.~Roncarelli, and M.~Rossetti.
\newblock {Hydrostatic mass profiles in X-COP galaxy clusters}.
\newblock 2018.

\bibitem[Nagai et~al.(2007)Nagai, Vikhlinin, and Kravtsov]{Nagai:2006sz}
Daisuke Nagai, Alexey Vikhlinin, and Andrey~V. Kravtsov.
\newblock {Testing X-ray Measurements of Galaxy Clusters with Cosmological
  Simulations}.
\newblock \emph{Astrophys. J.}, 655:\penalty0 98--108, 2007.
\newblock \doi{10.1086/509868}.

\bibitem[Brun et~al.(2014)Brun, McCarthy, Schaye, and Ponman]{Brun:2013yva}
Amandine M. C.~Le Brun, Ian~G. McCarthy, Joop Schaye, and Trevor~J. Ponman.
\newblock {Towards a realistic population of simulated galaxy groups and
  clusters}.
\newblock \emph{Mon. Not. Roy. Astron. Soc.}, 441\penalty0 (2):\penalty0
  1270--1290, 2014.
\newblock \doi{10.1093/mnras/stu608}.

\bibitem[Lieu et~al.(2016)]{Lieu:2015pit}
Maggie Lieu et~al.
\newblock {The XXL Survey IV. Mass-temperature relation of the bright cluster
  sample}.
\newblock \emph{Astron. Astrophys.}, 592:\penalty0 A4, 2016.
\newblock \doi{10.1051/0004-6361/201526883}.

\bibitem[Sereno and Ettori(2015)]{Sereno:2014pfa}
Mauro Sereno and Stefano Ettori.
\newblock {Comparing masses in literature (CoMaLit) ? I. Bias and scatter in
  weak lensing and X-ray mass estimates of clusters}.
\newblock \emph{Mon. Not. Roy. Astron. Soc.}, 450\penalty0 (4):\penalty0
  3633--3648, 2015.
\newblock \doi{10.1038/ncomms8173, 10.1093/mnras/stv810}.

\bibitem[Sun et~al.(2009)Sun, Voit, Donahue, Jones, and Forman]{Sun:2008eh}
M.~Sun, G.~M. Voit, M.~Donahue, C.~Jones, and W.~Forman.
\newblock {Chandra studies of the X-ray gas properties of galaxy groups}.
\newblock \emph{Astrophys. J.}, 693:\penalty0 1142--1172, 2009.
\newblock \doi{10.1088/0004-637X/693/2/1142}.

\bibitem[Vikhlinin et~al.(2009)]{Vikhlinin:2008cd}
A.~Vikhlinin et~al.
\newblock {Chandra Cluster Cosmology Project II: Samples and X-ray Data
  Reduction}.
\newblock \emph{Astrophys. J.}, 692:\penalty0 1033--1059, 2009.
\newblock \doi{10.1088/0004-637X/692/2/1033}.

\bibitem[Gonzalez et~al.(2013)Gonzalez, Sivanandam, Zabludoff, and
  Zaritsky]{Gonzalez:2013awy}
Anthony~H. Gonzalez, Suresh Sivanandam, Ann~I. Zabludoff, and Dennis Zaritsky.
\newblock {Galaxy Cluster Baryon Fractions Revisited}.
\newblock \emph{Astrophys. J.}, 778:\penalty0 14, 2013.
\newblock \doi{10.1088/0004-637X/778/1/14}.

\bibitem[Vogelsberger et~al.(2014)Vogelsberger, Genel, Springel, Torrey,
  Sijacki, Xu, Snyder, Nelson, and Hernquist]{Vogelsberger:2014dza}
Mark Vogelsberger, Shy Genel, Volker Springel, Paul Torrey, Debora Sijacki,
  Dandan Xu, Gregory~F. Snyder, Dylan Nelson, and Lars Hernquist.
\newblock {Introducing the Illustris Project: Simulating the coevolution of
  dark and visible matter in the Universe}.
\newblock \emph{Mon. Not. Roy. Astron. Soc.}, 444\penalty0 (2):\penalty0
  1518--1547, 2014.
\newblock \doi{10.1093/mnras/stu1536}.

\bibitem[van Daalen and Schaye(2015)]{vanDaalen:2015msa}
Marcel~P. van Daalen and Joop Schaye.
\newblock {The contributions of matter inside and outside of haloes to the
  matter power spectrum}.
\newblock \emph{Mon. Not. Roy. Astron. Soc.}, 452\penalty0 (3):\penalty0
  2247--2257, 2015.
\newblock \doi{10.1093/mnras/stv1456}.

\bibitem[Schneider et~al.(2012)Schneider, Smith, Maccio, and
  Moore]{Schneider:2011yu}
Aurel Schneider, Robert~E. Smith, Andrea~V. Maccio, and Ben Moore.
\newblock {Nonlinear Evolution of Cosmological Structures in Warm Dark Matter
  Models}.
\newblock \emph{Mon. Not. Roy. Astron. Soc.}, 424:\penalty0 684, 2012.
\newblock \doi{10.1111/j.1365-2966.2012.21252.x}.

\bibitem[Kilbinger et~al.(2013)]{Kilbinger:2012qz}
Martin Kilbinger et~al.
\newblock {CFHTLenS: Combined probe cosmological model comparison using 2D weak
  gravitational lensing}.
\newblock \emph{Mon. Not. Roy. Astron. Soc.}, 430:\penalty0 2200--2220, 2013.
\newblock \doi{10.1093/mnras/stt041}.

\bibitem[Heymans et~al.(2013)]{Heymans:2013fya}
Catherine Heymans et~al.
\newblock {CFHTLenS tomographic weak lensing cosmological parameter
  constraints: Mitigating the impact of intrinsic galaxy alignments}.
\newblock \emph{Mon. Not. Roy. Astron. Soc.}, 432:\penalty0 2433, 2013.
\newblock \doi{10.1093/mnras/stt601}.

\bibitem[{Kaiser}(1992)]{Kaiser:1992aaa}
N.~{Kaiser}.
\newblock {Weak gravitational lensing of distant galaxies}.
\newblock \emph{Astrophys. J.}, 388:\penalty0 272--286, April 1992.
\newblock \doi{10.1086/171151}.

\bibitem[Ade et~al.(2016)]{Ade:2015xua}
P.~A.~R. Ade et~al.
\newblock {Planck 2015 results. XIII. Cosmological parameters}.
\newblock \emph{Astron. Astrophys.}, 594:\penalty0 A13, 2016.
\newblock \doi{10.1051/0004-6361/201525830}.

\bibitem[Hildebrandt et~al.(2017)]{Hildebrandt:2016iqg}
H.~Hildebrandt et~al.
\newblock {KiDS-450: Cosmological parameter constraints from tomographic weak
  gravitational lensing}.
\newblock \emph{Mon. Not. Roy. Astron. Soc.}, 465:\penalty0 1454, 2017.
\newblock \doi{10.1093/mnras/stw2805}.

\bibitem[Mccarthy et~al.(2018)Mccarthy, Bird, Schaye, Harnois-Deraps, Font, and
  Van~Waerbeke]{McCarthy:2017csu}
Ian~G. Mccarthy, Simeon Bird, Joop Schaye, Joachim Harnois-Deraps, Andreea~S.
  Font, and Ludovic Van~Waerbeke.
\newblock {The BAHAMAS project: the CMB large-scale structure tension and the
  roles of massive neutrinos and galaxy formation}.
\newblock \emph{Mon. Not. Roy. Astron. Soc.}, 476\penalty0 (3):\penalty0
  2999--3030, 2018.
\newblock \doi{10.1093/mnras/sty377}.

\bibitem[Joudaki et~al.(2017{\natexlab{a}})]{Joudaki:2016mvz}
Shahab Joudaki et~al.
\newblock {CFHTLenS revisited: assessing concordance with Planck including
  astrophysical systematics}.
\newblock \emph{Mon. Not. Roy. Astron. Soc.}, 465\penalty0 (2):\penalty0
  2033--2052, 2017{\natexlab{a}}.
\newblock \doi{10.1093/mnras/stw2665}.

\bibitem[Joudaki et~al.(2017{\natexlab{b}})]{Joudaki:2016kym}
Shahab Joudaki et~al.
\newblock {KiDS-450: Testing extensions to the standard cosmological model}.
\newblock \emph{Mon. Not. Roy. Astron. Soc.}, 471\penalty0 (2):\penalty0
  1259--1279, 2017{\natexlab{b}}.
\newblock \doi{10.1093/mnras/stx998}.

\bibitem[Hikage et~al.(2018)]{Hikage:2018qbn}
Chiaki Hikage et~al.
\newblock {Cosmology from cosmic shear power spectra with Subaru Hyper
  Suprime-Cam first-year data}.
\newblock 2018.

\bibitem[Battaglia et~al.(2017)Battaglia, Ferraro, Schaan, and
  Spergel]{Battaglia:2017neq}
Nicholas Battaglia, Simone Ferraro, Emmanuel Schaan, and David Spergel.
\newblock {Future constraints on halo thermodynamics from combined
  Sunyaev-Zel'dovich measurements}.
\newblock \emph{JCAP}, 1711\penalty0 (11):\penalty0 040, 2017.
\newblock \doi{10.1088/1475-7516/2017/11/040}.

\bibitem[Aguirre et~al.(2018)]{Ade:2018sbj}
James Aguirre et~al.
\newblock {The Simons Observatory: Science goals and forecasts}.
\newblock 2018.

\bibitem[Komatsu et~al.(2011)]{Komatsu:2010fb}
E.~Komatsu et~al.
\newblock {Seven-Year Wilkinson Microwave Anisotropy Probe (WMAP) Observations:
  Cosmological Interpretation}.
\newblock \emph{Astrophys. J. Suppl.}, 192:\penalty0 18, 2011.
\newblock \doi{10.1088/0067-0049/192/2/18}.

\bibitem[Dubois et~al.(2014)]{Dubois:2014lxa}
Y.~Dubois et~al.
\newblock {Dancing in the dark: galactic properties trace spin swings along the
  cosmic web}.
\newblock \emph{Mon. Not. Roy. Astron. Soc.}, 444\penalty0 (2):\penalty0
  1453--1468, 2014.
\newblock \doi{10.1093/mnras/stu1227}.

\bibitem[{Dubois} et~al.(2016){Dubois}, {Peirani}, {Pichon}, {Devriendt},
  {Gavazzi}, {Welker}, and {Volonteri}]{Dubois:2016aaa}
Y.~{Dubois}, S.~{Peirani}, C.~{Pichon}, J.~{Devriendt}, R.~{Gavazzi},
  C.~{Welker}, and M.~{Volonteri}.
\newblock {The HORIZON-AGN simulation: morphological diversity of galaxies
  promoted by AGN feedback}, December 2016.

\bibitem[Teyssier(2002)]{Teyssier:2001cp}
Romain Teyssier.
\newblock {Cosmological hydrodynamics with adaptive mesh refinement: a new high
  resolution code called ramses}.
\newblock \emph{Astron. Astrophys.}, 385:\penalty0 337--364, 2002.
\newblock \doi{10.1051/0004-6361:20011817}.

\bibitem[Pillepich et~al.(2018{\natexlab{a}})]{Pillepich:2017fcc}
Annalisa Pillepich et~al.
\newblock {First results from the IllustrisTNG simulations: the stellar mass
  content of groups and clusters of galaxies}.
\newblock \emph{Mon. Not. Roy. Astron. Soc.}, 475:\penalty0 648,
  2018{\natexlab{a}}.
\newblock \doi{10.1093/mnras/stx3112}.

\bibitem[Pillepich et~al.(2018{\natexlab{b}})]{Pillepich:2017jle}
Annalisa Pillepich et~al.
\newblock {Simulating Galaxy Formation with the IllustrisTNG Model}.
\newblock \emph{Mon. Not. Roy. Astron. Soc.}, 473\penalty0 (3):\penalty0
  4077--4106, 2018{\natexlab{b}}.
\newblock \doi{10.1093/mnras/stx2656}.

\bibitem[Springel(2010)]{Springel:2010aaa}
Volker Springel.
\newblock {E pur si muove: Galilean-invariant cosmological hydrodynamical
  simulations on a moving mesh}.
\newblock \emph{Mon. Not. Roy. Astron. Soc.}, 401:\penalty0 791--851, January
  2010.
\newblock \doi{10.1111/j.1365-2966.2009.15715.x}.

\bibitem[{Budzynski} et~al.(2012){Budzynski}, {Koposov}, {McCarthy}, {McGee},
  and {Belokurov}]{Budzynski:2012aaa}
J.~M. {Budzynski}, S.~E. {Koposov}, I.~G. {McCarthy}, S.~L. {McGee}, and
  V.~{Belokurov}.
\newblock {The radial distribution of galaxies in groups and clusters}.
\newblock \emph{Mon. Not. Roy. Astron. Soc.}, 423:\penalty0 104--121, June
  2012.
\newblock \doi{10.1111/j.1365-2966.2012.20663.x}.

\end{thebibliography}


\appendix
\section{Comparison with other hydrodynamical simulations}\label{hydrosims}
The matter power spectrum has been predicted by several different hydrodynamical simulations which all show similar trends but do not agree with one another at the quantitative level (see e.g. Fig.~\ref{fig:PShydro}). The discrepancies are a consequence of different implementations of AGN feedback in simulations. While some implementations result in merely heating the gas to prevent excessive star formation, others push the gas far out into the intergalactic medium. In this Appendix, we compare the \emph{baryonic correction} (BC) model with several hydrodynamical simulations from the literature. We show that in general, the BC model is able to predict the power spectra of a variety of different hydrodynamical simulations to the level of a few percent provided the gas and stellar fractions of the simulations are known.

\begin{figure}[tbp]
\center{
\includegraphics[width=.49\textwidth,trim={0.8cm 0.8cm 1.0cm 0.1cm,clip}]{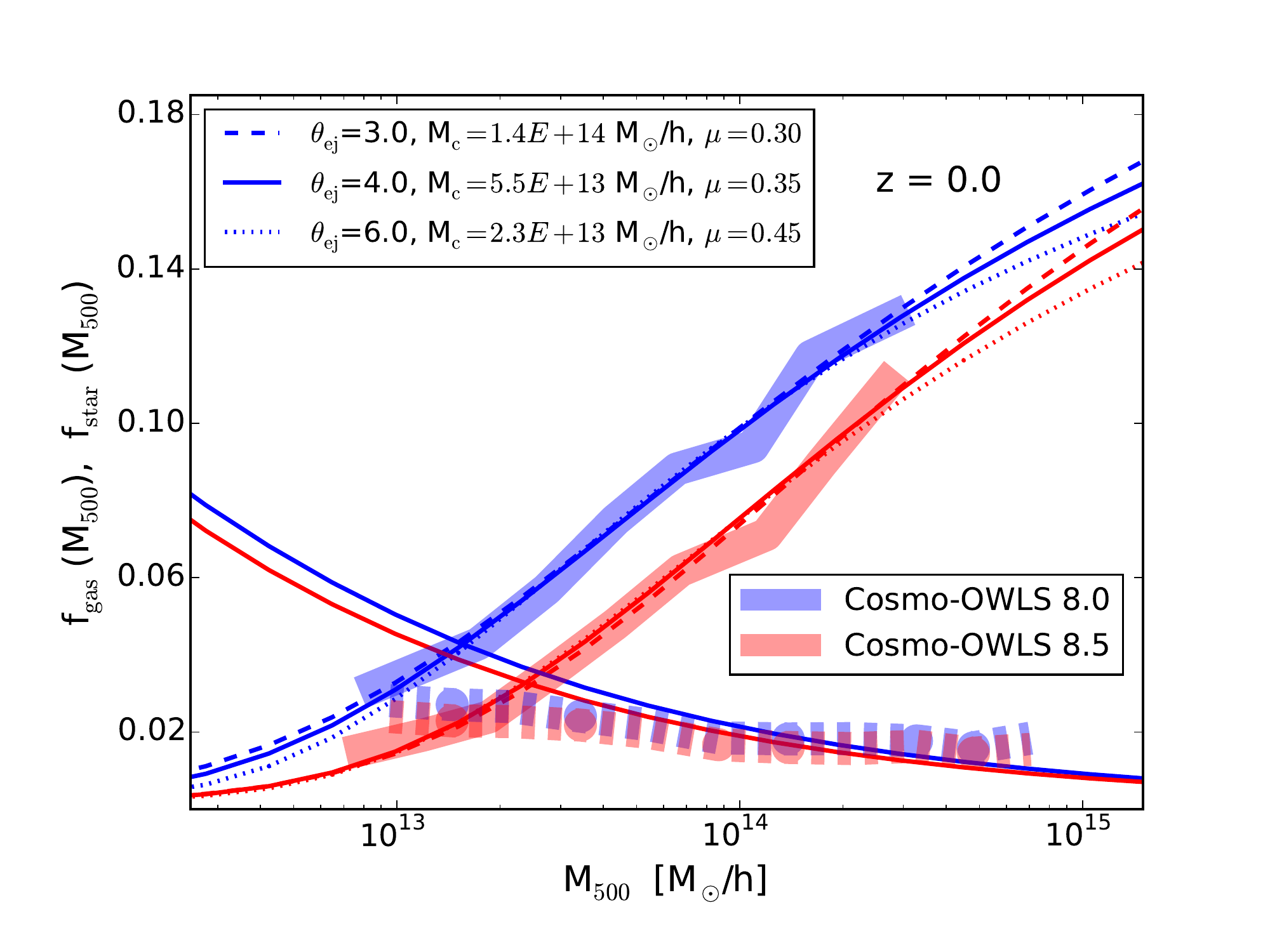}
\includegraphics[width=.49\textwidth,trim={0.8cm 0.8cm 1.0cm 0.1cm,clip}]{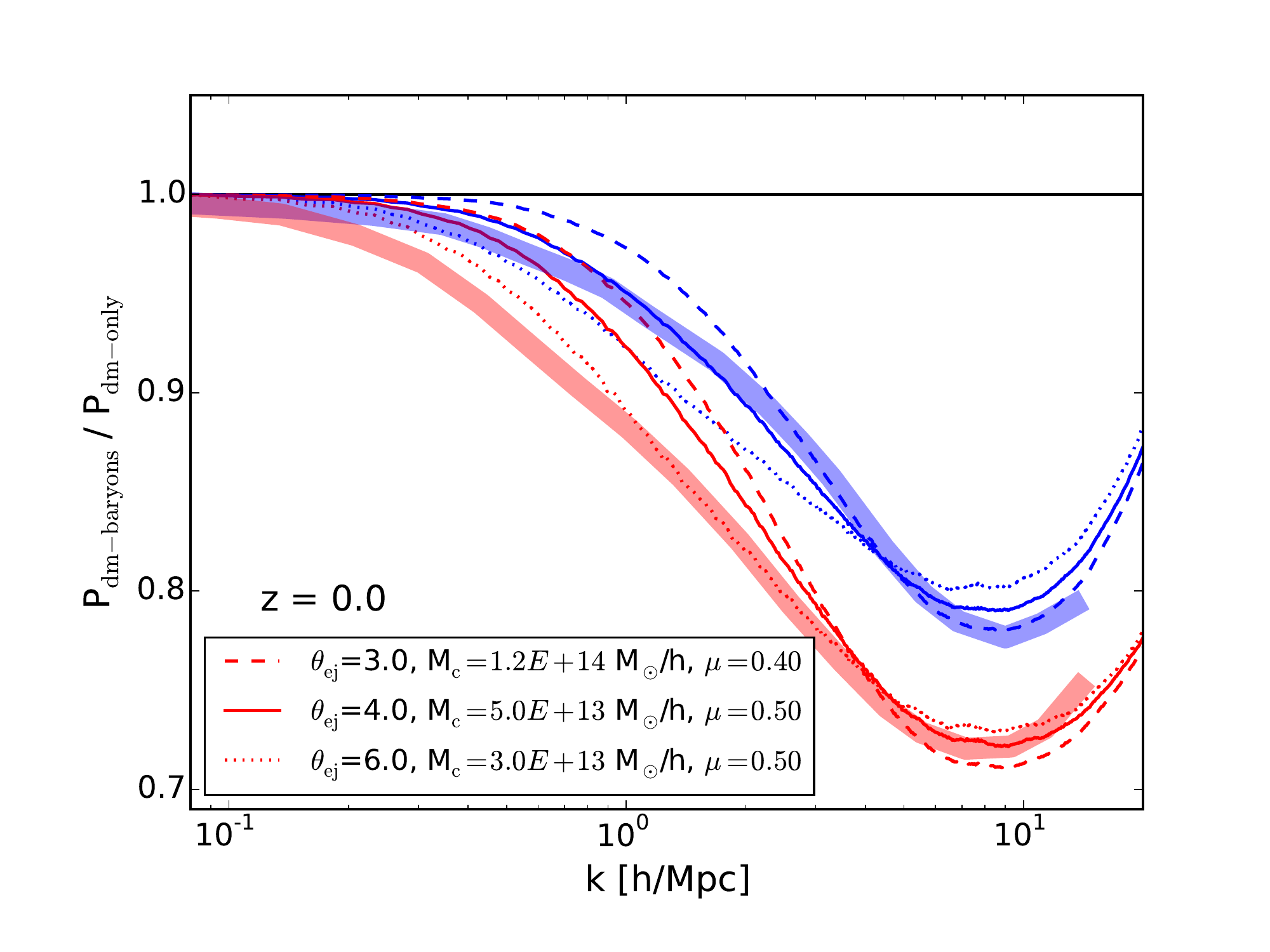}\\
\includegraphics[width=.49\textwidth,trim={0.8cm 0.8cm 1.0cm 0.1cm,clip}]{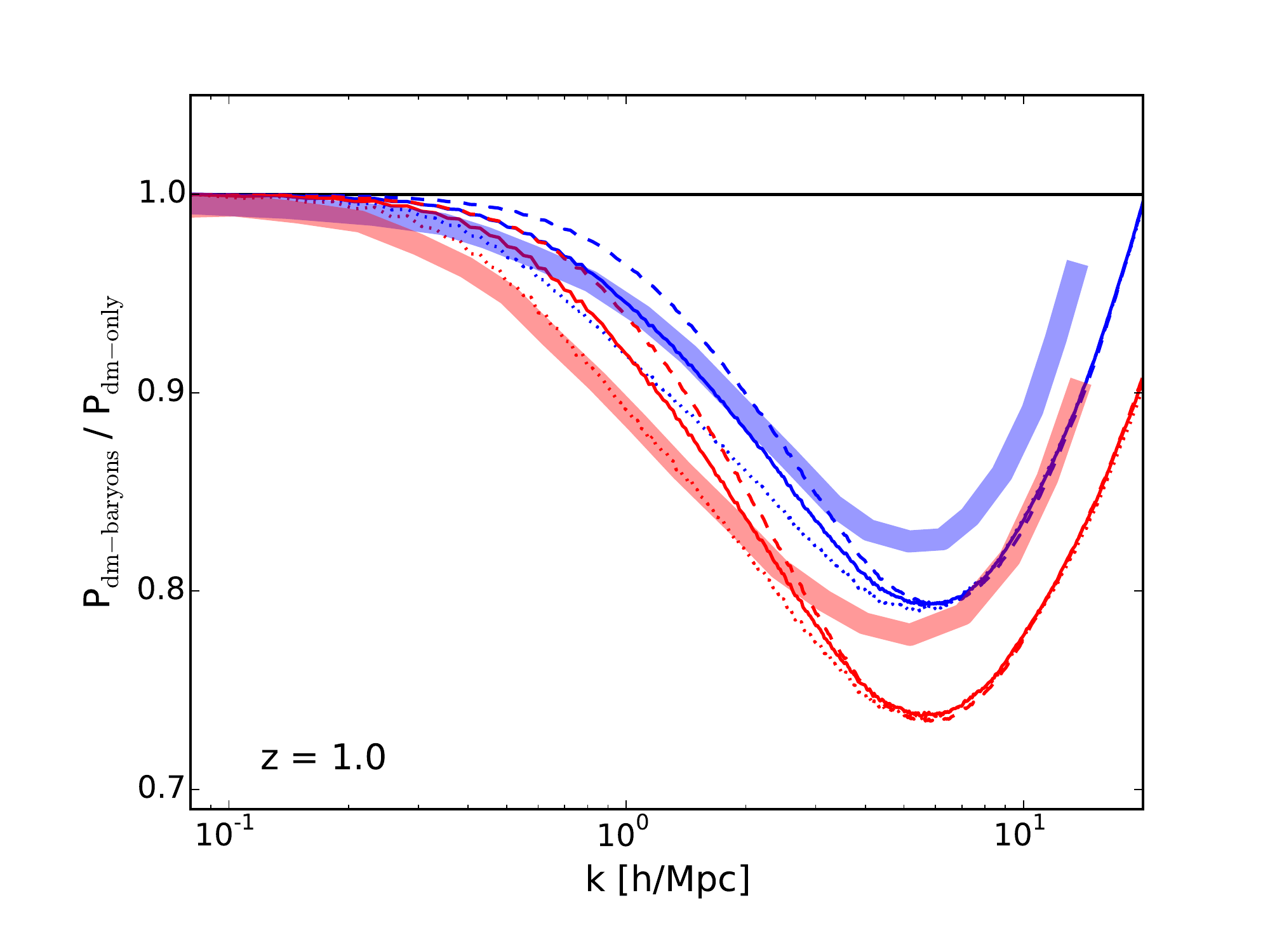}
\includegraphics[width=.49\textwidth,trim={0.8cm 0.8cm 1.0cm 0.1cm,clip}]{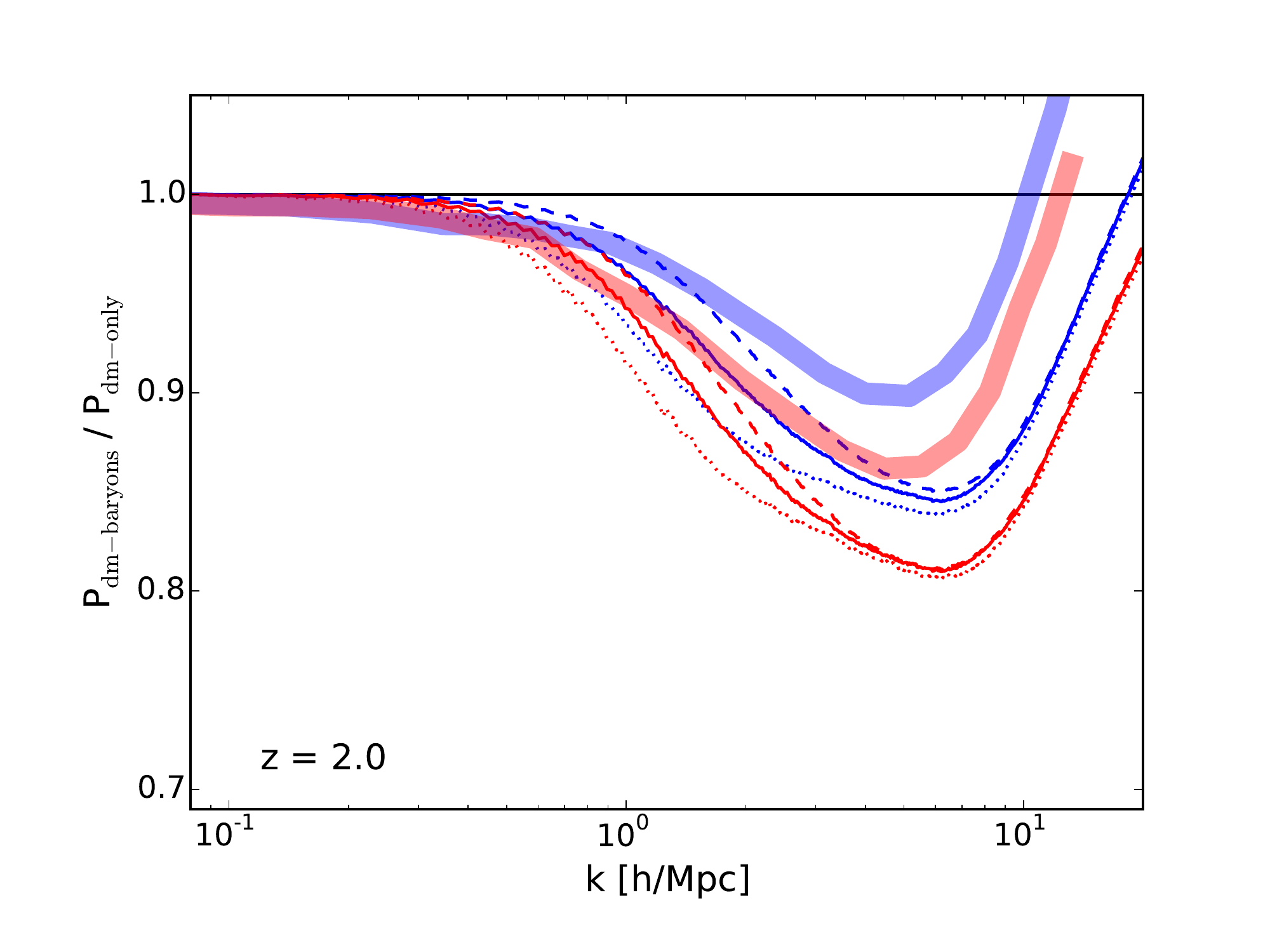}
\caption{\label{fig:CosmoOWLS}Comparison between the baryonic correction (BC) model and the cosmo-OWLS runs \citep{Brun:2013yva,Mummery:2017lcn} with two different AGN heating temperatures $\Delta T_{\rm heat}=10^8$ K (blue) and  $\Delta T_{\rm heat}=10^{8.5}$ K (red). The BC parameters $M_c$ and $\mu$ are tuned to match the gas and stellar fraction of Cosmo-OWLS shown in the top-left panel (solid and dashed band). Regarding the gas ejection parameter, we assume three different cases, $\theta_{\rm ej}=3,4,6$. Note that the fits deteriorate substantially for $\theta_{\rm ej}$ below 3 or above 6. The stellar parameters are set to $\eta_{\rm star}=0.3$, $\eta_{\rm cga}=0.5$ for the blue and $\eta_{\rm star}=0.3$, $\eta_{\rm cga}=0.6$ for the red case. The other panels show a comparison of the power spectra predicted by the BC model and measured from Cosmo-OWLS at redshifts 0, 1, and 2. The growing differences towards small scales and high redshifts are most likely due an evolving stellar fraction in the cosmo-OWLS runs.}}
\end{figure}

\subsection{Cosmo-OWLS}
The cosmo-OWLS hydrodynamical runs \citep{Brun:2013yva} are built upon the original OWLS covering a larger simulation volume and relying on updated cosmological parameters from {\tt WMAP7} \citep{Komatsu:2010fb} and {\tt Planck} \citep{Planck:2015xua}. One of the main additional benefits of Cosmo-OWLS suite is that it investigates different AGN heating temperatures, resulting in runs with different gas fractions that can be compared to X-ray observations.

The goal of this appendix is to establish how well the BC model is able to predict the power spectrum of the cosmo-OWLS runs, provided the model takes the cosmo-OWLS baryon fractions as an input. In the top-left panel of Fig.~\ref{fig:CosmoOWLS} we show both the gas and total stellar fractions of cosmo-OWLS for two different AGN heating temperatures $\Delta T_{\rm heat}=10^8$ K (blue) and $\Delta T_{\rm heat}=10^{8.5}$ K (red) as broad solid and dashed bands. The best-fitting BC models for the values $\theta_{\rm ej}=3,4,6$ are shown as blue and red lines. We have checked that choosing values for $\theta_{\rm ej}$ smaller than 3 or larger than 6 leads to an overall decrease of the quality of the fit.

The predicted power spectra of the BC model are shown in the other panels of Fig.~\ref{fig:CosmoOWLS}. At redshift zero (top right) they match the power spectra of cosmo-OWLS up to a few percent. This confirms the good agreement obtained with OWLS (see main text). The bottom panels compare BC model predictions with Cosmo-OWLS for the redshifts 1 and 2. While at wave modes below $k\sim5$ h/Mpc the agreement is similar to the case at redshift zero, it becomes worse for higher wave numbers. This is most probably a consequence of growing stellar fractions in Cosmo-OWLS towards high redshifts, leading to a more pronounced upturn of the power spectrum at high wave numbers.

Finally, note that the power suppression of the cosmo-OWLS run with $\Delta T_{\rm heat}=10^8$ K is not as strong as the one from OWLS (see Fig.~\ref{fig:OWLS}) although both simulations are based on the same AGN feedback implementation. The most likely reason for the apparent difference is that OWLS relies on {\tt WMAP3} and cosmo-OWLS on {\tt WMAP7} parameters. The higher cosmic baryon fraction $f_b=\Omega_b/\Omega_m$ of {\tt WMAP3} is likely to result in a significantly stronger suppression effect.

\begin{figure}[tbp]
\center{
\includegraphics[width=.49\textwidth,trim=0.3cm 0.6cm 1.0cm 0.9cm,clip]{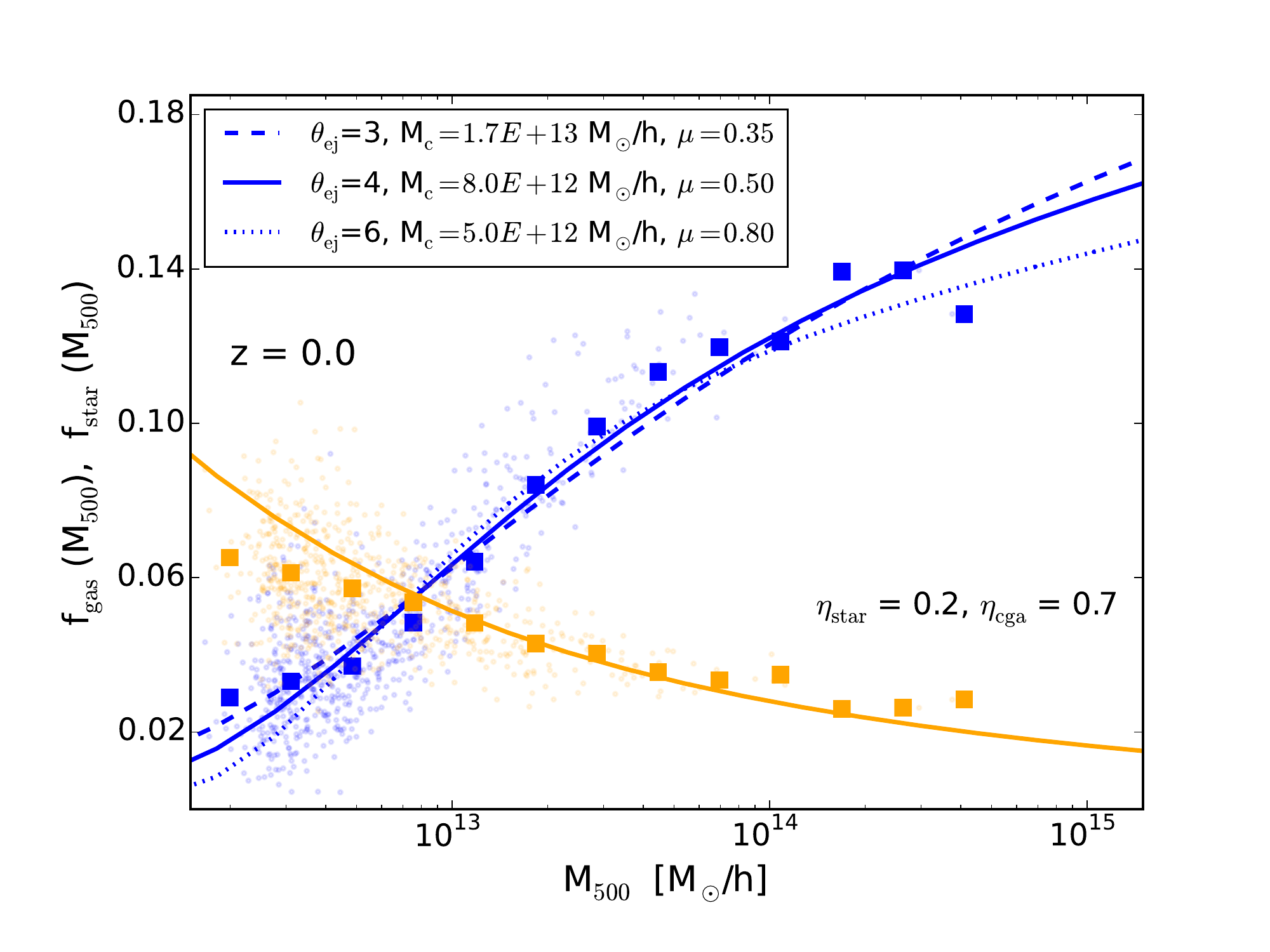}
\includegraphics[width=.49\textwidth,trim=0.3cm 0.6cm 1.0cm 0.9cm,clip]{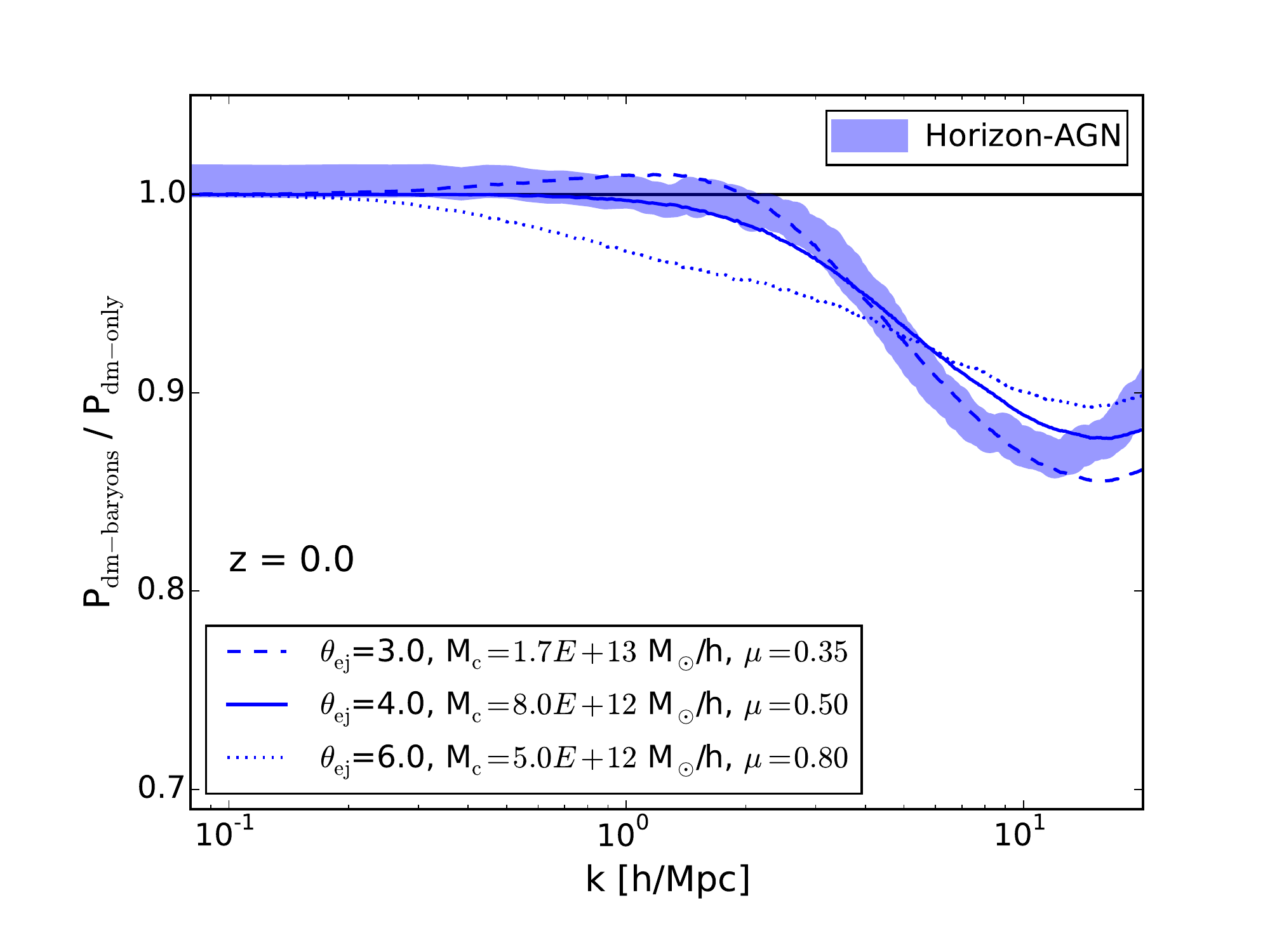}\\
\includegraphics[width=.49\textwidth,trim=0.3cm 0.6cm 1.0cm 0.9cm,clip]{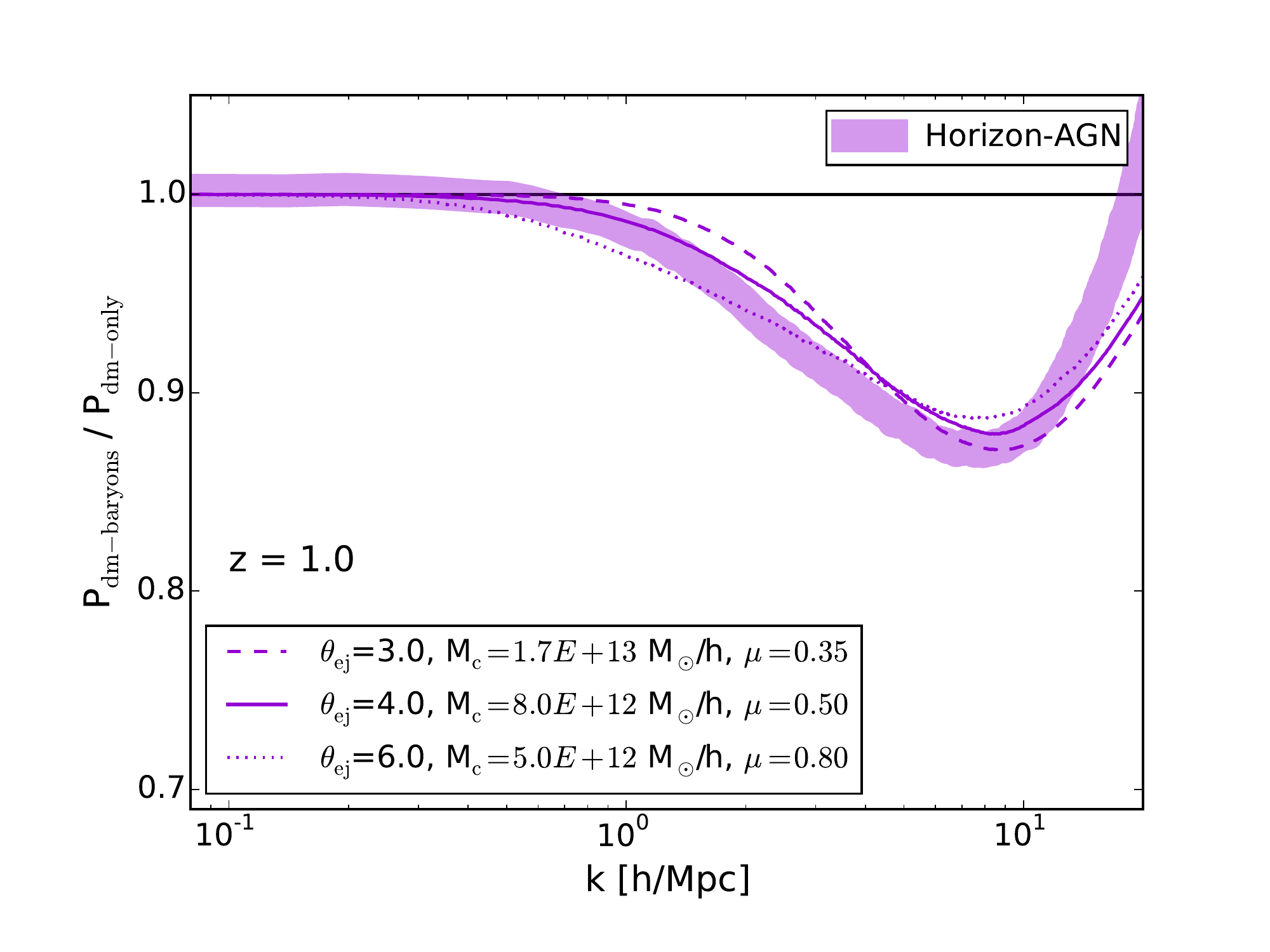}
\includegraphics[width=.49\textwidth,trim=0.3cm 0.6cm 1.0cm 0.9cm,clip]{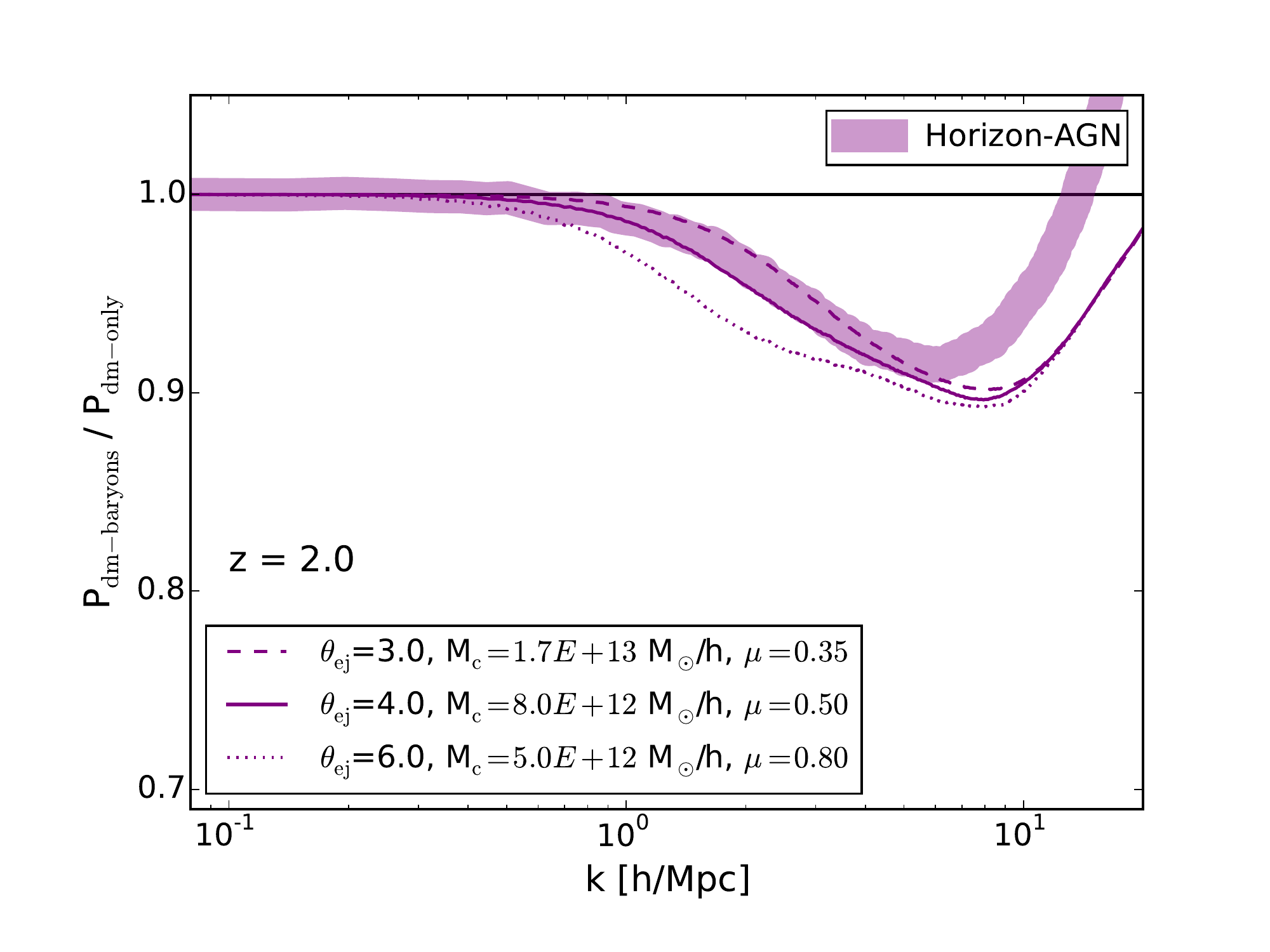}
\caption{\label{fig:HAGN}Comparison between the BC model and the Horizon-AGN simulation \citep{Dubois:2014lxa,Dubois:2016aaa,Chisari:2018prw}. The BC parameters $M_c$ and $\mu$ are tuned to match the gas and stellar fraction of Horizon-AGN shown in the top-left panel (blue and yellow dots). Regarding the gas ejection parameter, we assume three different cases, $\theta_{\rm ej}=3,4,6$. Note that the fits deteriorate substantially for $\theta_{\rm ej}$ below 3 or above 6. The other panels show a comparison of the power spectra predicted by the BC model and measured from Horizon-AGN at redshifts 0, 1, and 2.}}
\end{figure}

\subsection{Horizon-AGN}
The Horizon-AGN simulations \citep{Dubois:2014lxa,Dubois:2016aaa} are performed with the adaptive mesh code {\tt RAMSES} \citep{Teyssier:2001cp} and consist of full hydrodynamical runs with $L=100$ Mpc/h and with cosmological parameters from {\tt WMAP7} \citep{Komatsu:2010fb}. Recently, \citet{Chisari:2018prw} studied the impact of baryons on the power spectrum of the Horizon-AGN runs with and without AGN feedback.

Fig.~\ref{fig:HAGN} illustrates how the BC model compares to the Horizon-AGN simulations. In the top-left panel, we show the gas and stellar fractions from Horizon-AGN, where the small blue and orange dots are results from individual haloes while the large symbols correspond to binned average values. The best fitting BC model for the parameter values $\theta_{\rm ej}=3,4,6$ are shown as blue and orange lines. We have checked that models with $\theta_{\rm ej}$ below 3 or above 6 do not match the gas and stellar fractions of Horizon-AGN equally well.

The predicted power spectra of the BC models are shown in the top-right and bottom panels of Fig.~\ref{fig:HAGN}. At redshift zero very good agreement is obtained for the case of $\theta_{\rm ej}=3$ compared to the simulation results. At higher redshifts, the results do not match equally well. This could be due to a redshift dependence of the gas and stellar fractions of the Horizon-AGN simulations that is not accounted for in the BC model. However, note that the BC model predictions are within a few percent from the simulation at all redshifts investigated.

\subsection{Illustris-TNG}
The Illustris-TNG suite \citep{Springel:2017tpz,Pillepich:2017fcc,Pillepich:2017jle} consists of magneto-hydrodynamical simulations performed with the {\tt AREPO} code \citep{Springel:2010aaa} assuming {\tt Planck} cosmological parameters. The simulations cover several different setups in terms of resolution and box-length. Here we focus on the $L=100$ Mpc/h run since this is the only one with both published gas fraction and matter power spectrum.

\begin{figure}[tbp]
\center{
\includegraphics[width=.32\textwidth,trim=0.95cm 0.5cm 2.0cm 0.5cm,clip]{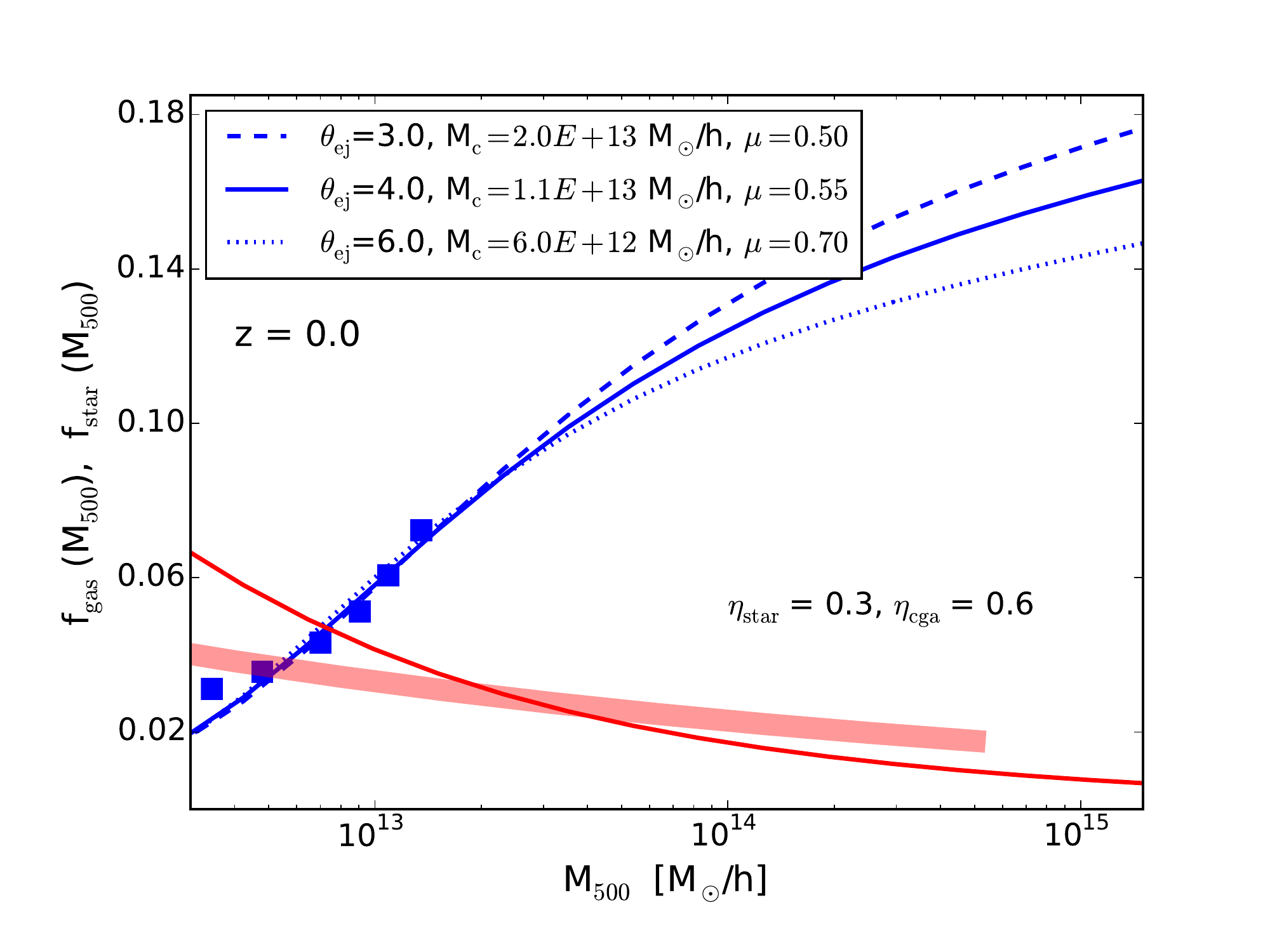}
\includegraphics[width=.32\textwidth,trim=0.95cm 0.5cm 2.0cm 0.5cm,clip]{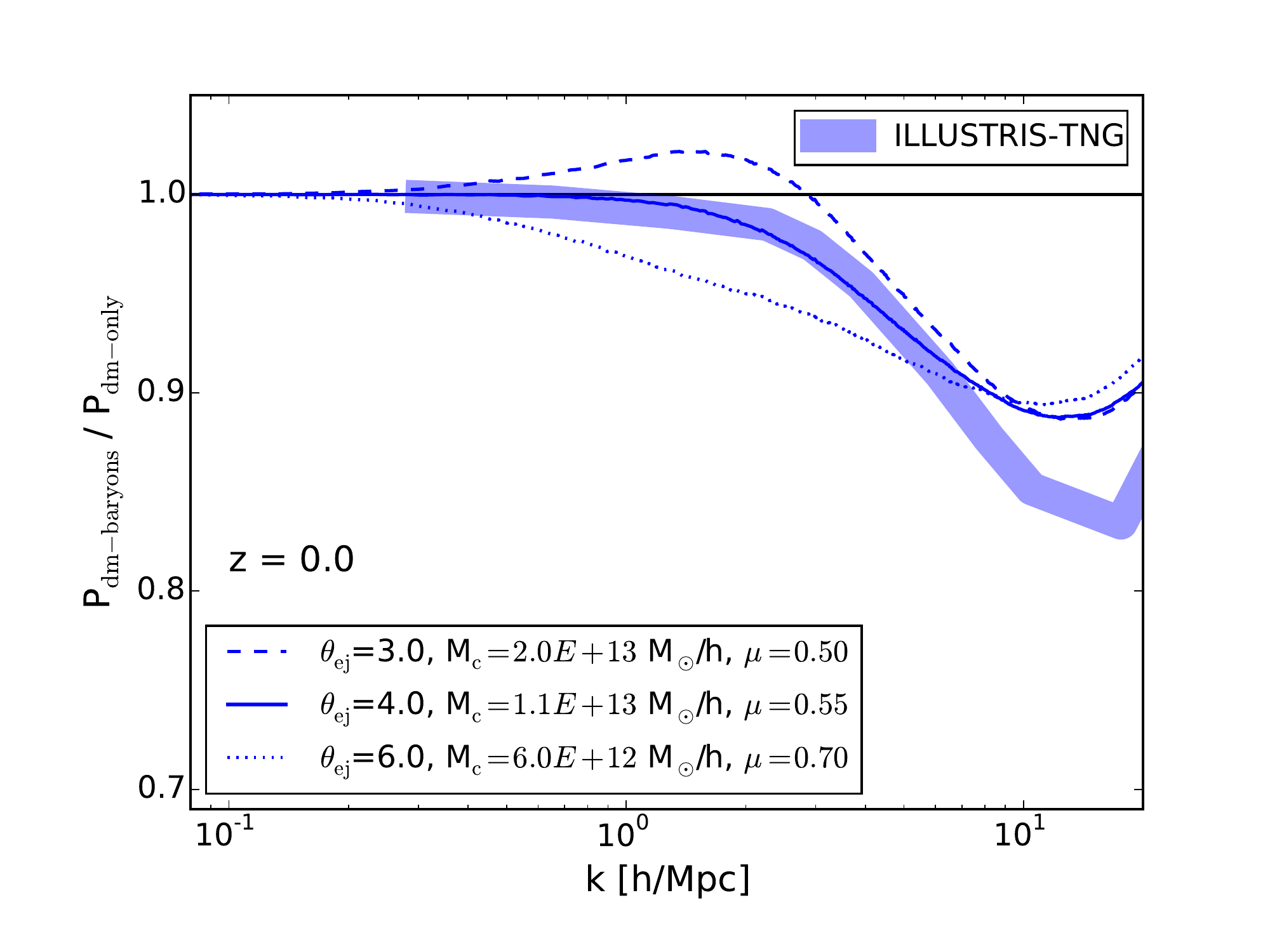}
\includegraphics[width=.32\textwidth,trim=0.95cm 0.5cm 2.0cm 0.5cm,clip]{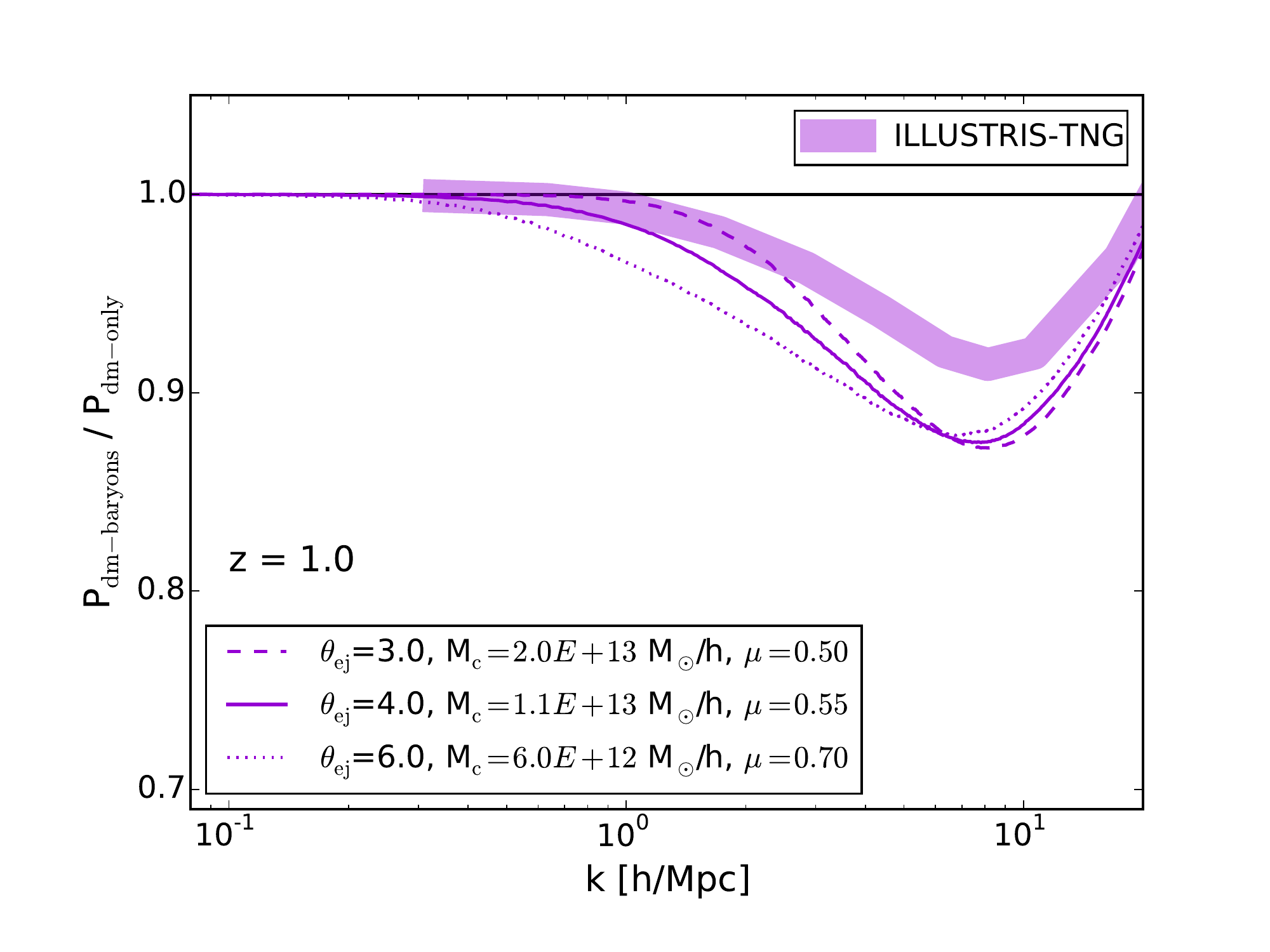}
\caption{\label{fig:TNG}Comparison between the BC model and the Illustris-TNG simulation \citep{Pillepich:2017fcc}. The BC parameters $M_c$ and $\mu$ are tuned to match the gas and stellar fraction of Illustris-TNG shown in the leftmost panel (blue dots and red band). Regarding the gas ejection parameter, we assume three different cases, $\theta_{\rm ej}=3,4,6$. Note that the fits deteriorate substantially for $\theta_{\rm ej}$ below 3 or above 6. The other panels show a comparison of the power spectra predicted by the BC model and measured from Illustris-TNG at redshifts 0 and 1.}}
\end{figure}

The left-hand-side panel of Fig.~\ref{fig:TNG} shows the gas and total stellar fractions from Illustris-TNG (blue squares and red band). Although the latter consists of a fit to the Illustris-TNG simulation with $L=300$ Mpc/h, it is shown in Ref.~\citep{Pillepich:2017fcc} to be very close to the results from the $L=100$ Mpc/h run. The BC model with best fitting values for $\theta_{\rm ej}=3,4,6$ are shown as blue and red lines.

The central and right-hand-side panels of Fig.~\ref{fig:TNG} show the matter power spectra of the Illustris-TNG simulation with $L=100$ Mpc/h for redshift zero and one \citep[blue and purple band, see Ref.][]{Springel:2017tpz}. The corresponding BC model (with parameters obtained by fitting to the gas and stellar fractions shown in the panel to the left) are plotted for comparison (blue and purple lines). The difference between the simulations and the BC model is less than five percent over all scales considered.

\section{Potential systematic uncertainties of the baryon correction model}\label{systematics}
In this appendix, we discuss potential systematic uncertainties regarding the \emph{parametrisation} and the \emph{method} of the baryonic correction (BC) model. We focus on the matter power spectrum as our prime statistic. Note that while some uncertainties are straight-forward to assess, others are much more difficult to quantify. This is due to the fact that the exact signature of baryon processes on the large-scale structure is inherently unknown.

\subsection{Systematics related to the parametrisation}
The BC model is based on a parametrisation of the stellar, gas, and dark matter components. We now specifically discuss potential systematics due to different choices regarding this parametrisation:

\begin{itemize}
\item The gas profile (defined in Eq.~\ref{rhogas}) consists of a central part of the BC model. Modifying the gas profile directly affects the shape of the power spectrum. This becomes evident when comparing the present results with ST15 \citep{Schneider:2015wta}, where the gas profile was assumed to follow a different shape. In this paper we apply a functional form that is motivated by stacked X-ray observations of Refs.~\citep{Eckert:2012aaa,Eckert:2015rlr} shown in Sec.~\ref{Xrayprofiles}. The resulting gas profile corresponds to a power law with an inner core ($r_{\rm co}$) and a truncation at the ejection radius ($r_{\rm ej}$). While $r_{\rm ej}$ consists of a free model parameter, the core radius has been fixed to the value $r_{\rm co}=0.1\times r_{200}$ (see Eq.~\ref{thco}). This is a somewhat arbitrary choice since the X-ray profiles of Sec.~\ref{Xrayprofiles} are roughly equally well fitted for values within the range $r_{\rm co}\in\left[0.05,0.15\right]\times r_{200}$ (while the fits degrade substantially outside of this range). In the left panel of Fig.~\ref{fig:systematics}, we show that reducing the core radius to $0.05\times r_{200}$ (solid red line) or increasing it to $0.15\times r_{200}$ (solid brown line) without modifying any of the other model parameters leads to changes in the power spectrum of 3-5 percent at maximum. It is important to note, however, that this shift is not directly related to the core radius itself, but it is due to the fact that increasing $\theta_{\rm co}$ leads to a \emph{decrease} of the gas fraction within $M_{500}$. Indeed, if we simultaneously modify the model parameters $M_c$ and $\mu$ so that the original gas fraction is recovered, then the resulting change of the power spectrum becomes smaller than one percent over the full range in $k$. This is shown by the dashed red and brown lines in Fig.~\ref{fig:systematics}. We therefore conclude that fixing the core radius to $0.1\times r_{200}$ does not significantly affect the large-scale clustering signal.

\item Another important element of the BC model is the truncated NFW profile (defined in Eq.~\ref{rhoNFW}) which is used to describe both the initial total profile as well as the final profile of the collisionless component. Regarding the truncation radius, we have assumed a fixed fractional value of $\varepsilon=r_{\rm tr}/r_{200}=4$ independent of halo mass and redshift. This choice is inspired by the work of \citet{Oguri:2011aaa} who applied fits to haloes from $N$-body simulations and found $\varepsilon$ to be in the range $3.6-4.3$ with little mass and redshift dependence\footnote{The numbers stated here correspond to the medium values obtained in Ref.~\citep{Oguri:2011aaa} based on fits to haloes assuming free and fixed concentrations (see Fig. 1 and subsequent text in Ref.~\citep{Oguri:2011aaa}; we have adopted the original numbers to account for the different definition regarding the virial radius).}. In the central panel of Fig.~\ref{fig:systematics}, we show that varying $\varepsilon$ within the range given above leads to a shift in the matter power spectrum of a little more than one percent with respect to the default result (see solid lines). We have checked that this shift is strongly degenerate with the parameter $\theta_{\rm ej}$. For example, the power spectra for $\varepsilon=3.6$, $\theta_{\rm ej}=3.75$ and $\varepsilon=4.3$, $\theta_{\rm ej}=4.2$ differ by less than half a percent below $k=10$ h/Mpc (see dashed lines in Fig.~\ref{fig:systematics}). This means that uncertainties related to $\varepsilon$ can be fully absorbed by allowing for small variations of the gas ejection radius.

\item Furthermore, the truncated NFW profile is modified to account for the effects of adiabatic relaxation (see Eq.~\ref{ACmodel}). We tested two different adiabatic relaxation models (as mentioned in Sec.~\ref{sec:clm}) both of them providing results that differ by less than half a percent in terms of the matter power spectrum. We therefore conclude that the adiabatic relaxation model does not induce any substantial systematical uncertainties.

\begin{figure}[tbp]
\center{
\includegraphics[height=.305\textwidth,trim=0.3cm 0.5cm 1.5cm 0.5cm,clip]{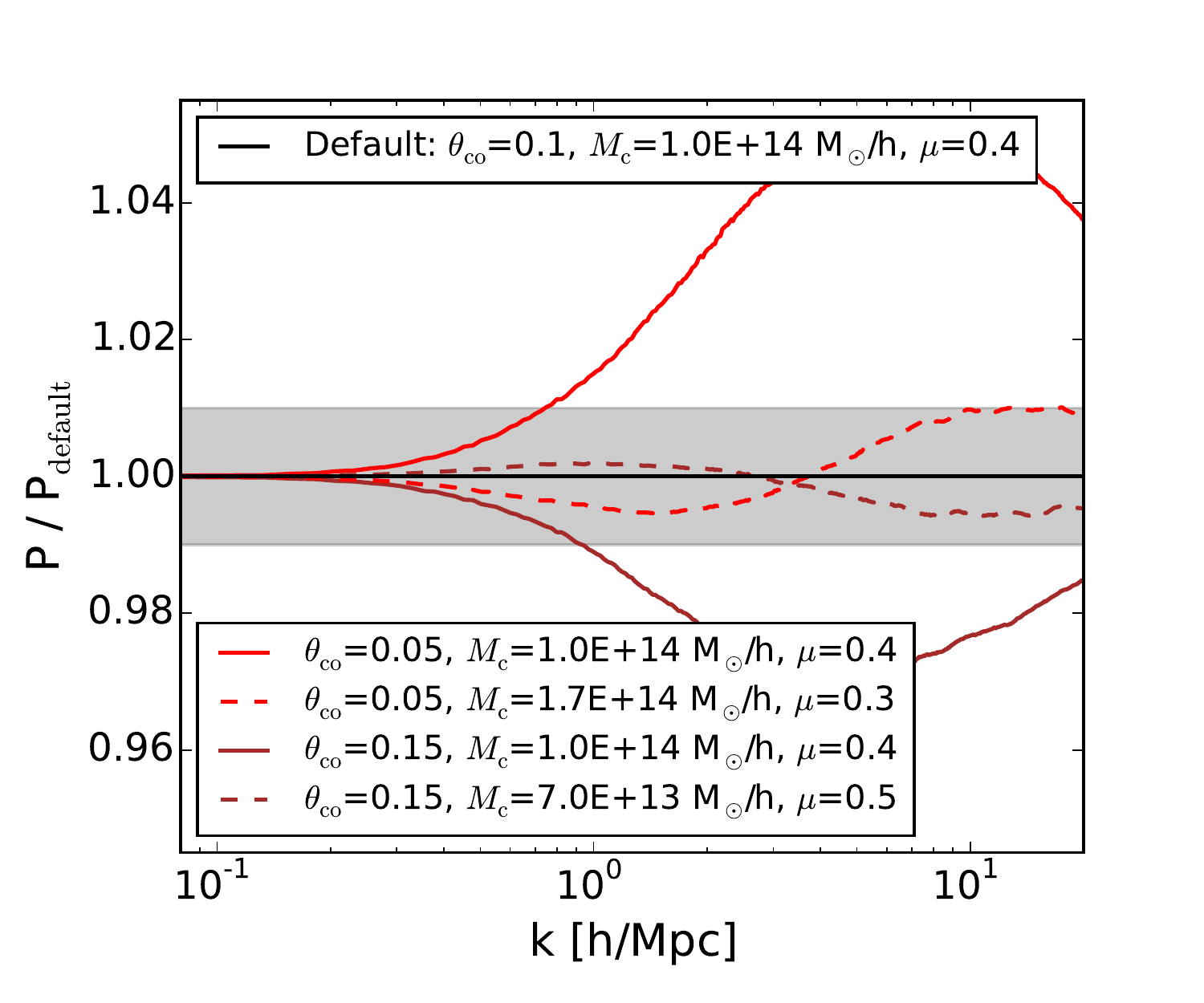}
\includegraphics[height=.305\textwidth,trim=2.25cm 0.5cm 1.5cm 0.5cm,clip]{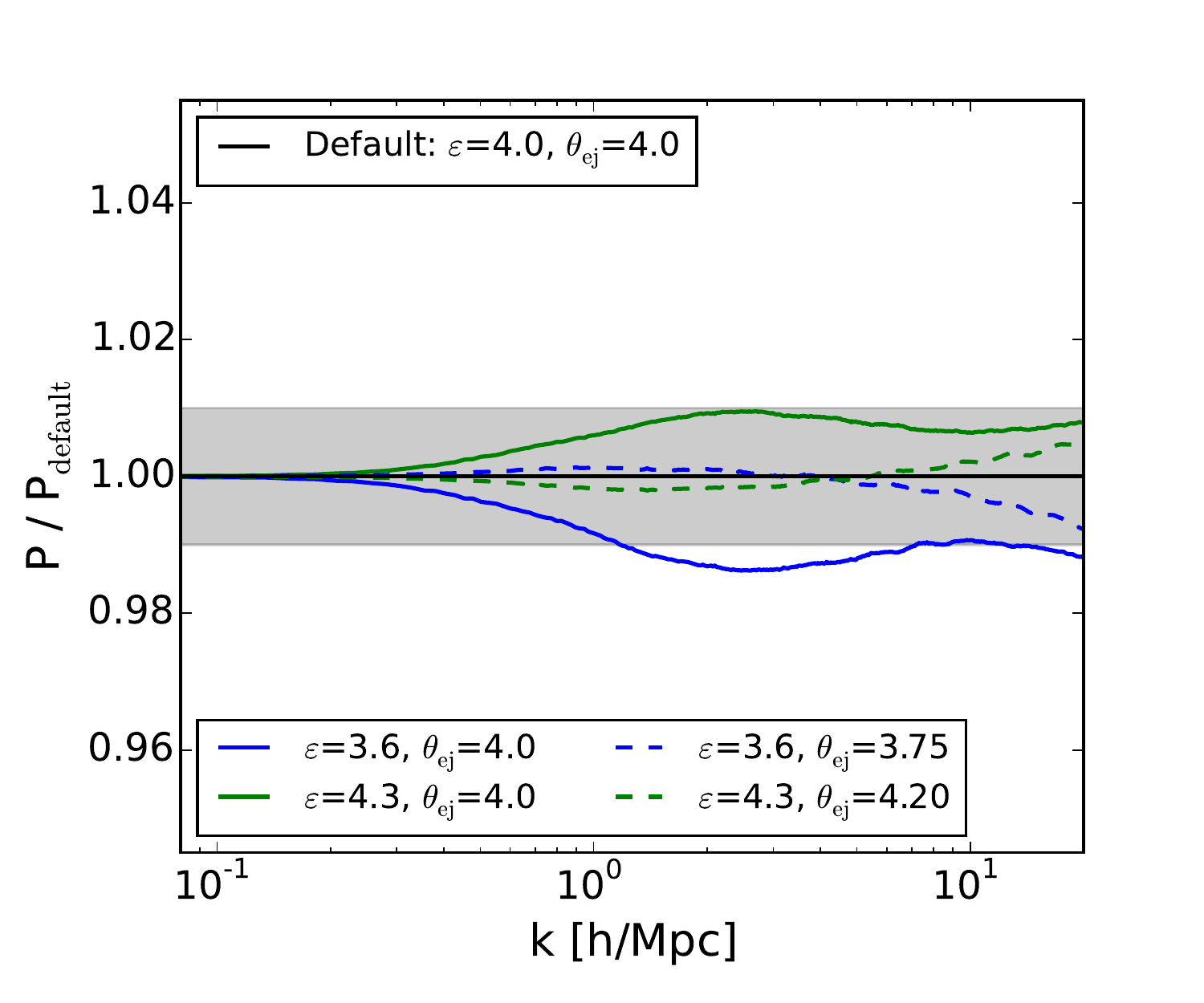}
\includegraphics[height=.305\textwidth,trim=2.25cm 0.5cm 1.5cm 0.5cm,clip]{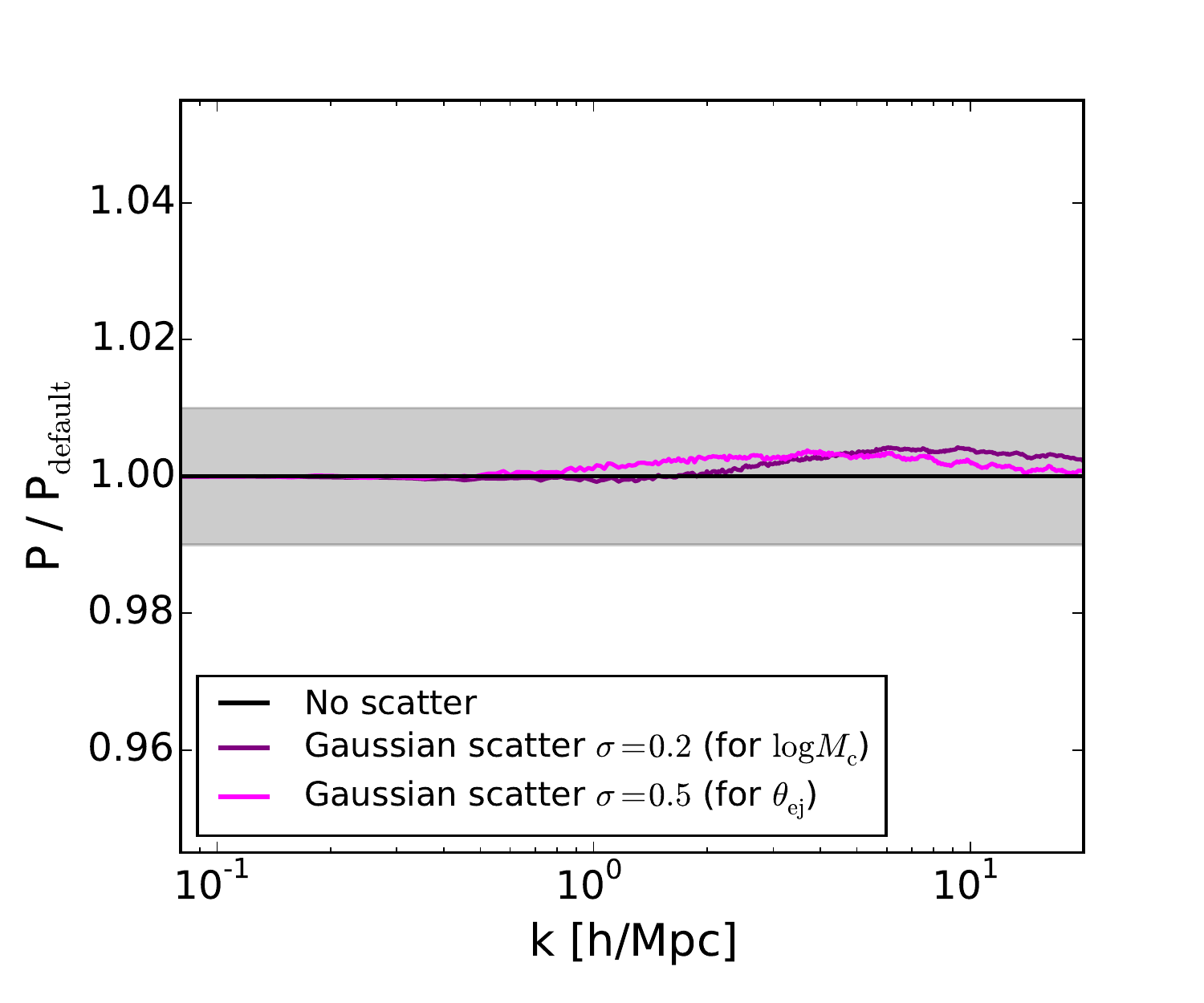}
\caption{\label{fig:systematics}Systematic uncertainties of the matter power spectrum caused by varying the core radius of the gas, the truncation radius of the NFW profile, and by introducing scatter on key model parameters. \emph{Left}: Impact on the power spectrum from changing the core radius of the gas profile (see Eq.~\ref{rhogas}) when all other parameters are kept constant (solid lines) and when $M_c$ and $\mu$ are modified in order to recover the original gas fraction at $M_{500}$ (dashed lines). \emph{Centre}: Impact on the power spectrum from varying the truncation radius of the NFW profile (parametrised by $\varepsilon$, see  Eq.~\ref{rhoNFW}) within the range of uncertainties. Solid lines refer to the case when all other parameters are kept constant, while dashed lines show the case of simultaneously adopting the maximum gas ejection radius ($\theta_{\rm ej}$). \emph{Right}: Impact on the power spectrum from including additional Gaussian scatter on the key model parameters $M_c$ and $\theta_{\rm ej}$ of the gas profile. The grey bands indicate the one percent uncertainty range.}}
\end{figure}

\item The BC model assumes satellite galaxies to follow a truncated NFW profile with identical concentration parameter than that of the dark matter component. While this might not be completely accurate, the approximation has no influence on the large-scale clustering signal because satellites consist of only a small fraction of the total collisionless matter component. Note furthermore, that adiabatic relaxation leads to a effective \emph{decrease} of the halo concentrations. This is in qualitative agreement with previous findings that satellite galaxies follow NFW-like profiles with reduced values for the concentration \citep[see Ref.][]{Budzynski:2012aaa}.

\item The stellar profile of the central galaxy is modelled using an exponentially truncated power-law (see Eq.~\ref{rhocga}), which consists of a relatively crude approximation. This is, however, not an issue, since the stellar profile does not influence the large-scale structure statistics. We have verified this by modifying the half-light radius $R_h$, effectively changing the size of the central galaxy. A two-times smaller or larger value of $R_{h}$ compared to the default value of $R_{h}=0.015\times r_{200}$ has no visible effect on the power spectrum below $k=10$ h/Mpc.

\item The stellar fractions of both the central and the satellite galaxies have a visible influence on the matter power spectrum as shown in Fig.~\ref{fig:PSvarparams}. The power-law nature of the stellar fractions (described in Eq.~\ref{stellarfraction}) is motivated by results from \citet{Moster:2012fv} and shows good agreement with results from abundance matching (see Fig.~\ref{fig:stellarfraction}). We have nevertheless tested an alternative model based on the fitting function of \citet{Behroozi:2012iw} which differs from a pure power law at large halo masses. This alternative description yields nearly indistinguishable results in terms of the matter power spectrum.

\item In the BC model, all halo profiles are uniquely determined by the corresponding halo mass and concentration. However, from both hydrodynamical simulations and direct observations we know that the stellar and gas distributions vary within haloes of the same mass and concentration. This is not surprising as each halo is subject to its individual merger history. It is straight-forward to include such effects in the BC model by simply allowing for a Gaussian scatter on individual model parameters. We specifically test the effects of scatter for $M_c$ and $\theta_{\rm ej}$, since changes in these parameters have the strongest influence on the matter clustering signal (see Fig.~\ref{fig:PSvarparams}). In the right panel of Fig.~\ref{fig:systematics}, we show that a Gaussian scatter of $\sigma_{\log_{10}M_c}=0.2$ and $\sigma_{\theta_{\rm ej}}=0.5$ leads to a shift of the power spectrum that is smaller than 0.5 percent. Hence, we conclude that adding (reasonable) scatter of the model parameters does not significantly affect the results. This finding is in agreement with ST15 who showed that including realistic scatter for the gas fractions of haloes has no influence on the clustering signal.
\end{itemize}

\subsection{Systematics related to the method and the underlying simulations}
Next to the issues related to the parametrisation, potential systematics could be induced by the method of displacing particles or by the underlying outputs of $N$-body simulations. We now discuss the largest potential uncertainties related to the model implementation:

\begin{itemize}
\item By construction the BC model only allows for spherically symmetric changes of the halo shapes. This does not mean that haloes (or any other large-scale structures) lose their original non-spherical shapes, but only that additional effects from baryons are included in a spherical symmetric way. Note that similar limitations may also be present in hydrodynamical simulations that only have thermal AGN feedback and do not account for non-spherical energy ejection. In principle, it is possible that initially strongly non-spherical feedback (such as momentum-driven AGN jets) does not fully randomise, leaving an imprint on cosmological scales.

\item The process of displacing particles can lead to slight distortions of the initial shapes of haloes within the maximum ejection radii of another halo. While this systematic is unlikely to affect the matter power spectrum, it could have some influence on higher-order statistics. Quantifying this effect is tricky and beyond the scope of the present paper.

\item Regarding the underlying $N$-body simulations, we rely on relatively small boxes with length $L=256$ Mpc/h and particle numbers $N=512^3$. Both box-size and resolution are too small to guarantee an absolute power spectrum free of systematics \citep[see e.g. Ref.][for a discussion on resolution and finite box-size effects]{Schneider:2015yka}. However, since we only use ratios of matter power spectra and weak lensing statistics, both box-size and resolution effects cancel out, leaving no systematics above the percent level. This statement relies on the investigation performed in ST15 (see their Fig.~5).
\end{itemize}

\noindent The analysis described in Appendix~\ref{systematics} reveils that potential sources of systematics related to the BC model are small compared to the current observation uncertainties. As a consequence, the predictions provided in this paper are limited by the poorly known total mass of observed X-ray clusters and galaxy groups.

\end{document}